\documentclass[a4paper,11pt,titlepage]{scrbook}

\usepackage[T1]{fontenc}
\usepackage[ansinew]{inputenc}
\usepackage{graphicx}
\usepackage{units}  
\usepackage{amsmath,amssymb,amstext,amsfonts} 
\usepackage{dsfont} 
\usepackage{slashed} 
\usepackage{multirow,multicol}
\usepackage{booktabs} 
\usepackage{float} 
\usepackage{mathrsfs} 
\usepackage{bbm}
\usepackage{feynmp} 
\usepackage{longtable} 
\usepackage{mathpazo} 
\usepackage{pifont} 

\usepackage{lscape}
\usepackage{psfrag}

\usepackage{titlesec}
\titleformat{\chapter}{\bf\Huge}{\thechapter\quad}{0em}{} 

\usepackage{pst-all}
\usepackage{color}

\usepackage{fancyhdr}
\usepackage[format=hang, labelfont=bf]{caption,subfig} 

\usepackage{hyperref}  
\hypersetup{
	pdfauthor={Roland Schieren},
	pdftitle={Discrete Symmetries in the MSSM},
	pdfkeywords={Discrete Symmetries, MSSM, Anomalies},
	colorlinks=false,   
  pdfborder={0 0 0}, 
  plainpages = false, 
  pdfdisplaydoctitle = true, 
  pdfpagelayout = {TwoColumnRight} 
}

\usepackage[numbers,sort&compress]{natbib}
\usepackage{hypernat}

\newcommand{\I}{\mathrm{i}}
\newcommand{\D}{\mathrm{d}}
\newcommand{\SU}[1]{\ensuremath{\mathrm{SU}(#1)}}
\newcommand{\SO}[1]{\ensuremath{\mathrm{SO}(#1)}}
\newcommand{\E}[1]{\ensuremath{\mathrm{E}_{#1}}} 
\newcommand{\Eqref}[1]{equation~\eqref{#1}}
\newcommand{\U}[1]{\ensuremath{\mathrm{U}(#1)}}
\newcommand{\Z}[1]{\ensuremath{\mathbbm{Z}_{#1}}} 
\newcommand{\Aut}[1]{\ensuremath{\text{Aut}(#1)}}

\newcommand{\rep}[1]{\ensuremath\boldsymbol{#1}}
\newcommand{\crep}[1]{\ensuremath\overline{\boldsymbol{#1}}}

\newcommand{\Yu}{\ensuremath{\boldsymbol{Y}_{\!\!u}}}
\newcommand{\Yd}{\ensuremath{\boldsymbol{Y}_{\!\!d}}}
\newcommand{\Ye}{\ensuremath{\boldsymbol{Y}_{\!\!e}}}

\DeclareMathOperator{\tr}{tr}
\DeclareMathOperator{\re}{Re}

\DeclareMathOperator{\rank}{rank\,}
\DeclareMathOperator{\diag}{diag}

\newcommand{\comment}[1]{}

\begin{document}

\pagestyle{empty}
\pagenumbering{Roman}
\cleardoublepage
\thispagestyle{empty}
 \phantom{x}
 \vfill

  \begin{center}
  \Huge{\textbf{Discrete Symmetries in the MSSM}}
  \end{center}
  \vfill
 
  \vspace{1cm}
  
\begin{center}
{\huge Roland Schieren}\\
\vspace{1cm} 
\textit{\large Physik Department T30, Technische Universität München,\\ James-Franck-Straße, 85748 Garching, Germany} 
\end{center}
  \vfill
\begin{center}
{\bf \noindent Abstract\\}
\end{center}
The use of discrete symmetries, especially abelian ones, in physics beyond the standard model of particle physics is discussed. A method is developed how a general, abelian, discrete symmetry can be obtained via spontaneous symmetry breaking. In addition, anomalies are treated in the path integral approach with special attention to anomaly cancellation via the Green-Schwarz mechanism. All this is applied to the minimal supersymmetric standard model. A unique $\Z4^R$ symmetry is discovered which solves the $\mu$-problem as well as problems with proton decay and allows to embed the standard model gauge group into a simple group, i.e.\ the $\Z4^R$ is compatible with grand unification. Also the flavor problem in the context of minimal flavor violation is addressed. Finally, a string theory model is presented which exhibits the mentioned $\Z4^R$ symmetry and other desirable features.
\vfill

\newpage
\mbox{   }

\pagestyle{fancy}
\tableofcontents
\newpage
\thispagestyle{empty}
\phantom{x}
\newpage

\pagenumbering{arabic}
\setcounter{page}{1}

\chapter{Introduction}

\section{Motivation}
The standard model of particle physics (SM) treats the elementary particles and their interactions on energy scales up to around \unit[100]{GeV} with high accuracy. Nevertheless, we know that the SM does not offer a complete description of the physical world. First of all, neutrinos are massless in the SM in contrast to observation.
In addition, the SM neither contains a dark matter candidate nor does it explain the baryon asymmetry in the universe. From a theoretical point of view we can ask how gravity can be merged with the SM and why the parameters of the SM take such peculiar values.

Supersymmetry~\cite{Wess:1992cp,Weinberg:2000cr} is a prominent scenario for physics beyond the SM. Especially the minimal supersymmetric standard model (MSSM) has received much attention due to the fact that
\begin{itemize}
\item it solves the hierarchy problem, i.e.\ it offers an explanation for the weakness of gravity in comparison with the weak force,
\item it exhibits gauge coupling unification~\cite{Langacker:1991an}, thereby supporting the idea of grand unification~\cite{Ross:1985ai},
\item it features radiative electroweak symmetry breaking~\cite{Ibanez:1982fr}, i.e.\ the non-vanishing vacuum expectation value of the Higgs field in the SM has a dynamical origin,
\item it provides a stable dark matter candidate,
\item it incorporates gravity if local supersymmetry is considered.
\end{itemize}
The MSSM contains roughly twice the number of degrees of freedom in comparison to the SM. Despite the mentioned good features, these new fields cause problems because they
\begin{itemize}
\item[\ding{172}] reintroduce the hierarchy problem via the $\mu$-problem,
\item[\ding{173}] violate baryon and lepton number which leads to yet unobserved proton decay,
\item[\ding{174}] contribute to flavor changing processes (flavor problem),
\item[\ding{175}] violate $CP$.
\end{itemize}
A solution of these problems seems to require more than just supersymmetry. In this thesis we will address problems \ding{172}-\ding{174} with the aid of discrete symmetries.

\section{Goals and Outline of this Study}
This thesis is organized as follows: In chapter~\ref{chap:abelian} we discuss a possible origin of abelian, discrete symmetries. We give reasons why discrete symmetries should not be imposed on a theory by hand but should have a dynamical origin. A method is developed, how a general, abelian, discrete symmetry can be obtained by spontaneous symmetry breaking~\cite{Petersen:2009ip}. Chapter~\ref{chap:anomalies} deals with anomalies, i.e.\ with the phenomenon that symmetries are broken by quantum effects. We will discuss anomalies in modern terms that is in terms of the path integral. Again, special attention is paid to anomalies of discrete symmetries. Anomaly cancellation via the Green-Schwarz mechanism is highlighted. In chapter~\ref{chap:application_MSSM} we apply the Green-Schwarz mechanism to searches for symmetries which cure some problems of the minimal supersymmetric standard model (MSSM), for example the $\mu$-problem and problems with dimension five proton decay operators. We find a unique $\Z4^R$ symmetry~\cite{Lee:2010gv,Lee:2011dy} which solves these problems and allows for grand unification by means of \SO{10}. In chapter~\ref{chap:MFV} we address another problem of the MSSM, namely the flavor problem, in the context of minimal flavor violation (MFV)~\cite{Paradisi:2008qh}. We discuss the evolution of the MFV parameters under the renormalization group, and show that they exhibit a fixed-point behavior, which relaxes the SUSY flavor problem. In chapter~\ref{chap:application_string} we apply all above tools to string theory, especially orbifold compactifications of the heterotic string~\cite{Kappl:2010yu}. We present a model with the exact spectrum of the MSSM and the previously mentioned $\Z4^R$. In the final chapter, we summarize our results.

Parts of this thesis have been published in the following articles:
\begin{itemize}
 \item P.~Paradisi, M.~Ratz, R.~Schieren, and C.~Simonetto,\\
 \emph{{Running Minimal Flavor Violation}}, \\
 Phys. Lett. \textbf{B668} (2008) 202--209, [\href{http://arxiv.org/abs/0805.3989}{arXiv:\texttt{0805.3989}}]
 \item B.~Petersen, M.~Ratz, and R.~Schieren, \\
 \emph{{Patterns of Remnant Discrete Symmetries}},\\
  JHEP \textbf{08} (2009) 111, [\href{http://arxiv.org/abs/0907.4049}{arXiv:\texttt{0907.4049}}]
 \item H.~Lee, S.~Raby, M.~Ratz, G.~Ross, R.~Schieren, K.~Schmidt-Hoberg, and P.~Vaudrevange,\\
 \emph{A unique} $\Z4^R$  \emph{Symmetry for the MSSM}, \\
 Phys. Lett. \textbf{B694} (2011) 491--495, [\href{http://arxiv.org/abs/1009.0905}{arXiv:\texttt{1009.0905}}]
 \item R.~Kappl, B.~Petersen, M.~Ratz, R.~Schieren, and P.~Vaudrevange,\\
 \emph{{String-derived MSSM vacua with residual $R$-symmetries}},\\
 \href{http://arxiv.org/abs/1012.4574}{arXiv:\texttt{1012.4574}}
 \item H.~Lee, S.~Raby, M.~Ratz, G.~Ross, R.~Schieren, K.~Schmidt-Hoberg, and P.~Vaudrevange,\\
 \emph{Discrete $R$ symmetries for the MSSM and its singlet extensions}, \\
 \href{http://arxiv.org/abs/1102.3595}{arXiv:\texttt{1102.3595}}
\end{itemize}

\clearpage
\thispagestyle{empty}

\chapter{Abelian Discrete Symmetries}
\label{chap:abelian}

\section{Origin of Discrete Symmetries}
In many theories beyond the standard model, discrete symmetries are imposed as global symmetries. Here we give some general arguments why discrete symmetries should have a dynamical origin.


\subsection{Gravity vs. Global Symmetries}
One motivation to contemplate the origin of discrete symmetries is of speculative nature. It has been argued that (quantum) gravity violates global symmetries~\cite{Kallosh:1995hi,Abbott:1989jw}. The argument is based on the no-hair theorem of black holes~\cite{Ruffini:1971xx,Bekenstein:1972ky}, which states that black holes do not carry any other charges than those associated with local symmetries. For example, if ordinary matter falls into a black hole, only its mass, electric charge and angular momentum are conserved but not baryon or lepton number. Afterwards the black hole can evaporate via Hawking radiation~\cite{Hawking:1974sw} which is supposed to be exactly thermal, and baryon and lepton number are violated. This issue is linked to the information-loss problem of black holes~\cite{Hawking:2005kf} to which a rigorous solution has not been found yet. 

In summary, there is some evidence that global symmetries are violated by gravity effects although a profound resolution of this matter requires a theory of quantum gravity which has not been formulated up to now. 

\subsection{Domain Walls}
\label{sec:domain_walls}
If a discrete symmetry is spontaneously broken down to a non-trivial subgroup, the space of vacua is not connected, i.e.\ the different ground states of the theory cannot be continuously deformed into each other. This is, for example, in contrast to the usual electroweak Higgs potential where the space of vacua is a circle at the bottom of the famous Mexican hat potential. Hence, during the phase transition from the unbroken to the broken phase, different regions in space will pick different vacua. These regions are separated by so-called domain walls~\cite{Kibble:1976sj,Abel:1995wk}. 


It was realized by Zeldovich et al.~\cite{Zeldovich:1974uw} that domain walls lead to severe cosmological problems. Domain walls greatly influence the isotropy and homogeneity of the universe due to their large energy density. This contradicts observation, e.g.\ the extreme isotropy and homogeneity of the cosmic microwave background~\cite{Komatsu:2010fb}, at least if the domain walls are produced after inflation. In addition, the equation of state of domain walls is $p=-\frac{2}{3}\rho$, i.e.\ their energy density evolves as $\rho\sim a^{-1}$~\cite{Zeldovich:1974uw} where $a$ is the scale factor of the Friedman-Robertson-Walker metric. This means that domain walls will eventually, at some stage, dominate the energy content of the universe, because the energy density of matter and radiation drops faster than $a^{-1}$, which contradicts standard big bang cosmology.

One way to avoid the problems associated with domain walls is to consider a non-exact discrete symmetry. This can be done by soft-breaking or by anomalies~\cite{Preskill:1991kd} which lead to instable domain walls. Another, possibly more attractive way to get rid of these cosmological problems, is to embed the discrete symmetry into a continuous one~\cite{Lazarides:1982tw,Preskill:1991kd}. During the phase transition from the continuous to the discrete symmetry strings, carrying a flux, form. These destabilize the domain walls which develop when the discrete symmetry is broken (cf.\ \cite{Vilenkin:1982ks} for a specific realization). 

Note that almost all internal, discrete symmetries in models beyond the standard model must be ultimately broken because they would lead to unobserved degeneracies in the particle spectrum. Hence, the present discussion of domain walls tells us that imposing an exact, global, discrete symmetry is, generically, not consistent with cosmological observations.

\section{Obtaining Abelian Discrete Symmetries}
\label{sec:obtaining_abelian_discrete_symmetries}
As discussed in the previous section, apart from esthetic reasons, one can argue that discrete symmetries should have a dynamical origin. In this section we will develop a method for obtaining a general, abelian, discrete symmetry by spontaneously breaking a $\U1^N$ symmetry~\cite{Petersen:2009ip}. In appendix~\ref{app:package} we present a \texttt{Mathematica} package which contains a routine that automatizes the subsequent calculations.

\subsection[Review: $\U1\rightarrow\Z{q}$]{Review: $\boldsymbol{\U1\rightarrow\Z{q}}$}
\label{sec:abelian_breaking_one_dim}
Consider a model with gauge group \U1 and two scalar fields $\phi$ and $\psi$ (cf.\ \cite{Krauss:1988zc}). Under the \U1 the fields transform according to 
\begin{subequations}
\begin{align}
\phi\: &\rightarrow\: e^{\I \, q\, \alpha(x)} \phi\:,\\
\psi\: &\rightarrow\: e^{\I\, \alpha(x)} \psi \: ,\label{eq:abelian_easy_traf_of_psi}
\end{align}
\end{subequations}
with $q\in\mathbb{N}$ and $\alpha(x)$ parameterizing the \U1 transformation. Hence, $\phi$ has charge $q$ and $\psi$ has charge one. Suppose now that $\phi$ acquires a vacuum expectation value (VEV). To determine which is the unbroken subgroup, we have to calculate the subgroup which leaves the VEV invariant that is
\begin{equation}
e^{\I\, q\, \alpha(x)} \phi = \phi \qquad\Leftrightarrow\qquad \alpha = 2\pi \frac{n}{q} \quad\text{with}\quad n\in\Z{}\:. 
\label{eq:single_U1_breaking_condition}
\end{equation}
Substituting this result into \Eqref{eq:abelian_easy_traf_of_psi} we see that $\psi$ transforms as
\begin{equation}
\psi\: \rightarrow\: e^{2\pi \I\, n/q} \psi \quad\text{with}\quad n\in\Z{}\:. \label{eq:single_U1_trafo_of_psi}
\end{equation}
Thus, by giving a VEV to a field with charge $q$, the \U1 gets broken down to a \Z{q}. In the following we will generalize this result.

\subsection{The General Case}
\label{sec:abelian_breaking_general_case}
Let us now consider the general case of a $\U1^N$ gauge theory with $M$ scalar fields $\phi^{(i)}\:(1\leq i \leq M)$, which will acquire VEVs, and $K$ other fields $\psi^{(j)}\:(1\leq j \leq K)$ (cf.\ \cite{Petersen:2009ip}). We will denote the charge of a field w.r.t.\ the $j^\text{th}$ \U1 factor by $q_j(\phi^{(i)})$ and $q_j(\psi^{(i)})$ respectively. Without loss of generality, we will take all \U1 charges to be integers. Accordingly, the $\phi^{(i)}$ transform as
\begin{equation}
\phi^{(i)}\: \rightarrow\: \exp\left(\I\sum_{j=1}^N q_j(\phi^{(i)})\, \alpha_j(x) \right)\phi^{(i)}
\end{equation}
where the functions $\alpha_j(x)$ parameterize the \U1 transformations. $q(\phi^{(i)})$ can be thought of as an $N$-dimensional charge vector and we can define an $M\times N$ charge matrix $Q$ of all $\phi$ fields by
\begin{equation}
Q_{ij}=q_j(\phi^{(i)})\:. \label{eq:definition_charge_matrix}
\end{equation}
Suppose now that $N>\rank Q$. In this case, there will be unbroken \U1 factors. Then we can rotate the \U1 directions by an orthogonal transformation such that all $\phi^{(i)}$ will be uncharged under $(N-\rank Q)$ \U1 factors that is after a transformation $Q\rightarrow Q O$  with $O\in\SO{N}$ the last $(N-\text{rank}\, Q)$ rows in $Q$ are zero. The corresponding \U1 factors will not be affected by the VEVs of the $\phi$ fields and we do not have to consider them any further. Therefore, without loss of generality, we will from now on consider the case $N\leq\rank Q$. Notice that in supersymmetric theories the rank of the charge matrix $Q$ cannot be maximal as $D$-flatness \cite{Wess:1992cp} requires a non-trivial solution of $\sum_i n_i \, q(\phi^{(i)})=0$ with $n_i\in\mathbb{N}_0$.

\subsubsection{The Breaking Condition}
To identify the remnant discrete symmetries after spontaneous symmetry breaking, consider the generalization of \Eqref{eq:single_U1_breaking_condition} in our simple example in the previous section,
\begin{subequations}
\begin{alignat}{2}
& \exp\left(\I\sum_{j=1}^N q_j(\phi^{(i)})\, \alpha_j(x) \right)\phi^{(i)}&&=\phi^{(i)}\\
\Leftrightarrow  \quad & \sum_{j=1}^N q_j(\phi^{(i)})\, \alpha_j(x)&&=2\pi n^{(i)} \quad\text{with}\quad n^{(i)}\in\Z{}\:. \label{eq:general_case_breaking_condition}
\end{alignat}
\end{subequations}
Equation~(\ref{eq:general_case_breaking_condition}) can be recast in matrix notation,
\begin{equation}
Q\,\alpha = 2\pi\, n \quad\text{with}\quad n\in\Z{}^M\:. \label{eq:breaking_condition_matrix_form}
\end{equation}
To solve this matrix equation over \Z{}, we will introduce a mathematical tool called the Smith normal form of $Q$.

\subsubsection{The Smith Normal Form}
\label{sec:Smith_normal_form}
Remember that $Q$ is an $M\times N$ matrix with entries in \Z{}. Such a matrix can always be diagonalized by two unimodular transformations, i.e.\ invertible matrices over \Z{}. Concretely, there exist $A\in\text{GL}(M,\Z{})$ and $B\in\text{GL}(N,\Z{})$ such that 
\begin{equation}
A\,Q\,B=D=\diag'(d_1,\ldots,d_N)=\left(
\begin{array}{cccc}
d_1 & 0 & \cdots & 0\\
0 & d_2 & & \vdots \\
\vdots & & \ddots& 0 \\
0 & & &  d_N \\
\multirow{2}{*}{\vdots} & & & 0 \\
 & & & \vdots\\
0 & 0 & \cdots &  0
\end{array}
\right)\:,
\label{eq:Smith_normal_form}
\end{equation}
and $d_i$ divides $d_{i+1}$. Here diag$'$ means that $D$ is an $M\times N$ matrix whose only non-zero elements $d_i$ are the components $D_{ii}$. There are $N$ non-zero entries due to $N\leq\text{rank}\, Q$. D is called Smith normal form~\cite{Smith1860,WA03} of $Q$. Note that the matrices $A$ and $B$ are not unique.

Note that unimodular matrices have two important properties:
\begin{enumerate}
	\item They have determinant $\pm 1$. Since we work over \Z{} (which is a ring rather than a field) the only invertible elements are $\pm 1$. For a matrix to be invertible, its determinant has to be invertible because the determinant is a homomorphism.
	\item The greatest common divisor of all matrix elements in a single row is 1. If this were not the case, the determinant would not be $\pm 1$. One can see this by looking at the Laplace expansion. The same applies to each column.
\end{enumerate}

\subsubsection{Determination of the Unbroken Subgroup}
Transformation (\ref{eq:Smith_normal_form}) allows us to rewrite \Eqref{eq:breaking_condition_matrix_form} as
\begin{equation}
A^{-1} \, D \, B^{-1}\, \alpha = 2\pi\, n\:.
\end{equation}
Due to the second property of unimodular matrices, $n'=A\, n$ still takes all values in $\Z{}^M$ if $n$ does. Hence, we can multiply this equation by $A$. Defining $\alpha'=B^{-1}\alpha$ we arrive at
\begin{equation}
\alpha'_j = 2\pi\frac{n'^{(j)}}{d_j} \qquad \text{with} \qquad 0 \leq n'^{(j)} \leq d_j-1\:. \label{eq:abelian_general_redefinition_alpha}
\end{equation} 
Hence, we see that the remnant discrete symmetry is $\Z{d_1}\times\ldots\times\Z{d_N}$. This is exactly the invariant factor form of a finite, abelian group as described in appendix \ref{sec:structure_of_finite_abelian_groups}. If some $d_i=1$, the corresponding factors are trivial and can be omitted.

Finally, we calculate the transformation of the $\psi$ fields analogously to \Eqref{eq:single_U1_trafo_of_psi} 
\begin{equation}
\psi^{(j)} \rightarrow \exp\left(\I\sum_{k,l} q_k(\psi^{(j)}) \, B_{kl} \alpha'_l \right) \psi^{(j)}= \exp\left(2\pi\I\sum_{k=1}^N q'_k(\psi^{(j)})\frac{n'^{(k)}}{d_k} \right)\psi^{(j)}
\end{equation}
with new charges 
\begin{equation}
q'_k(\psi^{(j)})=\sum_{i=1}^N q_i(\psi^{(j)}) \, B_{ik} \label{eq:abelian_general_case_new_charges}
\end{equation}
which are defined modulo $d_k$. That is, we can choose $q'_k(\psi^{(j)})\in\{0,\ldots,d_{k-1} \}$.

\subsection{An Example with two \U1 Factors}
\label{sec:example_abelian_breaking}
Let us illustrate the above procedure by a non-trivial example. Consider a $\U1_1\times\U1_2$ gauge theory with three fields obtaining VEVs and two other fields. That is, we have $N=2$, $M=3$ and $K=2$. The charges are given in table \ref{tab:example_abelian_breaking_charges}. 
\begin{table}[htbp]
\centering
\subfloat{
\begin{tabular}{ccc}
\toprule[1.3pt]
 & $\U1_1$ &$ \U1_2$ \\ \cmidrule{2-3}
$\phi^{(1)}$ & 8 & -2\\
$\phi^{(2)}$ & 4 & 2 \\ 
$\phi^{(3)}$ & 2 & 4 \\
\bottomrule[1.3pt]
\end{tabular}
}
\quad
\subfloat{
\begin{tabular}{ccc}
\toprule[1.3pt]
 & $\U1_1$ &$ \U1_2$ \\ \cmidrule{2-3}
$\psi^{(1)}$ & 1 & 3\\
$\psi^{(2)}$ & 1 & 5 \\ 
\bottomrule[1.3pt]
\end{tabular}
}
\pnode(0,0){nodeA}
\pnode(3,0){nodeB}
\ncline[linewidth=1.5pt]{->}{nodeA}{nodeB}
\naput{$\phi^{(i)}$ get VEVs}
\hspace{3cm}
\subfloat{
\begin{tabular}{ccc}
\toprule[1.3pt]
 & \Z2 & \Z6 \\ \cmidrule{2-3}
$\psi^{(1)}$ & 1 & 1\\
$\psi^{(2)}$ & 1 & 3 \\ 
\bottomrule[1.3pt]
\end{tabular}
}
\caption{Charges of the fields before and after breaking the two \U1s.}
\label{tab:example_abelian_breaking_charges}
\end{table}
In this example, we only consider scalar fields, such that we do not have to worry about anomalies. The charge matrix (cf.\ \Eqref{eq:definition_charge_matrix}) of the VEV fields is
\begin{equation}
Q = \left(
\begin{array}{cc}
8 & -2\\
4 & 2 \\ 
2 & 4 \\
\end{array}
\right)\:.
\end{equation}
The Smith normal form $D$ of $Q$ and the transformation matrices $A$ and $B$ as defined in \Eqref{eq:Smith_normal_form} are 
\begin{equation}
D=  \left(
\begin{array}{cc}
2 & 0\\
0 & 6 \\ 
0 & 0 \\
\end{array}
\right)\:,
\qquad
A=  \left(
\begin{array}{ccc}
0 & 0 & 1\\
0 & -1 & 2\\ 
0 & -3 & 2 \\
\end{array}
\right)
\qquad\text{and}\qquad
B=\left(
\begin{array}{cc}
1 & -2\\
0 & 1 
\end{array}
\right)\:.
\end{equation}
Hence, the diagonal elements of $D$ tell us that we are left with a $\Z2\times\Z6$ symmetry. The charges of the $\psi$-fields can be determined by multiplying their charge matrix by $B$ from the right (cf.\ \Eqref{eq:abelian_general_case_new_charges}),
\begin{equation}
\left(
\begin{array}{cc}
1 & 3\\
1 & 5 \\ 
\end{array}
\right)
\left(
\begin{array}{cc}
1 & -2\\
0 & 1 
\end{array}
\right)
=
\left(
\begin{array}{cc}
1 & 1\\
1 & 3 
\end{array}
\right)\:.
\end{equation}
The new charges of the $\psi$ fields are given by the rows of this matrix. 

\section{Visualization of the Breaking}
We will now provide a simple, geometrical way of envisaging the symmetry breaking pattern. First, notice that an abelian group $\Z{d_1}\times\ldots\times\Z{d_N}$ can be represented by an $N$-dimensional lattice. Each group element can be thought of as a point in the fundamental region (unit cell) of the lattice.

The breaking of a $\U1^N$ can then be visualized as follows. The \U1 charges of the VEV fields $\phi^{(i)}$ constitute an $N$-dimensional lattice. The lattice points are given by 
\begin{equation}
 \sum_{i=1}^M n_i \, q(\phi^{(i)}) \qquad\text{with}\qquad n_i\in\Z{}\:,
\end{equation}
that is by integer, linear combinations of the \U1 charges. In the broken phase, a coupling 
\begin{equation}
 \prod_{i=1}^K (\psi^{(i)})^{m_i} \qquad\text{with}\qquad m_i\in\Z{}
\end{equation}
is allowed by the discrete symmetry if $\sum_{i=1}^K m_i q(\psi^{(i)})$ is a lattice point.

The non-trivial step, in more than one dimension, is to identify the group to which the lattice corresponds. This problem is solved by the Smith normal form. In the following we will present two examples of this procedure.

\subsection{A one-dimensional Example}
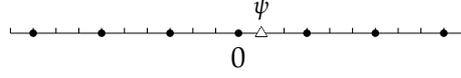
\begin{figure}[htb]
\centering
\begin{pspicture}(-3,-0.5)(3,0.5)
\psset{xunit=3mm}
\psset{yunit=3mm}
\psset{dotsize=1mm}
\psaxes[ticksize=2pt,linewidth=0.1mm,labels=none,arrowsize=0.1]{}(0,0)(-10,0)(10,0)
\pnode(0,0){zero}
\nput{270}{zero}{0}
\multido{\i=-9+3,}{7}{\psdot(\i,0)}
\pnode(1,0){psi}
\nput{90}{psi}{\footnotesize $\psi$}
\psdot[dotstyle=triangle,dotsize=0.15](1,0)
\end{pspicture}
\caption{Visualization of the breaking $\U1\to\Z{3}$. The dots represent the lattice defined by the charge of $\phi$ which is three.}
\label{fig:visualization_discrete_breaking_easy}
\end{figure}

In this subsection we will discuss the example of one \U1 breaking down to a \Z{q} (cf.\ section~\ref{sec:abelian_breaking_one_dim}). To be concrete, we will take $q=3$, i.e.\ we have a field $\phi$ with \U1 charge 3 which will obtain a VEV, and another field $\psi$ with charge 1. As mentioned above, the charge of $\phi$ defines a lattice which is, in this case, one-dimensional. The lattice is depicted by the dots in figure~\ref{fig:visualization_discrete_breaking_easy}.
The lattice spacing tells us that we have a \Z{3} unbroken subgroup. The charge of field $\psi$, whose position is marked by a triangle in figure~\ref{fig:visualization_discrete_breaking_easy}, is one. For example, this tells us that a coupling $\psi^3$ is now allowed while it was forbidden by the original \U1.

\subsection{A two-dimensional Example}
For the example in section~\ref{sec:example_abelian_breaking} this lattice is depicted in figure~\ref{fig:visualization_discrete_breaking}~\subref{fig:visualization_discrete_breaking1}. Since we have three VEV fields but only two \U1s, the charges of the $\phi$-fields are linear dependent over \Z{} and we choose $\phi^{(2)}$ and $\phi^{(3)}$ to be the basis vectors for the lattice. The lattice points are given by the dots. The charges of the $\psi$-fields are represented by triangles. 
\begin{figure}[htb]
\centering
\subfloat[Original lattice.\label{fig:visualization_discrete_breaking1}]{
\begin{pspicture}(-3,-3.5)(3,3.5)
\psset{xunit=3mm}
\psset{yunit=3mm}
\psset{dotsize=1mm}
\psset{labelsep=0.1}
\psaxes[ticksize=2pt,linewidth=0.1mm,labels=none,arrowsize=0.1]{->}(0,0)(-10,-10)(10,10)
\psdot(10,-10)
\multido{\i=4+4,\n=-10+2}{2}{\psdot(\i,\n)}
\multido{\i=-2+4,\n=-10+2}{4}{\psdot(\i,\n)}
\multido{\i=-8+4,\n=-10+2}{5}{\psdot(\i,\n)}
\multido{\i=-10+4,\n=-8+2}{6}{\psdot(\i,\n)}
\multido{\i=-8+4,\n=-4+2}{5}{\psdot(\i,\n)}
\multido{\i=-10+4,\n=-2+2}{6}{\psdot(\i,\n)}
\multido{\i=-8+4,\n=2+2}{5}{\psdot(\i,\n)}
\multido{\i=-10+4,\n=4+2}{4}{\psdot(\i,\n)}
\multido{\i=-8+4,\n=8+2}{2}{\psdot(\i,\n)}
\psdot(-10,10)
\pnode(4,2){phi2}
\nput{30}{phi2}{\footnotesize $\phi^{(2)}$}
\pnode(2,4){phi3}
\nput{30}{phi3}{\footnotesize $\phi^{(3)}$}
\pnode(10,0){xAxes}
\nput{300}{xAxes}{\begin{footnotesize}$\U1_1$\end{footnotesize}}
\pnode(0,10){yAxes}
\nput{130}{yAxes}{\begin{footnotesize}$\U1_2$\end{footnotesize}}
\psdot[dotstyle=triangle](1,3)
\psdot[dotstyle=triangle,fillcolor=black](1,5)
\end{pspicture}
}
\hspace{2cm}
\subfloat[Orthogonal lattice.\label{fig:visualization_discrete_breaking2}]{
\begin{pspicture}(-3,-3.5)(3,3.5)
\psset{xunit=3mm}
\psset{yunit=3mm}
\psset{dotsize=1mm}
\psset{labelsep=0.1}
\psaxes[ticksize=2pt,linewidth=0.1mm,labels=none,arrowsize=0.1]{->}(0,0)(-10,-10)(10,10)
\multido{\i=-10+2}{11}{\psdot(\i,6)}
\multido{\i=-10+2}{11}{\psdot(\i,0)}
\multido{\i=-10+2}{11}{\psdot(\i,-6)}
\dotnode[dotstyle=triangle](1,1){psi1}
\nput{10}{psi1}{\footnotesize $\psi^{(1)}$}
\dotnode[dotstyle=triangle,fillcolor=black](1,3){psi2}
\nput{70}{psi2}{\footnotesize $\psi^{(2)}$}
\end{pspicture}
}
\caption{Visualization of the breaking mechanism. The $\phi$-fields define a lattice \subref{fig:visualization_discrete_breaking1} which is orthogonalized \subref{fig:visualization_discrete_breaking2} by calculating the Smith normal form.}
\label{fig:visualization_discrete_breaking}
\end{figure}
\noindent Our procedure described in section~\ref{sec:abelian_breaking_general_case} amounts to transforming the original lattice into an orthogonal one. \comment{transformation noch genauer erklären.}

\section{Generalization}
The starting point of the method we described in section~\ref{sec:obtaining_abelian_discrete_symmetries} is a $\U1^N$ theory. This assumption can be relaxed to other cases which we treat below.

\subsection{Breaking \Z{N}}
In this subsection we will discuss the case where we break a general product of abelian groups $G=\Z{d_1}\times\cdots\times\Z{d_M}\times\U1^N$. The important trick is now to write $G$ as $\U1^{M+N}=\U1_1\times\cdots\times\U1_{M+N}$. Additionally, we introduce an auxiliary field $\Phi^{(i)}$ for each $\Z{d_i}$ which is only charged under $\U1_i$ with charge $d_i$, i.e.\ $q_j(\Phi^{(i)})=d_i\delta_{ij}$. The $\Phi^{(i)}$-fields are then treated as if they would get a VEV.

Now it is possible to use the method of section~\ref{sec:obtaining_abelian_discrete_symmetries} to calculate the unbroken subgroup. We give an example of this below.

\subsubsection{Example}
Consider a model with a $\Z6\times\U1$ symmetry. We will assume a field $\phi$ having \Z6 charge two and \U1 charge two to obtain a VEV. As described above, we enlarge the symmetry group to $\U1_1\times\U1_2$ and introduce an auxiliary VEV field $\Phi$ which is only charged under the first \U1 with charge six. Thus, we have a charge matrix (cf.\ \Eqref{eq:definition_charge_matrix})
\begin{equation}
 Q=\begin{pmatrix}6&0\\2&2\end{pmatrix} \quad \text{whose Smith normal form is} \quad \begin{pmatrix}6&0\\0&2\end{pmatrix}\:.
\end{equation}
From the Smith normal form we can read off that we are left with a $\Z6\times\Z2$ symmetry.

\subsection{Breaking $R$-symmetries}
In $\mathcal{N}=1$ supersymmetry one can have a discrete or continuous, abelian symmetry $G_R$ which does not commute with the supersymmetry generators. The superspace coordinates $\theta$ transform with charge $q_\theta$ under $G_R$, i.e.\ $\theta\to e^{\I\,q_\theta\,\alpha}\theta$. Hence, the superpotential $\mathscr{W}$ carries $R$-charge $2\,q_\theta$ which follows from $\mathscr{L}\supset\int\D^2\theta\,\mathscr{W}$. A possible origin of an $R$-symmetry is Lorentz invariance in compactified dimensions~\cite{Lebedev:2007hv}.

As in the previous subsection, we introduce an auxiliary field $\Omega$. To this field we assign the charge of the superpotential, e.g.\ in the case $G_R=\U1$ $\Omega$ has charge $2\,q_\theta$. If some fields which are charged under $G_R$ attain VEVs, we can go through the procedure described in section~\ref{sec:obtaining_abelian_discrete_symmetries}, treating $\Omega$ as not getting a VEV. Then $\Omega$ keeps track of the transformation properties of the superpotential, and we can figure out which remnant symmetries are $R$-symmetries.

\subsubsection{Example}
We discuss a $\Z4^R\times\U1$ symmetry. Suppose the superpotential has charge two under $\Z4^R$. Hence, we introduce an auxiliary field $\Omega$ which is uncharged under the \U1 and has charge two under $\Z4^R$. 

Now fields with charge (1,3) and (2,2) obtain VEVs. This breaks $\Z4^R\times\U1\to\Z4$ where $\Omega$ has charge two. Thus, the remnant symmetry is a $\Z4^R$ under which the superpotential transforms with charge two.


\section{Elimination of Redundancies}
\label{sec:redundandies}
When dealing with discrete, abelian symmetries, it can happen that the symmetry is smaller than one would naively expect, depending on the field content. This phenomenon does not occur for continuous symmetries.

Exemplary, consider a \Z{6} symmetry with just one field $\phi$ having charge 4. That is, a \Z{6} transformation looks like (cf.\ \Eqref{eq:general_ZN_trafo})
\begin{equation}
 \phi \to e^{2\pi\I \, 4\,  \frac{n}{6}} \phi = e^{2\pi\I \, 2\,  \frac{n}{3}} \phi\:.
\label{eq:redundancies_easy_example}
\end{equation}
Naively, one would expect the parameter $n$ to take values $0,\ldots,5$ corresponding to the six elements of \Z{6}. However, the last term in \Eqref{eq:redundancies_easy_example} teaches that only $n=0,1,2$ give rise to distinct transformations. Hence, the true symmetry of the system is a \Z{3} where $\phi$ has charge 2.

\subsection{The General Case I}
\label{sec:redundancies_general_case_I}
Consider now a discrete, abelian symmetry $G=\Z{d_1}\times\ldots\times\Z{d_N}$. We denote the field content by fields $\psi^{(i)}$ and their \Z{d_j}-charges by  $q_j(\psi^{(i)})$. 

As described above, depending on the field content, it may happen that the true symmetry of the system is smaller than $G$ because not all symmetry transformations are independent. This situation can be identified by looking at equivalent charge assignments of the fields $\psi^{(i)}$ under $G$. To do so, it is most convenient to bring $G$ in the the elementary divisor form, i.e.\ $d_i=p_i^{e_i}$ with some prime $p_i$ (see appendix \ref{sec:structure_of_finite_abelian_groups} for details). How to obtain these different charge assignments is explained in appendix \ref{sec:automorphisms_abelian_groups}, and in appendix~\ref{app:package} a \texttt{Mathematica} package is presented which calculates all possible charge assignments. There are two cases which can occur if $G$ is too large:
\begin{enumerate}
\item In one \Z{d_j} subgroup the GCD of all charges $q_j(\psi^{(i)})$ and $d_j$ is greater than one. In this case, we can divide all charges and $d_j$ by this GCD and end up with a smaller symmetry.
\item There are two subgroups \Z{d_j} and \Z{d_k} with $d_j=d_k$ and $q_j(\psi^{(i)})=q_k(\psi^{(i)})$ for all  $\psi^{(i)}$, i.e.\ they are equal. In this case, we can delete one of these subgroups.
\end{enumerate}

\subsubsection{Examples}
Let us illustrate this procedure with two examples. The first example is given in table~\ref{tab:example_redundancies1}. Table~\subref{tab:example_redundancies1a} shows the starting point while~\subref{tab:example_redundancies1b} displays an equivalent charge assignment. The setup in~\subref{tab:example_redundancies1b} belongs to the first case since we can divide the charges of the \Z4 symmetry by two and arrive at a \Z2. The result is given in~\subref{tab:example_redundancies1c}.
\begin{table}[htbp]
\centering
\subfloat[]{\label{tab:example_redundancies1a}
\begin{tabular}{ccc}
\toprule[1.3pt]
& \Z4 & \Z8 \\ \cmidrule{2-3}
$\psi^{(1)}$ & 2 & 4\\
$\psi^{(2)}$ & 3 & 3 \\ 
\bottomrule[1.3pt]
\end{tabular}
}
\qquad
\subfloat[]{\label{tab:example_redundancies1b}
\begin{tabular}{ccc}
\toprule[1.3pt]
& \Z4 & \Z8 \\ \cmidrule{2-3}
$\psi^{(1)}$ & 2 & 4\\
$\psi^{(2)}$ & 2 & 3 \\ 
\bottomrule[1.3pt]
\end{tabular}
}
\qquad
\subfloat[]{\label{tab:example_redundancies1c}
\begin{tabular}{ccc}
\toprule[1.3pt]
& \Z2 & \Z8 \\ \cmidrule{2-3}
$\psi^{(1)}$ & 1 & 4\\
$\psi^{(2)}$ & 1 & 3 \\ 
\bottomrule[1.3pt]
\end{tabular}
}
\caption{An example of a symmetry group which is actually too large. All three tables show physically equivalent symmetry groups. However,~\subref{tab:example_redundancies1c} shows the smallest possible choice.}
\label{tab:example_redundancies1}
\end{table}

\noindent Table~\ref{tab:example_redundancies2} shows a second example which is the outcome of the example in section~\ref{sec:example_abelian_breaking}. Subtable~\subref{tab:example_redundancies2a} gives an abelian symmetry in the invariant factor form (cf.\ appendix~\ref{sec:structure_of_finite_abelian_groups}). The elementary divisor form is presented in~\subref{tab:example_redundancies2b} from which we infer that we are dealing with case two. Hence we can forget about one \Z2 factor in~\subref{tab:example_redundancies2b} and arrive at a smaller symmetry.
\begin{table}[htbp]
\centering
\subfloat[]{\label{tab:example_redundancies2a}
\begin{tabular}{ccc}
\toprule[1.3pt]
& \Z2 & \Z6 \\ \cmidrule{2-3}
$\psi^{(1)}$ & 1 & 1\\
$\psi^{(2)}$ & 1 & 3 \\ 
\bottomrule[1.3pt]
\end{tabular}
}
\qquad
\subfloat[]{\label{tab:example_redundancies2b}
\begin{tabular}{cccc}
\toprule[1.3pt]
& \Z2 & \Z2 & \Z3 \\ \cmidrule{2-4}
$\psi^{(1)}$ & 1 & 1 & 2\\
$\psi^{(2)}$ & 1 & 1 & 0\\ 
\bottomrule[1.3pt]
\end{tabular}
}
\qquad
\subfloat[]{\label{tab:example_redundancies2c}
\begin{tabular}{cc}
\toprule[1.3pt]
 & \Z6 \\ \cmidrule{2-2}
$\psi^{(1)}$  & 1\\
$\psi^{(2)}$  & 3 \\ 
\bottomrule[1.3pt]
\end{tabular}
}
\caption{The example of section~\ref{sec:example_abelian_breaking} continued. All three tables show physically equivalent symmetry groups. However,~\subref{tab:example_redundancies2c} shows the smallest possible choice.}
\label{tab:example_redundancies2}
\end{table}

\subsection{The General Case II}
The preceding subsection illustrated the elimination of redundancies in terms of equivalent charge assignments. Although this is a good way to think about these redundancies, for practical purposes it is not very efficient because it involves scanning over the automorphisms of the symmetry group which can be very large (see \Eqref{eq:number_automorphisms}). In this subsection we will present an alternative way of bringing a given discrete, abelian symmetry into its smallest form~\cite{Petersen:2009ip}. In addition, the procedure will bring any symmetry in a canonical form, namely the invariant factor form.

As before, consider a discrete, abelian symmetry $G=\Z{d_1}\times\ldots\times\Z{d_N}$. We do not make any assumptions about the $d_i$. The field content is denoted by $K$ fields $\psi^{(i)}$, $i=1,\ldots,K$, and their \Z{d_j}-charges by $q_j(\psi^{(i)})$. We define a charge matrix by $(Q_\psi)_{ij}=q_j(\psi^{(i)})$ which is a $K\times N$ matrix with integer entries. The general idea is to look at the Smith normal form of $Q_\psi$. However, to get a meaningful result, we need to bring $G$ into the form $\Z{d}\times\ldots\times\Z{d}$ where $d$ is the LCM of the $d_i$. \comment{Noch genauer erklären warum das funtioniert??} Enlarging the symmetry implies a rescaling of the charge matrix as $(Q'_\psi)_{ij}=\frac{d}{d_j}\,q_j(\psi^{(i)})$. The fields now transform according to
\begin{equation}
 \psi^{(i)}\to \exp\left( 2\pi\I\sum_{j=1}^N (Q'_\psi)_{ij}\, \alpha_j \right) \psi^{(i)} \quad\text{where}\quad \alpha_j=\frac{n^{(j)}}{d} \qquad \text{with} \quad 0\leq n^{(j)} \leq d-1\:. \label{eq:abelian_redundancies_trafo}
\end{equation}
$Q'_\psi$ can be brought into Smith normal form $S$ by unimodular transformations $E\in\text{GL}(K,\Z{})$ and $F\in\text{GL}(N,\Z{})$ (cf.\ \Eqref{eq:Smith_normal_form})
\begin{equation}
 E\, Q'_\psi\, F = S = \diag'(s_1,\ldots,s_k) \quad \text{where} \quad k=\min(K,N)\:.
\end{equation}
Recall that $s_j$ divides $s_{j+1}$. If $\rank Q'_\psi < k$, some $s_j$ might vanish. The corresponding rows of $Q'_\psi$ do not have to be considered further. This yields the transformation behavior
\begin{align}
 \psi^{(i)} \to & \exp\left( 2\pi\I\sum_{j,m,p}E^{-1}_{ij} \, S_{jm} \, F^{-1}_{mp} \, \alpha_p \right) \psi^{(i)} \nonumber\\
 = & \exp\left( 2\pi\I \sum_{j,p} E^{-1}_{ij} \, s_j \, F^{-1}_{jp} \frac{n^{(p)}}{d} \right) \psi^{(i)} \:.
\end{align}
As in \Eqref{eq:abelian_general_redefinition_alpha}, we are allowed to define $n'^{(j)}=\sum_i F^{-1}_{ji}\,n^{(i)}$. Hence, we have rewritten \Eqref{eq:abelian_redundancies_trafo} as
\begin{equation}
 \psi^{(i)} \to \exp\left(2\pi\I \sum_j E^{-1}_{ij} \frac{n^{(j)}}{d/s_j} \right) \psi^{(i)}\:.
\end{equation}
This tells us that we have rearranged $G$ as $\Z{d'_1}\times\ldots\times\Z{d'_k}$ where the $d'_i$ is the numerator of the (reduced) fraction $\frac{d}{s_i}$ and $d'_{i+1}$ divides $d'_{i}$. If a $d'_i$ is equal to one, this factor can be omitted. The new charges of the fields, $q''_j(\psi^{(i)})$, are encoded in the matrix $E^{-1}$ 
\begin{equation}
 q''_j(\psi^{(i)})= E^{-1}_{ij}\:.
\end{equation}
An obvious consequence of our discussion is that after the simplification there are at most as many \Z{d_i} factors as fields. In appendix~\ref{app:package} we present a \texttt{Mathematica} package which contains a routine that automatizes the above described method.

\subsubsection{Examples}
With this alternative method we will recalculate the examples in section~\ref{sec:redundancies_general_case_I}. Let us start with the field content of a $\Z4\times\Z8$ symmetry defined in table~\ref{tab:example_redundancies1a}. The charge matrix of the enlarged $\Z8\times\Z8$ and the Smith normal form are
\begin{equation}
 Q'_\psi=\begin{pmatrix}4&4\\6&3\end{pmatrix} = 
\begin{pmatrix}4&1\\3&1\end{pmatrix}
\begin{pmatrix}1&0\\0&12\end{pmatrix}
\begin{pmatrix}-2&1\\1&0\end{pmatrix}
=E^{-1}\, S\, F^{-1}\:. \label{eq:example_redundacies_1}
\end{equation}
From this we can read off $s_1=1$ and $s_2=12$ from which we infer $d'_1=8$ and $d'_2=2$. Hence, we have a $\Z8\times\Z2$ symmetry. The charges are given by the matrix $E^{-1}$ and equal the ones we obtained earlier in table~\ref{tab:example_redundancies1c}.

The second example in section~\ref{sec:redundancies_general_case_I} is a $\Z2\times\Z6$ symmetry with field content given in~\ref{tab:example_redundancies2a}. In this case, we obtain for the enlarged $\Z6\times\Z6$ symmetry
\begin{equation}
 Q'_\psi=\begin{pmatrix}3&1\\3&3\end{pmatrix} = 
\begin{pmatrix}1&0\\3&-1\end{pmatrix}
\begin{pmatrix}1&0\\0&6\end{pmatrix}
\begin{pmatrix}3&1\\1&0\end{pmatrix}
=E^{-1}\, S\, F^{-1}\:, \label{eq:example_redundacies_2}
\end{equation}
which implies $s_1=1$, $s_2=6$ and $d'_1=6$, $d'_2=1$. Because $d'_2=1$, we are left with a $\Z6$ symmetry. The charges of the $\psi$-fields are given by the first column of $E^{-1}$ and coincide with the ones in table~\ref{tab:example_redundancies2c}.

\section{Identifying Subgroups}
If one wants to apply the above described breaking mechanism in models of particle physics, one usually has some symmetry in mind which one wants to obtain. However, if the remnant discrete symmetry is of large order, it is non-trivial to see, whether the wanted group is actually present. 

The most important discrete symmetry is probably $R$-parity~\cite{Farrar:1978xj} or equivalently matter parity~\cite{Dimopoulos:1981dw} in the minimal supersymmetric standard model~\cite{Martin:1997ns} which is a \Z2 symmetry. We will discuss how to identify matter parity. The generalization to other groups is straight forward.

Under matter parity all SM superfields are odd while the two Higgs doublets are even (see table~\ref{tab:MSSM_discrete_symmetries}). 

\subsection{Identifying Matter Parity}
\label{sec:identifying_matter_parity}
Consider the minimal supersymmetric standard model together with an abelian, discrete symmetry $G$. We want to investigate, if $G$ has matter parity as a subgroup. 

A finite abelian group can always be written as a direct product
$G=H_{p_1}\times\ldots\times H_{p_n}$ where the $p_i$ are pairwise distinct
primes and $H_p=\Z{p^{e_1}}\times\ldots\times\Z{p^{e_m}}$ with $e_1\leq \cdots
\leq e_m$ positive integers (cf.\ appendix~\ref{sec:structure_of_finite_abelian_groups}). Of course, if $G$ does not contain a \Z{2}
subgroup, there is no matter parity. Hence, let us assume that $H_2$ is
non-trivial and has a \Z{2} subgroup, i.e.\ $e_1=1$.

If $H_2$ consists of more than one factor, there are many equivalent charge
assignments some of which will make a matter parity obvious while others will
conceal its existence. This freedom corresponds to the automorphisms of $G$ which are described in appendix~\ref{sec:automorphisms_abelian_groups}. An important fact is that the automorphism group
factorizes in the way we have written $G$, i.e.\ $\text{Aut}(H_{p_1}\times
H_{p_2})\cong \text{Aut}(H_{p_1})\times \text{Aut}(H_{p_2})$. Thus, we only need
to look at $H_2$.

To see whether a matter parity is present, we scan over all possible charge assignments of $H_2$ and look for a
\Z{2} subgroup under which all SM matter fields are odd. In appendix~\ref{app:package} a \texttt{Mathematica} package is presented which assists in performing the subsequent calculations.

To illustrate this procedure, let us look at a simple example in table
\ref{tab:example_matter_parity}. The two charge assigments are equivalent. While
in \subref{tab:example_matter_parity1} it is not obvious that this symmetry contains a matter
parity, in the second charge assignment in \subref{tab:example_matter_parity2} all fields are odd w.r.t.\ the \Z2
subgroup.

\begin{table}[ht]
\centerline{
\subfloat[\label{tab:example_matter_parity1}]{
\begin{tabular}{ccc}
\toprule[1.3pt]
& \Z2 &  \Z4\\ \cmidrule{1-3}
$\phi_1$ & 1 & 2 \\
$\phi_2$ & 0 & 3 \\
$\phi_3$ & 0 & 1 \\
\bottomrule[1.3pt]
\end{tabular}
}\qquad
\subfloat[\label{tab:example_matter_parity2}]{
\begin{tabular}{ccc}
\toprule[1.3pt]
& \Z2 &  \Z4\\ \cmidrule{1-3}
$\phi_1$ & 1 & 0 \\
$\phi_2$ & 1 & 1 \\
$\phi_3$ & 1 & 3\\
\bottomrule[1.3pt]
\end{tabular}
}
}
\caption{An example for a hidden matter parity. The two charge assignments are equivalent.}
\label{tab:example_matter_parity}
\end{table}

\clearpage
\thispagestyle{empty}
\chapter{Anomalies}
\label{chap:anomalies}
In this chapter we will derive consequences and constraints originating from anomalies. The Fujikawa method~\cite{Fujikawa:1979ay,Fujikawa:1980eg}, which rests on the non-trivial transformation of the path integral measure under a symmetry transformation, is nowadays the standard way of deriving anomaly constraints. For reviews see~\cite{Bertlmann:1996xk,Weinberg:1996kr,Araki:2008ek}.


In the whole chapter we will work in Euclidean space and assume that all correlation functions can be analytically continued to Minkowski space. 

\section{The Setup}
\label{sec:anomalies_setup}
Consider a gauge theory with gauge group $G_\text{gauge}$ and a set of left-handed\footnote{If one wants to include right-handed fermions, one can treat their left-handed anti-particles as the relevant degrees of freedom.} Weyl fermions $\Psi=(\psi^{(1)},\ldots,\psi^{(m)})$ with $\psi^{(i)}$ transforming in an irreducible representation of all non-abelian symmetries. A symmetry group $G$, which can be a global symmetry group or a subgroup of $G_\text{gauge}$, acts on $\Psi$ via
\begin{equation}
\Psi\rightarrow U \, \Psi=\exp(\I \alpha)\, \Psi=\exp(\I \alpha_a \mathbf{t}^a)\, \Psi
\label{eq:anomaly_symmetry_trafo}
\end{equation}
where $U$ is a (reducible) representation of $G$ and $\mathbf{t}^a$ its generators. Let us look at an operator $\mathcal{O}$ which picks up a phase under the action of $G$
\begin{equation}
 \mathcal{O}\: \xrightarrow{G} \: e^{\I\gamma} \mathcal{O}\:. \label{eq:anomaly_field_product}
\end{equation}
Of course, a general operator will not transform only with a phase, e.g.\ if there are uncontracted indices. However, this case is not relevant for our discussion of anomalies as will be shown below.
If $\gamma=0$, the operator $\mathcal{O}$ is an allowed term in the Lagrangian. Consider now the correlation function of the field product in \Eqref{eq:anomaly_field_product} in the quantum theory given by the path integral
\begin{equation}
 \left\langle\mathcal{O}\right\rangle  = \int \mathcal{D}\Psi\,\mathcal{D}\bar\Psi \, e^{\I S[\psi]}\, \mathcal{O}
\end{equation}
where $S[\psi]$ is the action. Naively, one would expect this correlation function to vanish if $\gamma\neq 0$ because the action is, by definition, invariant under $G$. Fujikawa~\cite{Fujikawa:1979ay,Fujikawa:1980eg} has shown that the path integral measure picks up a phase under the action of $G$
\begin{equation}
 \mathcal{D}\Psi\,\mathcal{D}\bar\Psi \rightarrow \exp\left(\I\int\D^4x\: \mathcal{A}(x;\alpha)\right) \mathcal{D}\Psi\,\mathcal{D}\bar\Psi \label{eq:trafo_path_integral}
\end{equation}
where $\mathcal{A}(x;\alpha)$ is the anomaly function. Hence, we can have a non-trivial cancellation between the phase from the path integral measure and $\gamma$, leading to a non-zero expectation value even in the case $\gamma\neq 0$. This phenomenon is known as an anomaly~\cite{Bertlmann:1996xk,Weinberg:1996kr}. In other words, an anomaly is the breaking of a classical symmetry by quantum effects. 

Now we understand, why considering operators which transform with a phase according to \Eqref{eq:anomaly_field_product} is sufficient. Any other transformation cannot be canceled by the transformation of the path integral.

The anomaly function consists of two parts. They can be respectively evaluated to~\cite{AlvarezGaume:1984dr,AlvarezGaume:1983ig}
\begin{subequations}
\label{eq:anomaly_function}
\begin{align}
\mathcal{A}_{\text{gauge}}(x;\alpha)&=\frac{1}{32\pi^2}\, \tr \left[ \alpha\, F \widetilde{F} \right]\:, \label{eq:anomaly_function_gauge} \\
\mathcal{A}_{\text{gravity}}(x;\alpha)&= - \frac{1}{384\pi^2}\mathcal{R} \widetilde{\mathcal{R}}\,\tr\left[\alpha \right] \: .
\label{eq:anomaly_function_gravity}
\end{align}
\end{subequations}
We have suppressed index contractions, i.e.\ $F\widetilde{F} = F^{\mu\nu}\,\widetilde{F}_{\mu\nu}$ and $F_{\mu\nu}= F_{\mu\nu}^a\mathbf{t}^a$ is the gauge field strength. Its dual is $\widetilde{F}^{\mu\nu}=\frac{1}{2}\varepsilon^{\mu\nu\rho\sigma}F_{\rho\sigma}$. Similarly, $\mathcal{R}_{\mu\nu\rho\sigma}$ is the Riemann curvature tensor and $\mathcal{R} \widetilde{\mathcal{R}}=\frac{1}{2}\varepsilon^{\mu\nu\rho\sigma} \mathcal{R}_{\mu\nu}^{\phantom{\mu\nu}\lambda\gamma} \mathcal R_{\rho\sigma\lambda\gamma} $. '$\tr$' indicates the trace over all indices. The full anomaly function is the sum of both contributions.

\paragraph{Notation}
Anomalies originating from $\mathcal{A}_{\text{gauge}}(x;\alpha)$ are usually abbreviated by $G_\text{gauge}-G_\text{gauge}-G$ while anomalies coming from $\mathcal{A}_{\text{gravity}}(x;\alpha)$ are written as grav$-$grav$-G$. 

\section{Anomaly Constraints for Gauge Symmetries}
\label{sec:anomaly_constraints_gauge}
In this section we will discuss constraints coming from anomalies of type $G_\text{gauge}-G_\text{gauge}-G_\text{gauge}$ and grav$-$grav$-G_\text{gauge}$, i.e.\ the transformation parameter $\alpha(x)$ in \Eqref{eq:anomaly_symmetry_trafo} is a function of spacetime. One can show that a non-trivial Jacobian in \Eqref{eq:trafo_path_integral} leads to a non gauge invariant effective action. However, the effective action has to be gauge invariant in order to prove renormalizability and unitarity (e.g.\ decoupling of ghosts) of the gauge theory~\cite{AlvarezGaume:1983ig,Preskill:1990fr,Bilal:2008qx}. Hence, an anomaly is not acceptable for gauge theories and signals an inconsistency.

The most general form $G_\text{gauge}$ can take is
\begin{equation}
 G_\text{gauge} = G_1 \times \ldots \times G_M \times \U1_1 \times \ldots \times \U1_N
\label{eq:anomaly_general_gauge_group}
\end{equation}
with each $G_i$ being a simple Lie group. We denote the $\U1_i$ charge of $\psi^{(j)}$ by $Q_i^{(j)}$.

For the $G_\text{gauge}-G_\text{gauge}-G_\text{gauge}$ anomaly to be absent, the path integral measure has to transform trivially for all $\alpha(x)$. According to \Eqref{eq:anomaly_function_gauge} the anomaly function is proportional to $\tr\left[\mathbf{t}^a\mathbf{t}^b\mathbf{t}^c \right]$ where the indices $a,b,c$ belong to some simple or abelian subgroup of $G_\text{gauge}$ that is one of the factors in \Eqref{eq:anomaly_general_gauge_group}. Since $\tr\mathbf{t}^a=0$ for all simple groups, there are only three types of anomalies:
\begin{description}
 \item[$G_j-G_j-G_j$] This anomaly vanishes if $\Psi$ is in a real or pseudo-real representation and can be shown~\cite{Bertlmann:1996xk,Weinberg:1996kr} to exist only for $G_j=\SU{N}$ with $N\geq3$.
\item[$G_j-G_j-\U1_k$] This leads to the constraint
\begin{equation}
 \sum_{i} Q_k^{(f)} \ell(\mathbf{r}^{(f)})=0 \label{eq:u1_g_g_anomaly_condition}
\end{equation}
where the sum runs over all fermionic, irreducible representations $\mathbf{r}^{(f)}$ of $G_j$ with Dynkin index $\ell(\mathbf{r}^{(f)})$. 
 \item[$\U1_j-\U1_k-\U1_l$] with $j,k,l\in\{0,\ldots,N\}$. This anomaly vanishes if 
\begin{equation}
 \sum_{i}  Q_j^{(i)}Q_k^{(i)}Q_l^{(i)} = 0
\label{eq:U1_cubed_anomaly}
\end{equation}
where the sum runs over all fermions.
\end{description}
Next, we will treat the grav$-$grav$-G_\text{gauge}$ anomaly. According to \Eqref{eq:anomaly_function_gravity} this anomaly is proportional to $\tr\left[\mathbf{t}^a\right]$ which implies that the anomaly exists only for \U1 group factors of $G_\text{gauge}$. The constraint now reads for grav$-$grav$-\U1_j$
\begin{equation}
 \sum_i  Q_j^{(i)} =0 \label{eq:grav_u1_anomaly}
\end{equation}
where the sum runs over all fermions. Note that the case grav--grav--grav is excluded in four dimensions. Purely gravitational anomalies exist only in $4k+2$ dimensions~\cite{AlvarezGaume:1983ig}.

\subsection{Comment on Abelian Subgroups}
\label{sec:anomaly_only_abelian_subgroups_matter}
In this subsection we will show that the abelian subgroups of a simple group determine the anomaly.

Consider a theory with a simple gauge group $G_\text{gauge}$ having rank $r$, i.e.\ $G$ has a maximal, abelian subgroup $\U1^r=\U1_1\times\ldots\times\U1_r$. In addition, let there be a global $\U1_X$ symmetry. We will gather all fermions in the theory in a vector $\Psi=(\psi^{(1)},\ldots,\psi^{(n)})$ where each $\psi^{(i)}$ is a component of an irreducible representation of $G_\text{gauge}$. Then each $\psi^{(i)}$ transforms under $\U1^r$ with charge given by its Dynkin label. 
\comment{Anhang um das zu zeigen.}
We will denote the charge of $\psi^{(i)}$ under $\U1_j$ by $Q^{(i)}_j$ and the charge under $\U1_X$ by $Q^{(i)}_X$. There are two potential anomalies:
\paragraph {$G_\text{gauge}-G_\text{gauge}-G_\text{gauge}$} Above we have argued that this anomaly vanishes if $\Psi$ transforms in a real or pseudo-real representation of $G$. A representation is real or pseudo-real if to each state with Dynkin label $\Lambda$ there is a state with Dynkin label $-\Lambda$~\cite{Slansky:1981yr}. That is, for each state with $\U1^r$ charge $Q$ there is a state with $\U1^r$ charge $-Q$.

Let us presume that $\Psi$ is in a real or pseudo-real representation. There are potential anomalies of the form $\U1_j-\U1_k-\U1_l$ with $j,k,l \in \{ 1,\ldots,r \}$. The anomaly coefficient is given by (cf.\ \Eqref{eq:U1_cubed_anomaly})
\begin{align}
\sum_{i=1}^n Q^{(i)}_j  Q^{(i)}_k Q^{(i)}_l = \sum_{i=1}^{n/2} Q^{(i)}_j  Q^{(i)}_k Q^{(i)}_l + \left( -Q^{(i)}_j \right)\left( -Q^{(i)}_k \right)\left( -Q^{(i)}_l \right)=0
\end{align}
since for every state there is one with opposite charge. Hence, if $\Psi$ is in a (pseudo-)real representation, the anomaly is absent.
\paragraph{$G_\text{gauge}-G_\text{gauge}-\U1_X$} The anomaly coefficient is given in \Eqref{eq:u1_g_g_anomaly_condition} and involves the Dynkin index. The crucial point is to realize that the Dynkin index of a representation $\mathbf{r}$ is proportional to the sum of the squares of the $\U1^r$ charges, i.e.\
\begin{equation}
 \ell(\mathbf{r})= a \sum_{i=1}^{\text{dim}(\mathbf{r})} \left( Q^{(i)}_j \right)^2
\end{equation}
with a proportionality constant $a$ depending on the normalization of the generators. The sum runs over all states in $\mathbf{r}$ and the index $j=1,\ldots,r$ labels an arbitrary $\U1_j$ subgroup of $G$ that is the sum does not depend on the $\U1_j$ subgroup chosen. With this relation we can write
\begin{equation}
 \sum_{i} q_X^{(i)} \ell(\mathbf{r}^{(i)})=a\sum_{k=1}^{n} q_X^{(k)} \left( Q^{(k)}_j \right)^2 
\end{equation}
where $i$ runs over all irreducible representations of $G_\text{gauge}$ and $k$ runs over all fermions. The left hand side is the anomaly coefficient of the $G_\text{gauge}-G_\text{gauge}-\U1_X$ anomaly (cf.\ \Eqref{eq:u1_g_g_anomaly_condition}) whereas the right hand side is the anomaly coefficient of $\U1_X-\U1_j-\U1_j$. Hence, one can evaluate either anomaly and get the same answer. Note that it is sufficient to check a single $\U1_j$ subgroup of $G_\text{gauge}$.

Altogether, one can think of the anomalies as being controlled by the $\U1^r$ subgroup of $G$.

\paragraph{Example}
Let us illustrate all this with a concrete example. Consider an \SU{3} gauge theory with a global $\U1_X$ symmetry. The field content is given in table \ref{tab:example_abelian_anomalies}. Since \SU{3} has rank two, there are two \U1 subgroups. 

Consider two scenarios. First, suppose we have only the triplet in the theory. Since the $\mathbf{3}$ is a complex representation, there is an $\SU{3}-\SU{3}-\SU{3}$ anomaly. Additionally, we find an $\SU{3}-\SU{3}-\U1_X$ anomaly by evaluating \Eqref{eq:u1_g_g_anomaly_condition} which gives $\frac{1}{2}\neq 0$. Alternatively, we could just consider the $\U1_1\times\U1_2$ subgroup of \SU{3} and find that the theory has $\U1_1-\U1_1-\U1_2$, $\U1_1-\U1_2-\U1_2$ and $\U1_1-\U1_2-\U1_X$ anomalies.

Secondly, we consider both the triplet and the antitriplet. Since $\mathbf{3}+\mathbf{\bar{3}}$ is a real representation of \SU{3}, there is no $\SU{3}-\SU{3}-\SU{3}$ anomaly. Also, the $\SU{3}-\SU{3}-\U1_X$ anomaly now vanishes. Again, we would conclude anomaly freedom from the abelian $\U1_1\times\U1_2$ subgroup because all $\U1_j-\U1_k-\U1_l$ anomaly coefficients with $j,k,l\in\{1,2,X\}$ are zero.

\begin{table}[tb]
\centering
 \begin{tabular}{cccc}
\toprule[1.3pt]
 \SU{3} & $\U1_1$ & $\U1_2$ & $\U1_X$ \\ \midrule
\multirow{3}{*}{$\mathbf{3}$} & 1 & 0 & \multirow{3}{*}{1} \\
 & -1 & 1 & \\
 & 0 & -1 & \\ \cmidrule{2-4}
\multirow{3}{*}{$\mathbf{\bar{3}}$} & 0 & 1 & \multirow{3}{*}{-1} \\
 & 1 & -1 & \\
 & -1 & 0 & \\
\bottomrule[1.3pt]
\end{tabular}
\caption{The field content of a \SU{3} gauge theory with a global $\U1_X$ symmetry. With just one triplet the model is anomalous. With triplet and antitriplet there are no anomalies.}
\label{tab:example_abelian_anomalies}
\end{table}

\section{Anomaly Conditions for Global Symmetries}
\label{sec:anomaly_conditions_global}
In this section, we will work with a gauge group $G_\text{gauge}$ which is either simple or a single \U1. Additionally, let there be a global symmetry $G_\text{global}$, which can be continuous or discrete. $G_\text{global}$ being a global symmetry, means that $\alpha$ in \Eqref{eq:anomaly_symmetry_trafo} is independent of space-time. This implies that we can pull $\alpha$ out of the integration in \Eqref{eq:trafo_path_integral} and calculate the Jacobian explicitly with the aid of some powerful index theorems~\cite{Atiyah:1988ji,AlvarezGaume:1984dr} which state that
\begin{subequations}
\begin{alignat}{2}
 \int \D^4 x\, \frac{1}{32\pi^2}\, \epsilon^{\mu\nu\rho\sigma}\, F_{\mu\nu}^a \, F_{\rho\sigma}^a &= n_\text{gauge}  \quad &\in& \quad \Z{}\:, \label{eq:index_theorem}  \\
\frac{1}{2}\int \D^4 x \, \frac{1}{384\pi^2}\frac{1}{2} \, \epsilon^{\mu\nu\rho\sigma}\, \mathcal{R}_{\mu\nu}^{\phantom{\mu\nu}\alpha\beta}\,\mathcal{R}_{\rho\sigma\alpha\beta} & = n_{\text{grav}}  \quad &\in& \quad \Z{} \:. \label{eq:rohlin}
\end{alignat}
\end{subequations}
By combining \Eqref{eq:trafo_path_integral} and \Eqref{eq:anomaly_function} with these index theorems, we conclude that the path integral measure transforms according to
\begin{equation}
 \mathcal{D}\Psi \rightarrow  \exp\left(\I \alpha_a \left( \tr[\mathbf{t}^a] \tr\left[\mathbf{t}^b\mathbf{t}^b \right] n_\text{gauge} - 2\tr[\mathbf{t}^a]\, n_{\text{grav}}\right) \right) \mathcal{D}\Psi\:.
\label{eq:anomaly_trafo_path_integral_with_index}
\end{equation}
%
We will now discuss anomalies of type $G_\text{gauge}-G_\text{gauge}-G_\text{global}$ and grav$-$grav$-G_\text{global}$ and distinguish between the different types of which $G_\text{global}$ can be.

\subsection{Conditions for Continuous Symmetries}
Equation \eqref{eq:anomaly_trafo_path_integral_with_index} teaches that gauge and gravitational anomalies are proportional to $\tr[\mathbf{t}^a]$, i.e.\ they can only be non-zero if $G_\text{global}=\U1$. The anomaly equations are the same as in section~\ref{sec:anomaly_constraints_gauge}:
\paragraph{$G_\text{gauge}-G_\text{gauge}-\U1$}
This leads to the constraint
\begin{equation}
 \sum_{f} Q^{(f)} \ell(\mathbf{r}^{(f)})=0 
\label{eq:anomaly_global_g_g_U1}
\end{equation}
where the sum runs over all irreducible representations $\mathbf{r}^{(f)}$ of $G_\text{gauge}$ with Dynkin index $\ell(\mathbf{r}^{(f)})$ and $Q^{(f)}$ is the \U1 charge of $\psi^{(f)}$.
\paragraph{grav--grav--\U1}
 This anomaly vanishes if 
\begin{equation}
 \sum_i  Q^{(i)} =0
\end{equation}
where the sum runs over all fermions.

\subsection{Conditions for Abelian Discrete Symmetries}
\label{sec:anomaly_conditions_abelian}
Consider now an abelian discrete symmetry, i.e.\ $G_\text{global}=\Z{N}$. Under $G_\text{global}$ the fermions transform according to (see appendix~\ref{app:finite_abelian_groups})
\begin{equation}
 \psi^{(i)}\rightarrow \exp\left( 2\pi\I \, q^{(i)}\frac{n}{N}\right)  \psi^{(i)} \quad \Rightarrow \quad \alpha=\frac{2\pi n }{N}q^{(i)}\:.
\end{equation}
The \Z{N} charge $q^{(i)}$ and the transformation parameter $n$ are only defined modulo $N$ that is we can choose them to lie in the range $\{ 0,\,\ldots,\, N-1\}$. For the path integral measure to transform trivially (cf.\ \Eqref{eq:anomaly_trafo_path_integral_with_index}), we get the following conditions
\begin{subequations}
\label{eq:anomaly_conditions_ZN}
\begin{alignat}{2}
 G_\text{gauge}-G_\text{gauge}-\Z{N} &: \qquad  \sum_{f} q^{(f)} \ell(\mathbf{r}^{(f)}) = 0 &\mod \eta\:, \label{eq:anomaly_condition_gauge} \\
 \U1-\U1-\Z{N} &: \qquad \sum_i \left( Q^{(i)} \right)^2 q^{(i)}  =0 &\mod \eta \label{eq:anomaly_condition_U1}\:,\\
 \text{grav}-\text{grav}-\Z{N} &: \qquad \sum_i  q^{(i)} =0 &\mod \eta\:. \label{eq:anomaly_condition_gravity}
\end{alignat}
\end{subequations}
The summation in \Eqref{eq:anomaly_condition_gauge} runs over all fermionic, irreducible representations $\mathbf{r}^{(f)}$ of $G_\text{gauge}$ with Dynkin index $\ell(\mathbf{r}^{(f)})$ while the sums in equations \eqref{eq:anomaly_condition_U1} and \eqref{eq:anomaly_condition_gravity} run over all fermions.

We have defined
\begin{equation}
 \eta = \left\lbrace \begin{array}{ll}N, & \text{if} \: N \:\text{odd} \\ \frac{N}{2}, & \text{if} \: N \:\text{even}\end{array} \right. \: .  \label{eq:def_eta}
\end{equation}
That the condition in \Eqref{eq:anomaly_condition_gauge} has to be fulfilled only mod $\eta$ is due to our normalization of the Dynkin index which is $\ell(\mathbf{N})=\frac{1}{2}$ for \SU{N}/Sp$(N)$ and $\ell(\mathbf{N})=1$ for \SO{N}. For the grav--grav--\Z{N} anomaly, $\eta$ comes from the prefactor $\frac{1}{2}$ in \Eqref{eq:rohlin} whose origin is explained in section~\ref{sec:gravitational_instantons}. 


\paragraph{Comment on the $\boldsymbol{\U1}$-$\boldsymbol{\U1}$-$\boldsymbol{\Z{N}}$ Anomaly}

The anomaly condition in \Eqref{eq:anomaly_condition_U1} has a potential problem. In a general \U1 gauge group, one is free to scale the charges $Q^{(i)}$ by an arbitrary factor. However, this freedom renders the constraint in \Eqref{eq:anomaly_condition_U1} useless because it allows to fulfill the constraint by scaling the charges~\cite{Ibanez:1991hv}. The situation is different if one embeds the \U1 into a simple gauge group.

In the following we will discuss the most relevant case for our studies, the case where hypercharge $\U1_Y$ is embedded into \SU5. The $\U1_Y$ charges are scaled by a factor of $\sqrt{3/5}$ w.r.t. their standard values (cf.\ table~\ref{tab:MSSM_labels}). Since we have the freedom to shift the \Z{N} charge by integer multiples of $N$ (parameterized by the integers $k^{(i)}$), we can rewrite \Eqref{eq:anomaly_condition_U1} as
\begin{alignat}{2}
 & \quad \frac{3}{5}\sum_i \left( Q^{(i)} \right)^2 (q^{(i)}+k^{(i)}N)  & =0 &\mod \eta \\
 \Rightarrow & \quad \underbrace{\sum_i \frac{3}{5}\left( Q^{(i)} \right)^2 q^{(i)}}_{A_1} + \frac{3}{5} N \underbrace{\sum_i k^{(i)}\left( Q^{(i)}\right)^2}_{n}  & =0 &\mod \eta \:.
\end{alignat}
We have introduced the abbreviation $A_1$ for the anomaly coefficient with GUT charges. $n$ can be any integer because the right-handed electron carries hypercharge equal to one. It follows that the anomaly constraint reads
\begin{equation} 
 5A_1 = 0 \mod \eta \: . \label{eq:u1_u1_zn_gut_condition}
\end{equation}
The calculation can be easily generalized to other cases than hypercharge. Basically, the factor which shows up in front of $A_1$ in \Eqref{eq:u1_u1_zn_gut_condition} is the denominator of the (reduced) square factor by which the \U1 charges are scaled.

\paragraph{Comment on Anomaly Mixing}
Let us comment on an ambiguity in the anomaly coefficients in the presence of a \U1. Consider a $\U1\times\Z{N}$ symmetry. We will denote the $\U1$ charge by $Q^{(i)}$ and the \Z{N} charges by $q^{(i)}$, and we will assume all charge to be integers without loss of generality. Suppose we have a gauge group $G$. The anomaly coefficients as given in \Eqref{eq:anomaly_global_g_g_U1} and \Eqref{eq:anomaly_condition_gauge} read
\begin{subequations}
\begin{alignat}{2}
 G-G-\U1:&\quad \sum_f Q^{(f)} \ell(\mathbf{r}^{(f)})&& \equiv A \: , \\
 G-G-\Z{N}:&\quad \sum_f q^{(f)} \ell(\mathbf{r}^{(f)})&& \equiv B \: .
\end{alignat} 
\end{subequations}
We want to answer the question, under which circumstances it is possible to rotate the anomaly completely into the \U1.

We can redefine the $\Z{N}$ charge by shifting them by integer multiples of the \U1 charges as explained in detail in appendix~\ref{app:anomaly_mixing}. That is, we can define new \Z{N} charges $q'^{(i)}=q^{(i)}+nQ^{(i)}$. Then the new $G-G-\Z{N}$ anomaly coefficient is given by
\begin{equation}
 \sum_f q'^{(f)} \ell(\mathbf{r}^{(f)})=\sum_f (q^{(f)}+nQ^{(f)}) \ell(\mathbf{r}^{(f)})=B+nA
\end{equation}
with $n\in\Z{}$. Anomaly freedom corresponds to 
\begin{equation}
 \sum_f q'^{(f)} \ell(\mathbf{r}^{(f)})= 0 \mod \eta  \: .
\end{equation}
Hence, the \Z{N} can be made anomaly-free if there is a solution to 
\begin{equation} \label{eq:anomaly_mixing}
 B+nA = 0 \mod \eta   
\end{equation}
for $n\in\Z{}$.

\subsection[Conditions for $R$-symmetries]{Conditions for $\boldsymbol{R}$-Symmetries}
\label{sec:anomaly_R_symmetries}
In supersymmetric theories one can have symmetries which do not commute with the supersymmetry generators. These are called $R$-symmetries~\cite{Wess:1992cp}. We will limit ourselves to the case of $\mathcal{N}=1$ supersymmetry where the $R$-symmetry is abelian. Since the $R$-symmetry does not commute with supersymmetry, the superspace coordinate $\theta$ carries $R$-charge which we denote by $q_\theta$. The canonical choice is $q_\theta=\pm 1$. Hence, the superpotential $\mathscr{W}$ carries $R$-charge $2\,q_\theta$ which follows from $\mathscr{L}\supset\int\D^2\theta\,\mathscr{W}$.

Because $\theta$ carries non-trivial $R$-charge, different component fields of a superfield carry different $R$-charges. A chiral superfield $\Phi^{(i)}$, containing a complex scalar $\phi^{(i)}$ and a Weyl fermion $\psi^{(i)}$, and a vector superfield $V$, containing a vector field $A_\mu$ and a gaugino $\lambda$, can be written in component form as~\cite{Wess:1992cp}
\begin{subequations}
\begin{align}
 \Phi^{(i)} & = \phi^{(i)} + \sqrt{2}\, \theta \psi^{(i)} + \theta\theta F^{(i)} \: , \label{eq:chiral_superfield_components}\\
 V & = -\theta\sigma^\mu \bar\theta \,  A_\mu + \I\theta\bar\theta\bar\theta \bar \lambda - \I \bar\theta\bar\theta\theta \lambda + \frac{1}{2} \theta\theta\bar\theta\bar\theta D \: .
\end{align}
\end{subequations}
We are interested in the charges of the fermions $\psi$ and $\lambda$ because they contribute to anomalies. We will denote the $R$-charge of the superfield $\Phi^{(i)}$ by $R^{(i)}$. Then \Eqref{eq:chiral_superfield_components} teaches that $\psi^{(i)}$ has charge $R^{(i)}-q_\theta$.

To determine the charge of the gaugino $\lambda$, note that the real vector field $V$ cannot transform under the $R$-symmetry. Hence, $\lambda$ has $R$-charge $q_\theta$. 

For the gravitational anomaly there is an additional contribution. As is well known, in order to incorporate gravity in supersymmetric models, one has to consider local supersymmetry which is also known as supergravity. There the graviton, which does not carry $R$-charge, is accompanied by a spin-$\frac{3}{2}$ superpartner, the gravitino with $R$-charge $q_\theta$. The contribution of a spin-$\frac{3}{2}$ fermion is $-21$ the contribution of a spin-$\frac{1}{2}$ fermion~\cite{AlvarezGaume:1984dr,Christensen:1978gi,Nielsen:1978ex}.

Now we can apply \Eqref{eq:anomaly_conditions_ZN} and arrive at the following conditions for a $\Z{N}^R$ symmetry with $q_\theta=1$
\begin{subequations}
\label{eq:anomaly_R_symmetries}
\begin{alignat}{2}
 G_\text{gauge}-G_\text{gauge}-\Z{N}^R &: \qquad  \ell(\text{adj})  + \sum_{f} (R^{(f)}-1) \,  \ell(\mathbf{r}^{(f)}) = 0 &\mod \eta\:,  \\
\U1-\U1-\Z{N}^R &: \qquad \sum_{i} \left( Q^{(i)}\right)^2 (R^{(i)}-1)  =0 &\mod \eta \:,\\
 \text{grav}-\text{grav}-\Z{N}^R &: \qquad -21 + \text{dim}\,G + \sum_i  (R^{(i)}-1) =0 &\mod \eta\:.  \label{eq:grav_grav_ZNR_anomaly}
\end{alignat}
\end{subequations}
$\eta$ is defined in \Eqref{eq:def_eta}. Note that the Dynkin index in the adjoint representation $\ell(\text{adj})$ of $G_\text{gauge}$ is equal to the quadratic Casimir $c_2(G_\text{gauge})$. The summation index $f$ runs over all irreducible representations (only chiral superfields) of $G_\text{gauge}$ while $i$ runs over all chiral superfields. The term $\text{dim}\,G$ in \Eqref{eq:grav_grav_ZNR_anomaly} represents the dimension of the \textit{full} gauge group $G$ of the theory because all gauginos carry $R$-charge. For example, the dimension of \SU{M} is $M^2-1$ while the dimension of \U1 is 1. For the $\U1-\U1-\Z{N}^R$ anomaly the same comments as in the previous section apply.

\subsection{Conditions for non-abelian Discrete Symmetries}
\label{sec:anomaly_non_abelian_discrete}
Now we turn to the case where $G_\text{global}$ is a non-abelian, finite group. Consider a fixed element $g\in G_\text{global}$ of order $N$, i.e.\ $g^N=1$. Hence, $g$ generates a \Z{N} subgroup of $G_\text{global}$. The fermions transform according to
\begin{equation}
 \psi^{(i)}\rightarrow U^{(i)}_g\psi^{(i)}=\exp\left(\I \alpha \right) \psi^{(i)} \quad \Rightarrow \quad \alpha=-\I\, \text{Log}\left(  U^{(i)}_g \right) 
\end{equation}
with $\text{Log}(  U^{(i)}_g )$ being the matrix logarithm of the appropriate representation matrix of $g$. As in subsection \ref{sec:anomaly_conditions_abelian}, we look at the transformation of the path integral measure under the action of $g$ in \Eqref{eq:anomaly_trafo_path_integral_with_index}. We arrive at the following anomaly constraints
\begin{subequations}
\label{eq:anomaly_condition_non-abelian}
\begin{alignat}{2}
  G_\text{gauge}-G_\text{gauge}-G_\text{global} &: \qquad  \sum_{i,j} q^{(i)} \ell(\mathbf{r}^{(j)}) = 0 &\mod \frac{N}{2}\:, \\
  \text{grav}-\text{grav}-G_\text{global} &: \qquad \sum_i  q^{(i)} =0 & \mod \frac{N}{2}\:.
\end{alignat}
\end{subequations}
The summation index $i$ labels the irreducible representations of $G_\text{global}$ and $j$ refers to the irreducible representations of $G_\text{gauge}$.  The \Z{N} charge $q^{(i)}$ is defined by
\begin{equation}
 q^{(i)} = N\frac{\ln\det U^{(i)}_g}{2\pi\I}\:.
\end{equation}
Because $g$ is of order $N$, we know that $\det U^{(i)}_g$ is an $N$-th root of unity. Therefore, we introduced the factor of $N$ to make $q^{(i)}$ integer. 

Consider now two elements $g_1,g_2\in G_\text{global}$ for which equations \eqref{eq:anomaly_condition_non-abelian} hold. Then the anomaly constraints are also fulfilled for $g=g_1g_2$ and $g'=g_2g_1$. Hence, the anomaly conditions have to be checked only for the generators of $G_\text{global}$. In other words, just the abelian subgroups of $G_\text{global}$ contribute to the anomaly~\cite{Araki:2007zza}. This is in accordance with the findings in section \ref{sec:anomaly_only_abelian_subgroups_matter}.

\section{Anomalies and Instantons}
\label{sec:instantons}

\subsection{Explicit Instanton Solution}
Remember that we are working in Euclidean space, i.e.\ the metric is $g_{\mu\nu}=\delta_{\mu\nu}$. We will consider pure Yang-Mills theory with a simple gauge group $G_\text{gauge}$. Hence, the path integral looks like
\begin{equation}
 \int \mathcal{D}A\:e^{-S} \qquad \text{with action} \qquad S=\frac{1}{2g^2} \int \D^4 x \, \tr F^{\mu\nu} F_{\mu\nu} \: .
\end{equation}
The field strength is given by $F_{\mu\nu}=\partial_\mu A_\nu - \partial_\nu A_\mu + [A_\mu,A_\nu]$ and the matrix valued gauge field is $A_\mu=A_\mu^a\mathbf{t}^a$. The generators of $G_\text{gauge}$ $\mathbf{t}^a$ in a representation $\mathbf{r}$ define the Dynkin index by $\tr \mathbf{t}^a\mathbf{t}^b=\ell(\mathbf{r})\delta^{ab}$. We use the usual normalization of the Dynkin index that is $\ell(\mathbf{N})=\frac{1}{2}$ for \SU{N}.

The familiar approach to quantum field theory is to perform perturbation theory around stationary points of finite action. In many cases there exists only one such point where all fields vanish, at least in the absence of spontaneous symmetry breaking. However, Polyakov et al.~\cite{Belavin:1975fg} discovered that non-abelian gauge theories admit more than one field configuration with finite action. These field configurations are called instantons~\cite{Coleman:1978ae,Vandoren:2008xg}.

First, notice that for the action to be finite the field strength has to vanish at infinity, i.e.\ $F_{\mu\nu}(x)\to 0$ as $x\to\infty$. This is equivalent to the gauge field being a pure gauge, i.e.\ $A_\mu(x)=g(x)\,\partial_\mu g^{-1}(x)$ with $g(x)\in G_\text{gauge}$. We observe that a replacement $g(x)\to g(x) g^{-1}(x_1)$ with a fixed $x_1$ leaves $A_\mu$ unchanged. Thus, we can arrange for $g(x_1)=1$ in any direction $x_1$. Each gauge field with finite action therefore defines a mapping from the unit sphere $S_3$ in four dimensions into the gauge group. These mappings can be characterized by an integer due to a non-vanishing third homotopy group $\pi_3(G_\text{gauge})=\Z{}$. This integer is called the instanton number and is given by
\begin{equation}
 k = \frac{1}{32\pi^2}\epsilon^{\mu\nu\rho\sigma}\int\D^4 x \,\tr F_{\mu\nu} \, F_{\rho\sigma}\:.
\end{equation}
This coincides with the integer defined in \Eqref{eq:index_theorem}. There is an important theorem~\cite{Bott:1956tf} which states that every mapping $S_3\to G$ can be continuously deformed into a mapping $S_3\to\SU2$ where $\SU2$ is a subgroup of $G$. Since the instanton number $k$ is a topological invariant, i.e.\ it does not change under continuous deformations of the gauge field, $k$ does not change if we only consider an \SU2 subgroup of $G$. In this \SU2 subgroup, an instanton with $k=1$ is given by
\begin{equation}
 A_\mu^a(x)=2\eta_{\phantom{a}\mu\nu}^a \frac{(x-x_0)^\nu}{(x-x_0)^2+\rho^2}\:,
\label{eq:instanton_solution}
\end{equation}
where $x_0$ is the position of the instanton and $\rho$ its size. These parameters are sometimes called collective coordinates. The t'Hooft tensors~\cite{'tHooft:1976up,'tHooft:1976fv} are defined by
\begin{equation}
\eta_{a\mu\nu}=\left\lbrace 
\begin{array}{rcl}
 \epsilon_{a\mu\nu}&\text{for} & \mu,\nu=1,2,3\\  
 -\delta_{a\mu}&\text{for} & \mu=1,2,3 \, , \quad \nu=0 \\
 \delta_{a\nu}&\text{for} & \nu=1,2,3\, , \quad \mu=0\\
 0&\text{for} & \mu=\nu=0
\end{array}
\right. \:.
\end{equation}
The gauge field in \Eqref{eq:instanton_solution} has action 
\begin{equation}
 S=\frac{8\pi^2}{g^2}\:.
\end{equation}
In fact, \Eqref{eq:instanton_solution} is the only solution with $k=1$~\cite{Atiyah:1977pw}. Solutions with higher instanton number have action $S=\frac{8\pi^2}{g^2}|k|$ and their contribution is exponentially suppressed compared to the $k=1$ solution. 

Finally let us remark that in an instanton background the theta angle term, which is a total derivative, cannot be neglected in the Lagrangian and has to be included because
\begin{equation}
 \I\frac{\theta}{32\pi^2} \epsilon^{\mu\nu\rho\sigma}\int\D^4 x \,\tr F_{\mu\nu} \, F_{\rho\sigma} = \I\theta k\:.
\label{eq:theta_term}
\end{equation}

\subsection{Fermionic zero-modes in an Instanton Background}
Consider now a left-handed Weyl fermion $\psi$ in an instanton background of a simple gauge group $G_\text{gauge}$. The Langrangian reads
\begin{equation}
 \mathscr{L}=\I\bar{\psi}\slashed{D}\psi
\end{equation}
with Weyl operator $\slashed{D}=\bar\sigma^\mu(\partial_\mu+A_\mu^a\mathbf{t}^a)$ where $A^a_\mu$ is a fixed instanton solution with instanton number $k$ and $\mathbf{t}^a$ are the generators of the irreducible representation $\mathbf{r}^{(\psi)}$ in which $\psi$ transforms under $G_\text{gauge}$. $\bar\sigma^\mu=(-\I\mathds{1},\tau^a)$ are the $2\times 2$ Pauli matrices. Let us look at the normalizable eigenmodes of the Weyl operator\footnote{Strictly speaking, this eigenvalue problem is not well-defined because $\slashed{D}\psi$ is a right-handed Weyl spinor. One can make sense out of this by introducing a right-handed Weyl fermion which does not couple to the gauge fields. All equations are then well-defined by replacing $\slashed{D}$ by the Dirac operator and $\psi$ by the 4-component Dirac spinor. For a completely correct calculation see~\cite{Peccei:1977np}.}
\begin{equation}
 \I\slashed{D}\psi_i=\lambda_i\psi_i\:.
 \label{eq:eingenmodes_Weyl_operator}
\end{equation}
As we will see later, the zero-modes are important. For example, for the instanton solution in \Eqref{eq:instanton_solution} a zero-mode is given by
\begin{equation}
 \psi(x)=\frac{\rho}{((x-x_0)^2+\rho^2)^{-\frac{3}{2}}} u\:,
\end{equation}
where $u$ is a constant, left-handed Weyl spinor. There is an important theorem, namely the Atiyah-Singer index theorem~\cite{Atiyah:1988ji,AlvarezGaume:1984dr}, which makes a statement about the number of zero-modes in an instanton background~\cite{Vandoren:2008xg}, which is called the index of  $\I\slashed{D}$
\begin{equation}
 \text{Ind}\,\I\slashed{D}= 2 \left(\frac{1}{32\pi^2}\epsilon^{\mu\nu\rho\sigma}\int\D^4 x \, F_{\mu\nu}^a \, F_{\rho\sigma}^b \right) \tr \mathbf{t}^a\mathbf{t}^b = 2 k \,\ell(\mathbf{r}^{(\Psi)})\:.
\label{eq:anomaly_index_theorem}
\end{equation}
Hence, the Dynkin index counts the number of zero-modes. For example, a fermion in the fundamental representation of \SU{N} has a single zero mode for $k=1$, while a fermion in the adjoint representation has $2N$ zero-modes.

Finally, we will show that massive fermions do not have zero-modes. A massive fermion can be written in terms of a Dirac spinor $\Psi$. Then the eigenvalue equation becomes (cf.\ \Eqref{eq:eingenmodes_Weyl_operator})
\begin{equation}
 \I(\slashed{D}_\text{Dirac}-m)\Psi_i=(\lambda_i-\I m)\Psi_i
\end{equation}
where $m$ is the mass of the fermion. This can never have a zero eigenvalue because in Euclidean space the Dirac operator $\slashed{D}_\text{Dirac}$ is Hermitian and has therefore only real eigenvalues.

\subsection{Gravitational Instantons}
\label{sec:gravitational_instantons}
So far, we have only discussed instantons for gauge theories. There are instantons in gravity too, i.e.\ non-singular solutions to the Einstein equations~\cite{Hawking:1976jb}. In this case, the index theorem reads~\cite{AlvarezGaume:1984dr}
\begin{equation}
 \text{Ind}\,\I\slashed{D}=\int \D^4 x \, \frac{1}{384\pi^2}\frac{1}{2} \, \epsilon^{\mu\nu\rho\sigma}\, \mathcal{R}_{\mu\nu}^{\phantom{\mu\nu}\alpha\beta}\,\mathcal{R}_{\rho\sigma\alpha\beta}\:.
\end{equation}
The Weyl operator is $\slashed{D}=\bar\sigma^\mu(\partial_\mu+\omega_\mu)$ where $\omega_\mu$ is the spin connection. This integral differs by a factor of two from \Eqref{eq:rohlin} that is the index is always an even integer as a consequence of Rohlin's theorem~\cite{Rohlin:1951wd}. In other words, a Weyl fermion has an even number of zero modes in a gravitational instanton background (see also the discussion in~\cite{Banks:1991xj,Csaki:1997aw}).

An explicit example of an instanton solution with index two is the Taub-NUT metric~\cite{Newman:1963yy} while the Schwarzschild metric has index zero.

\subsection{Instantons in the Path Integral}
\label{sec:instantons_path_integral}
In this subsection, we will comment on how instantons show up in the path integral. As before, consider massless left-handed Weyl spinors $\Psi=(\psi^{(1)},\ldots,\psi^{(m)})$ in a gauge theory with simple gauge group $G_\text{gauge}$. Every $\psi^{(i)}$ transforms in an irreducible representation $\mathbf{r}^{(i)}$ of $G_\text{gauge}$. 

A Green's function of an operator $\mathcal{O}$ in an instanton background of $G_\text{gauge}$ with gauge coupling $g$ is given by the path integral~\cite{Peccei:1977np}
\begin{subequations}
\begin{align}
 \left\langle \mathcal{O} \right\rangle &
=\frac{1}{Z}\sum_k e^{i\theta k}\, e^{-\frac{8\pi^2}{g^2} |k|}\, Z_k\;,\quad\text{where}\\
 Z_k&= \int\mathcal{D}\Psi\,\mathcal{D}\bar{\Psi}\, \exp\left(-\int\D^4 x\, \I\bar{\Psi}\slashed{D}_k\Psi\right)  \mathcal{O} \:, \quad \text{and} \\
 Z &= \int\mathcal{D}\Psi\,\mathcal{D}\bar{\Psi}\, \exp\left(-\int\D^4 x\, \I\bar{\Psi}\slashed{D}\Psi\right)  \:.
\end{align}
\end{subequations}
The Weyl operator $\slashed{D}_k$ in $Z_k$ is understood to include a background gauge field with instanton number $k$. The parameter $\theta$ comes from the term in \Eqref{eq:theta_term} while the factor $\exp(-\frac{8\pi^2}{g^2} |k|)$ originates in the background action of the gauge fields. Correctly, one has to integrate $Z_k$ also over the collective coordinates, e.g.\ in the case $k=1$ the instanton size $\rho$ and position $x_0$ as has been done in~\cite{'tHooft:1976fv}. However, we will not discuss this complication because it does not change our qualitative analysis.

In usual perturbation theory one only calculates $Z_0$. The contributions of terms with non-zero instanton number are exponentially suppressed. In the following, we will just consider $Z_1$ since this gives the leading correction.

Let us expand the fields in eigenmodes of the Weyl operator (cf.\ \Eqref{eq:eingenmodes_Weyl_operator})
\begin{subequations}
\begin{align}
 \psi^{(k)}(x)&=\sum_i a^{(k)}_i \psi_i^{(k)}(x) + \sum_{i=1}^{2\ell(\mathbf{r}^{(k)})} b^{(k)}_i \chi_i^{(k)}(x) \label{eq:expansion_in_eigenmodes}\\
 \bar\psi^{(k)}(x)&=\sum_i \bar{a}^{(k)}_i \psi_i^{\ast(k)}(x)
\end{align}
\end{subequations}
where the $a^{(k)}_i$ and $b^{(k)}_i$ carry the anticommuting nature of the fermions. In \Eqref{eq:expansion_in_eigenmodes} we have explicitly used that, in an instanton background with instanton number one, a Weyl fermion has $2\ell(\mathbf{r}^{(k)})$ zero-modes $\chi_i^{(k)}$ (see \Eqref{eq:anomaly_index_theorem}). The eigenfunctions are orthonormal
\begin{equation}
 \int \D^4 x \, \psi_i^{\ast(k)} \psi_j^{(k)} = \delta_{ij}\:.
\end{equation}
With this we can write
\begin{equation}
 Z_1 = \int \prod_{i,i'} \D a^{(i')}_i \, \prod_{j,j'} \D \bar a^{(j')}_j \, \prod_{l,l'} \D b^{(l')}_l \, \exp\left( \sum_{n,n'}\bar a^{(n')}_n \lambda_n  a^{(n')}_n\right) \: \mathcal{O}\:.
\end{equation}
It is important to notice that the zero-modes $b^{(k)}_i$ do not appear in the exponential. Hence, $Z_1$ is zero unless $\mathcal{O}$ includes every zero-mode $b^{(k)}_i$ exactly once. This follows from the rules of integration over Grassmann variables. One possible form for $\mathcal{O}$ is
\begin{equation}
 \mathcal{O} = \prod_{i=1}^m \, \left(\psi^{(i)}\right)^{2\ell(\mathbf{r}^{(i)})}\:.
\end{equation}

One can include this instanton contribution to the path integral by inserting a term proportional to
\begin{equation}
 e^{-\frac{8\pi^2}{g^2}} \prod_{i=1}^m \, \left(\psi^{(i)}\right)^{2\ell(\mathbf{r}^{(i)})}
\label{eq:effective_instanton_operator}
\end{equation}
in the Lagrangian~\cite{'tHooft:1976up,'tHooft:1976fv}. This is the most important equation of this section. It allows us to interpret anomalies in the following way. The term in \Eqref{eq:effective_instanton_operator} is always present in a gauge theory. It potentially breaks any global symmetry, which acts chirally on the fields, unless the conditions in section~\ref{sec:anomaly_conditions_global} are fulfilled. 

For example, this term explicitly breaks all \U1s down to \Z{N} with $N=\sum_i\, 2 q^{(i)} \ell(\mathbf{r}^{(j)})$, unless the anomaly conditions in \Eqref{eq:anomaly_global_g_g_U1} are satisfied. We will give some examples in section~\ref{sec:consequences_of_anomalies}.

%

\section{Consequences of Anomalies}
\label{sec:consequences_of_anomalies}
In this section we will give some examples of the explicit breaking of global symmetries by anomalies. The breaking term is always of the same form as the one in \Eqref{eq:effective_instanton_operator}.

\subsection[An Anomalous \Z{6} Symmetry]{An Anomalous $\boldsymbol{\Z{6}}$ Symmetry}
Consider an \SU{N} gauge theory with a global \Z6 symmetry and field content given in table \ref{tab:discrete_anomaly_example}. Evaluating the anomaly condition in \Eqref{eq:anomaly_conditions_ZN} we get
\begin{subequations}
\begin{align}
 3\cdot\frac{1}{2} + 1\cdot\frac{1}{2} = 2 \neq 0 \mod 3\:,\\
 3\cdot 3 + 3 \cdot 1 = 12 = 0 \mod 3\:.
\end{align} 
\end{subequations}
Thus, this theory has a $\SU{N}-\SU{N}-\Z6$ anomaly. From \Eqref{eq:effective_instanton_operator} we know that instantons will generate a term $\psi_1\psi_2$ in the Lagrangian which breaks the \Z6 down to a \Z2.

\begin{table}
\centering
\subfloat{
\begin{tabular}{lll}
\toprule[1.3pt]
 & $\psi_1$ & $\psi_2$ \\ \midrule
\SU{N} & $\mathbf{N}$ & $ \overline{\mathbf{N}}$ \\ 
\Z6 & 3 & 1\\
\bottomrule[1.3pt]
\end{tabular}
}
\pnode(0,0){nodeA}
\pnode(2,0){nodeB}
\ncline[linewidth=1.5pt]{->}{nodeA}{nodeB}
\naput{anomaly}
\hspace{1.9cm}
\subfloat{
\begin{tabular}{lll}
\toprule[1.3pt]
 & $\psi_1$ & $\psi_2$ \\ \midrule
\SU{N} & $\mathbf{N}$ & $ \overline{\mathbf{N}}$ \\ 
\Z2 & 1 & 1\\
\bottomrule[1.3pt]
\end{tabular}
}
\caption{Field content of an \SU{N} gauge theory with an anomalous \Z6 symmetry which is broken down to a \Z2.}
\label{tab:discrete_anomaly_example}
\end{table}

\subsection{Baryon Number Violation in the Standard Model}
The SM has accidental symmetries $\U1_B$ and $\U1_L$ called baryon and lepton number. These symmetries prevent nucleons from decaying into light leptons. Under $\U1_B$ all quarks have charge $\frac{1}{3}$ while under $\U1_L$ all leptons have charge one. One can easily check that both symmetries are anomalous (cf.\ \Eqref{eq:anomaly_global_g_g_U1})
\begin{subequations}
\begin{align}
\SU{2}-\SU{2}-\U1_B & :\qquad N_g \cdot 3 \cdot \frac{1}{3} \cdot \frac{1}{2}\neq 0 \: , \\
\SU{2}-\SU{2}-\U1_L & :\qquad N_g \cdot 1 \cdot \frac{1}{2} \neq 0  \: ,
\end{align}
\end{subequations}
where $N_g$ is the number of generations. Let us only consider two generations. Then \Eqref{eq:effective_instanton_operator} implies that the correlator
\begin{equation}
 \left\langle qqqqqq\ell\ell \right\rangle 
\end{equation}
is non vanishing. This operator contributes to the decay of a proton and a neutron into two antileptons~\cite{'tHooft:1976up}
\begin{equation}
 p + n \to e^+ + \bar\nu_\mu \quad (\text{or  }\: \mu^+ + \bar\nu_e )\:.
\end{equation}
Thus, baryons are not completely stable in the standard model. However, the decay rate is exponentially suppressed by a factor $e^{-8\pi^2/g^2}$ where $g$ is the weak gauge coupling which makes the process unobservable. 

\section{Discrete Green-Schwarz Mechanism}
\label{sec:discrete_GS}
In supersymmetric theories, especially in string-derived models, there is a highly non-trivial way of cancelling anomalies. It was first discovered by Green and Schwarz~\cite{Green:1984sg} in 10 dimensions. We will first describe schematically how the mechanism works in ten-dimensions and then go to four dimensions.

\subsection{Ten Dimensions}
The ten-dimensional supergravity multiplet contains an antisymmetric tensor $B_{\mu\nu}$ whose field strength is given by\footnote{In this subsection we will simplify the notation by using differential forms, i.e.\ $dB=\frac{1}{2}\partial_\sigma B_{\mu\nu} \, dx^{\sigma}\wedge dx^{\mu}\wedge dx^{\nu}$.}
\begin{equation} \label{eq:field_strength_sugra}
 H=dB + \omega_Y - \omega_L \:.
\end{equation}
$\omega_Y$ and $\omega_L$ are Chern-Simons three-forms associated with the Yang-Mills and the Lorentz symmetry respectively. Their explicit form is given in~\cite{Green:1984sg}. The field strength satisfies
\begin{equation}
 dH= \tr F^2 - \tr \mathcal{R}^2 
\end{equation}
where $F$ is the field strength two-form of the gauge group and $\mathcal{R}$ is the Riemann curvature two-form. Under gauge and Lorentz transformations with parameter $\lambda$, $B$ shifts $B\to B + d\lambda$ while the field strength $H$ is invariant. Green and Schwarz worked out how this transformation of $B$ can be used to cancel all anomalies in the ten-dimensional supergravity theory. They discovered that anomaly freedom singles out the gauge groups \SO{32} and \E8 $\times$ \E8.

\subsection{Four Dimensions}
Witten has shown how the ten-dimensional supergravity action can be dimensionally reduced to $\mathcal{N}=1$  supergravity in four dimensions~\cite{Witten:1985xb}. It is important to observe that in four dimensions, an antisymmetric tensor has only one degree of freedom. This can be made explicit by defining a scalar field $a$ (axion) through
\begin{equation}
 \epsilon_{\mu\nu\rho\sigma}\partial^{\sigma} a \propto H_{\mu\nu\rho}\: ,
\end{equation}
where $H$ is the field strength defined in \Eqref{eq:field_strength_sugra}. In superfield notation, the axion becomes the imaginary part of the scalar component of the dilaton superfield $S$. As in ten-dimensions, a shift in $a$ can be used to cancel anomalies. 

\subsubsection{Continuous Case}
We will first discuss how the Green-Schwarz mechanism can be used to cancel the anomalies associated with an anomalous $\U1_{\text{anom}}$ gauge symmetry. Consider a gauge theory with gauge group $G_\text{gauge}=G\times \U1_{\text{anom}}$ where $G$ is a simple group or a \U1. The dilation superfield contains the axion $a$ in its scalar component
\begin{equation}
 S|_{\theta=0}=s+\I a\:.
\end{equation}
The part of the Lagrangian which depends on $S$ is given by~\cite{Kobayashi:1996pb}
\begin{equation}
\mathscr{L}_\text{S}=-\int \D^4 \theta \, \ln(S+\bar{S}-\delta_{\text{GS}} V_{\text{anom}}) + \int \D^2\theta\, \frac{S}{4}\left(W^\alpha W_{\alpha} + \text{h.c.} \right) 
\end{equation}
where $W^\alpha$ is the superfield strength. $V_{\text{anom}}$ is the gauge superfield corresponding to $\U1_{\text{anom}}$. The axion has the following couplings~\cite{Witten:1985xb,Witten:1984dg}
\begin{equation}
 \mathscr{L}
 ~\supset~
 -\frac{a}{8} F_\mathrm{anom} \widetilde{F}_\mathrm{anom} -\frac{a}{8} F^{a} \widetilde{F}^a  +
 \frac{a}{4}  \mathcal{R}\widetilde{\mathcal R}\;,
\end{equation}
where $F$ and $F_\mathrm{anom}$ denote the gauge field
strength of $G$ and $\U1_\mathrm{anom}$ respectively. Index contractions have been suppressed as explained below \Eqref{eq:anomaly_function}. These couplings allow us to cancel all anomalies associated with $\U1_{\text{anom}}$. A $\U1_{\text{anom}}$ gauge transformation is given by
\begin{subequations}
\begin{align}
 V_{\text{anom}} \quad &\to \quad V_{\text{anom}} + \frac{\I}{2}(\Lambda+\Lambda^\dagger) \:, \label{eq:TrafoVanom}\\
 S \quad &\to \quad S + \frac{\I}{2} \delta_{\text{GS}} \Lambda \:, \\
 \Phi^{(i)} \quad & \to \quad  e^{\I Q^{(i)}_\mathrm{anom}\Lambda}\,\Phi^{(i)} \:,
\end{align}
\end{subequations}
where $\Phi^{(i)}$ denotes the chiral superfields in the theory with $\U1_{\text{anom}}$ charge $Q^{(i)}_\mathrm{anom}$. The chiral superfield $\Lambda$ parameterizes the gauge transformation. Hence, under a
$\U1_\mathrm{anom}$ transformation with parameter $\alpha=\re \Lambda|_{\theta=0}$ the axionic Lagrangian shifts by
\begin{equation}
  \Delta\mathscr{L}_{\mathrm{axion}}  ~=~  
  -\frac{\alpha}{16}\delta_\mathrm{GS}\, F_\mathrm{anom} \widetilde{F}_\mathrm{anom} 
  -\frac{\alpha}{16}\delta_\mathrm{GS}\, F^{a} \widetilde{F}^a  
  + \frac{\alpha}{8}\delta_\mathrm{GS}\,  \mathcal{R}\widetilde{\mathcal R}
  \;.
\end{equation}
In addition, the anomaly leads to a shift in the Lagrangian (cf.\ section~\ref{sec:anomalies_setup})
\begin{eqnarray}
 \Delta\mathscr{L}_{\text{anomaly}} &= &  \frac{\alpha}{32\pi^2} F_\mathrm{anom} \widetilde{F}_\mathrm{anom}\, \,A_{\U1_\mathrm{anom}^3} \nonumber \\
&&{} + \frac{\alpha}{32\pi^2} F^{a} \widetilde{F}^a  \, A_{G-G-\U1_\mathrm{anom}} \nonumber \\
&&{} - \frac{\alpha}{384\pi^2}  \mathcal{R}\widetilde{\mathcal R}  \, A_{\text{grav}-\text{grav}-\U1_\mathrm{anom}}
\;.
\end{eqnarray}
The coefficients $A$ are the anomaly coefficients given by (see equations \eqref{eq:u1_g_g_anomaly_condition}, \eqref{eq:U1_cubed_anomaly} and \eqref{eq:grav_u1_anomaly})
\begin{subequations}
\label{eq:anomaly_coefficients}
\begin{align}
 A_{\U1_\mathrm{anom}^3} & = \frac{1}{3}\sum_i (Q_\mathrm{anom}^{(i)})^3 ~=~\frac{1}{3} \tr
 Q_\mathrm{anom}^3\;, \\
 A_{\text{grav}-\text{grav}-\U1_\mathrm{anom}} & = \sum_i Q_\mathrm{anom}^{(i)} ~=~ \tr Q_\mathrm{anom}\;, \\
 A_{G-G-\U1_\mathrm{anom}} & = \sum_{\boldsymbol{r}^{(f)}} \ell(\boldsymbol{r}^{(f)}) \, Q_\mathrm{anom}^{(f)} \:.
\end{align} 
\end{subequations}
The first two sums run over all left-handed Weyl fermions while the last sum runs over all irreducible representations $\boldsymbol{r}^{(f)}$ of $G$ and $\ell(\boldsymbol{r}^{(f)})$ is the Dynkin index.

The axion shift allows us to cancel the grav-grav-$\U1_{\mathrm{anom}}$,
$\U1_{\mathrm{anom}}^3$ and $G-G-\U1_{\mathrm{anom}}$ anomalies by demanding
$\Delta\mathscr{L}_{\mathrm{anomaly}} + \Delta\mathscr{L}_{\mathrm{axion}}=0$ which implies
\begin{equation}
 \delta_{\text{GS}}
 ~=~
 \frac{1}{48\pi^2}\tr Q_\mathrm{anom}
 ~=~
 \frac{1}{2\pi^2}\frac{1}{3}\tr Q_\mathrm{anom}^3 
 ~=~
 \frac{1}{2\pi^2}A_{G-G-\U1_\mathrm{anom}} \;.
\end{equation}
This is in agreement with~\cite{Lerche:1987sg,Lalak:1999bk,Dundee:2010sb,Maekawa:2001uk,Campbell:2000fd}. However in the literature an incorrect value, which differs by a factor of four, is often reported~\cite{ArkaniHamed:1998nu,Binetruy:1996uv,Kobayashi:1996pb}.

\subsubsection{Discrete Case}
The Green-Schwarz mechanism also works if we replace $\U1_{\mathrm{anom}}$ by a
discrete \Z{N}. In this case the parameter $\alpha$ is no longer continuous, but
$\alpha=\frac{2\pi}{N}$. Of course, there is no gauge
field associated with the \Z{N}, i.e.\ \Eqref{eq:TrafoVanom} does not apply
here. The discussion then goes as in the previous subsection. The fields transform according to
\begin{subequations}
\begin{align}
 S \quad &\to \quad S + \frac{\I}{2} \Delta_{\text{GS}} \:, \\
 \Phi^{(i)} \quad & \to \quad  e^{\I \frac{2\pi}{N} q^{(i)}}\,\Phi^{(i)} \:.
\end{align}
\end{subequations}
The \Z{N} charged are denoted by $q^{(i)}$. The Green-Schwarz
constant is now defined only modulo $\eta$
\begin{equation} \label{eq:discrete_GS_condition}
 \pi N \Delta_{\text{GS}}
 ~ \equiv ~
 \frac{1}{24}\,
 A_{\text{grav}-\text{grav}-\Z{N}} 
 ~ = ~
 A_{G-G-\Z{N}} 
 \mod \eta
 \;.
\end{equation}
The anomaly coefficients can be obtained from \Eqref{eq:anomaly_coefficients} by
replacing the $\U1_{\mathrm{anom}}$ charges $Q_\mathrm{anom}^{(m)}$ by the \Z{N}
charges $q^{(m)}$. Remember that $\eta$ is given by
\begin{equation}
 \eta = \left\lbrace \begin{array}{ll}N, & N \:\text{odd} \\ \frac{N}{2}, &  N \:\text{even}\end{array} \right. \: .
\end{equation}

\section{Anomaly Matching}
\label{sec:anomaly_matching}
The term anomaly matching refers to the fact that the calculation of anomaly coefficients is independent of the energy scale. This concept proofs very powerful when comparing a fundamental theory with its proposed low energy theory. For example, anomaly matching conditions put highly non-trivial constraints on the massless fermion content of a confining theory. Anomaly matching was first introduced by t'Hooft~\cite{'tHooft:1980xb} for continuous symmetries and was later generalized to discrete symmetries~\cite{Csaki:1997aw}.

Let us briefly summarize how anomaly matching works. Consider a high energy theory with simple gauge group $G_\text{gauge}$ and a global symmetry group $G_\text{global}$ which may be a direct product of discrete and continuous groups. Suppose that $G_\text{gauge}$ undergoes a change towards lower energies, e.g.\ it confines or gets spontaneously broken. Then all anomalies of type $G_\text{global}-G_\text{global}-G_\text{global}$ must have precisely the same value in both energy regimes. 

\paragraph{Example}
We will apply anomaly matching to a model presented in~\cite{Mohapatra:2007vd}. The model is a \SO{10} SUSY GUT with a global \Z{6} symmetry which is introduced in order to suppress proton decay. The field content before GUT breaking is given in table~\ref{tab:mohapatra_model1}. $\psi_m$ contains one generation of standard model matter that is there are $N_g=3$ of them. $H$ and $H'$ comprise the MSSM Higgs fields. The remaining fields are needed to ensure proper \SO{10} breaking. 
\begin{table}[H]
\centering
 \begin{tabular}{cccccccc}
\toprule[1.3pt]
& $\psi_m$ & $H$ & $H'$ & $\psi_H$ & $\overline{\psi}_H$ & $A$ & $S$ \\ \cmidrule{2-8}
\SO{10} & $\mathbf{16}$ & $\mathbf{10}$ & $\mathbf{10}$ & $\mathbf{16}$ & $\mathbf{\overline{16}}$& $\mathbf{45}$& $\mathbf{54}$\\
\Z{6} & 1 & 4 & 2 & 4 & 2 & 0 & 0\\
\bottomrule[1.3pt]
\end{tabular}
\caption{The field content of an \SO{10} gauge theory with a global \Z{6} symmetry.}
\label{tab:mohapatra_model1}
\end{table}
The model is anomaly free because by evaluating the anomaly coefficients in \Eqref{eq:anomaly_conditions_ZN} we get
\begin{subequations}
\begin{align}
\SO{10}-\SO{10}-\Z{6}&:\qquad \underbrace{N_g\cdot 1 \cdot 2}_{\psi_m} + \underbrace{4 \cdot 1}_{H} + \underbrace{2 \cdot 1}_{H'} + \underbrace{4\cdot 2}_{\psi_H}  + \underbrace{2 \cdot 2}_{\overline{\psi}_H}  = 0 \mod 3\:, \\
\text{grav}-\text{grav}-\Z{6} &: \qquad \underbrace{N_g\cdot 16 \cdot 1}_{\psi_m} + \underbrace{10 \cdot 4}_{H} + \underbrace{10\cdot 2}_{H'} + \underbrace{16 \cdot 4}_{\psi_H} + \underbrace{16 \cdot 2}_{\overline{\psi}_H} = 0 \mod 3\:.
\end{align}
\end{subequations}
Since \SO{10} has rank 5 there is a $\U1_X$ factor in addition to the SM gauge group. This $\U1_X$ gets broken when the appropriate components of the Higgs fields attain a VEV. To identify these components we look at the branching rules~\cite{Slansky:1981yr}
\begin{subequations}
\begin{align}
\SO{10}  & \supset \SU{5} \times \U1_X \\
\mathbf{16} & \rightarrow \mathbf{10}_{-1} + \mathbf{\overline{5}}_3 +\mathbf{1}_{-5}\\
\mathbf{45} & \rightarrow \mathbf{24}_{0} + \mathbf{\overline{10}}_{-4} +\mathbf{10}_{4}+ \mathbf{1}_{0}\\
\mathbf{54} & \rightarrow \mathbf{24}_{0} + \mathbf{\overline{15}}_{-4} +\mathbf{15}_{4}\:.
\end{align}
\end{subequations}
In order to break \SO{10} to the SM gauge group, the \SU5 singlets and the \SU5 adjoints (the $\mathbf{24}$) should get a VEV. We want to investigate what happens to the remaining $\U1_X\times \Z6$ in this case. The only VEV fields which are charged under $\U1_X\times \Z6$ are the \SU5 singlets in $\psi_H$ and $\overline{\psi}_H$. Hence, we have a situation where two fields with charges $(-5,4)$ and $(5,2)$ obtain a VEV. By the methods of section~\ref{sec:obtaining_abelian_discrete_symmetries} we see that the $\U1_X\times \Z6$ gets broken to a \Z{30}. The SM matter in $\psi_m$ and the SM Higgses in $\psi_H$, $\overline{\psi}_H$ have charges under this \Z{30} as displayed in table~\ref{tab:mohapatra_model2}.
\begin{table}[H]
\centering
 \begin{tabular}{cccccccc}
\toprule[1.3pt]
 & $q$ & $\bar{u}$ & $\bar{d}$ & $\ell$ & $\bar{e}$ & $h_u$ & $h_d$ \\\cmidrule{2-8}
\Z{30} & 23 & 23 & 1 & 1 & 23 & 14 & 6\\
$\Z{5}\times\Z6$ & (3,1) & (3,1) & (1,5) & (1,5) & (3,1) & (4,4) & (1,0)\\
\bottomrule[1.3pt]
\end{tabular}
\caption{The $\Z{30}\cong\Z5\times\Z6$ charges of the MSSM field content.}
\label{tab:mohapatra_model2}
\end{table}
\noindent Now this field content has an anomaly w.r.t. the standard model \SU{2}
\begin{equation}
 \SU2-\SU2-\Z{30}: \qquad \underbrace{N_g \cdot 23 \cdot \frac{1}{2}}_{q} \cdot 2 + \underbrace{N_g\cdot 1 \cdot \frac{1}{2}}_{\ell}  + \underbrace{14\cdot \frac{1}{2}}_{h_u} + \underbrace{6 \cdot \frac{1}{2}}_{h_d} \neq 0 \mod 15 \:. 
\end{equation}
This signals an inconsistency because anomaly matching tells us that by spontaneously breaking \SO{10}, we cannot generate an anomaly. Hence, either further light states in addition to the ones in table~\ref{tab:mohapatra_model2} have to be present, or the \Z{30} cannot be exact, i.e.\ we have to introduce further fields that attain VEVs.

\section{Discussion}
Anomalies of discrete symmetries were first discussed by Krauss and Wilczek~\cite{Krauss:1988zc}. Their treatment was based on black hole physics. Iba\~{n}ez and Ross~\cite{Ibanez:1991hv} wrote down explicit anomaly constraints for \Z{N} symmetries in term of charges, similar to the ones in section~\ref{sec:anomaly_conditions_abelian}. They derived their findings by embedding the \Z{N} in a \U1 gauge symmetry. This leads to constraints which are non-linear in the \Z{N} charges because in the underlying theory, the \U1-\U1-\U1 anomaly (cf.\ \Eqref{eq:U1_cubed_anomaly}) has to vanish. 

Banks and Dine~\cite{Banks:1991xj} pointed out that the non-linear constraints are useless from the low energy point of view because they depend on the chosen embedding. Additionally, they observed that the anomaly conditions have to be fulfilled only modulo $N/2$. Banks and Dine argued that, if $N$ is even, one can introduce fermions with charge $N/2$ which would contribute to the anomaly although they can be made arbitrarily heavy. 

The question about the correct anomaly conditions was answered by Araki et al.\ \cite{Araki:2008ek} who treated anomalies in the path integral following Fujikawa. They confirmed the findings of Banks and Dine that there are no non-linear constraints and that the anomaly constraints have to be fulfilled modulo $\eta$.

The connection between instantons and anomalies goes back to t'Hooft~\cite{'tHooft:1976up} who constructed an effective interaction induced by instantons. For the case of discrete anomalies the relation between instantons and anomalies (cf.\ section~\ref{sec:instantons_path_integral}) has not been worked out in the literature before, although~\cite{Preskill:1991kd} briefly touches upon this subject.

Another comment concerns the discrete Green-Schwarz mechanism. Although a couple of authors~\cite{Banks:1991xj,Ibanez:1992ji} mention the existence of a discrete analog of the Green-Schwarz mechanism, it has not been worked out with all numerical factors and not in a path integral approach as in section~\ref{sec:discrete_GS}. Especially, \Eqref{eq:discrete_GS_condition} has not been reported in the literature before, although an incorrect matching condition is given in~\cite[equation (16)]{Ibanez:1992ji}.

\clearpage
\thispagestyle{empty}

\chapter{Application to the MSSM}
\label{chap:application_MSSM}
In the next section we will apply the methods of chapters~\ref{chap:abelian} and \ref{chap:anomalies} to supersymmetric extensions of the SM.

\section{Problems of the MSSM}
Supersymmetric extensions of the SM, in its simplest version the minimal supersymmetric standard model (MSSM), solve the hierarchy problem. Additionally, in the MSSM the gauge couplings seem to unify at high energies ($\sim$ \unit[$10^{16}$]{GeV})~\cite{Langacker:1991an}, supporting the idea of grand unified theories (GUTs)~\cite{Ross:1985ai}.

However, the MSSM also introduces potential problems. Among them are the $\mu$-problem and baryon- and lepton number violation. Let us shortly review these issues. To clarify the notation, the labeling of the MSSM fields is given in table~\ref{tab:MSSM_labels}.
\begin{table}[htb]
\centering
\begin{tabular}{ccccccccc}
\toprule[1.3pt]
 Label & $q_i$ & $\bar u_i$ & $\bar d_i$  & $\ell_i$  & $\bar e_i$ & $h_u$ & $h_d$ \\ \midrule
$\SU3_C$ & $\rep{3}$ & $\crep{3}$  & $\crep{3}$ & $\rep{1}$  & $\rep{1}$ & $\rep{1}$  & $\rep{1}$ \\
 $\SU2_L$ & $\rep{2}$ & $\rep{1}$  & $\rep{1}$ & $\rep{2}$  & $\rep{1}$ & $\rep{2}$  & $\rep{2}$\\
 $\U1_Y$ & $\tfrac{1}{6}$ & -$\tfrac{2}{3}$ & $\tfrac{1}{3}$  & -$\tfrac{1}{2}$ & 1& $\tfrac{1}{2}$ & -$\tfrac{1}{2}$\\
\bottomrule[1.3pt]
\end{tabular}
\caption{The matter and Higgs content of the MSSM with their gauge quantum numbers. The index $i=\{1,2,3\}$ labels the generations.}
\label{tab:MSSM_labels}
\end{table}

The most general superpotential up to dimension five, allowed by the gauge symmetry, reads~\cite{Weinberg:1979sa,Sakai:1981pk}
\begin{subequations}
\label{eq:W_MSSM}
\begin{align}
 \mathscr{W} & = \mu \, h_u h_d + \mu_i' \, h_u \ell_i \label{eq:MSSM_muterm}\\
 & \quad + (Y_u)^{ij}\, q_i h_u \bar u_j + (Y_d)^{ij}\, q_i h_d \bar d_j + (Y_e)^{ij}\, \ell_i h_d \bar e_j \label{eq:MSSM_Yukawa_couplings} \\
 & \quad + \lambda^{(1)}_{ijk}\,\ell_i \ell_j \bar{e}_k + \lambda^{(2)}_{ijk}\, q_i \bar d_j \ell_k + \lambda^{(3)}_{ijk} \, \bar u_i \bar d_j \bar d_k \label{eq:MSSM_b+l_dim4}\\
 & \quad + \kappa^{(1)}_{ijkl} \, q_i q_j q_k \ell_l + \kappa^{(2)}_{ijkl} \, \bar u_i \bar u_j \bar d_k \bar e_l + \kappa^{(3)}_{ijk} \, q_i q_j q_k h_d + \kappa^{(4)}_{ijk}\, q_i \bar u_j \bar e_k h_d + \kappa^{(5)}_{ij} \ell_i \ell_j h_u h_u \label{eq:MSSM_b+l_dim5}
\end{align} 
\end{subequations}
where the indices $i,j,k,l$ label the generations. By phenomenology, the parameter $\mu$ is required to have a value near the scale of electroweak symmetry breaking $\mathcal{O}(\unit[100]{GeV})$. Since the $\mu$-term is allowed by supersymmetry there is, a priori, no reason why $\mu$ should be small compared to the Planck or GUT scale. Hence, $\mu$ reintroduces the hierarchy problem which is known as the $\mu$-problem~\cite{Kim:1983dt}. Also the parameter $\mu_i'$ has to be much smaller than the eletroweak scale because it contributes to neutrino masses~\cite{Hall:1983id}.

Another problem is caused by the baryon- and lepton number violating couplings in equation~\eqref{eq:MSSM_b+l_dim4} and \eqref{eq:MSSM_b+l_dim5}. The couplings $\lambda^{(2)}$, $\lambda^{(3)}$ and $\kappa^{(1)}$ have to be highly suppressed because they mediate proton decay~\cite{Hinchliffe:1992ad}, e.g.\ through the processes $p\to e^+ + \pi^0$ or $p\to K^+ + \bar \nu$ whose life time is known to be greater than \unit[$10^{32}$]{years}~\cite{Nakamura:2010zz}.

\subsection{Possible Solutions}
There are two well-known classes of solutions to the $\mu$-problem. The $\mu$-term can be generated by 
\begin{itemize}
\item[\ding{172}] a term $Y h_u h_d$ in the superpotential~\cite{Kim:1983dt} (Kim-Nilles mechanism),
\item[\ding{173}] a term $X^\dagger h_u h_d$ in the Kähler potential~\cite{Giudice:1988yz} (Giudice-Masiero mechanism). 
\end{itemize}
$X$ and $Y$ can be single fields or composite operators. The $\mu$-term is then generated if $X$ or $Y$ acquire a VEV.

To evade the strongest problems with proton decay, the MSSM is usually understood to be endowed with matter parity~\cite{Farrar:1978xj,Dimopoulos:1981zb,Dimopoulos:1981dw}. The charge assignment is given in table~\ref{tab:MSSM_discrete_symmetries}. Matter parity basically distinguishes the matter fields from the Higgses. Hence, matter parity forbids the $\mu'$ coupling in \Eqref{eq:MSSM_muterm}, all dimension four baryon- and lepton number violating operators in \Eqref{eq:MSSM_b+l_dim4} and the dimension five couplings $\kappa^{(3)}$ and $\kappa^{(4)}$ in \Eqref{eq:MSSM_b+l_dim5}. However, $\Z2^\mathcal{M}$ still allows for the dimension five proton decay operators $q\,q\,q\,\ell$ and $\bar u \,\bar u \,\bar d \, \bar e$.

In a subsequent study, Iba\~{n}ez and Ross~\cite{Ibanez:1991pr} found an anomaly free, abelian symmetry, called baryon triality $\Z3^\mathcal{B}$, which forbids all baryon number violating dimension four and five operators that is $\lambda^{(3)}=\kappa^{(1)}=\kappa^{(2)}=\kappa^{(3)}=0$.
\begin{table}[htb]
\centering
\begin{tabular}{ccccccccc}
\toprule[1.3pt]
  & $q_i$ & $\bar u_i$ & $\bar d_i$  & $\ell_i$  & $\bar e_i$ & $h_u$ & $h_d$ \\ \midrule
$\Z2^\mathcal{M}$ & 1 & 1 & 1 & 1 & 1 & 0 & 0 \\
 $\Z3^\mathcal{B}$ & 0 & 2 & 1 & 2 & 2 & 1 & 2\\
 $\Z6^\mathcal{P}$ & 0 & 1 & 5 & 4 & 1 & 5 & 1\\
\bottomrule[1.3pt]
\end{tabular}
\caption{The charge assignment of matter parity $\Z2^\mathcal{M}$, baryon triality $\Z3^\mathcal{B}$ and proton hexality $\Z6^\mathcal{P}$.}
\label{tab:MSSM_discrete_symmetries}
\end{table}

The direct product of matter parity and baryon triality has been rediscovered in~\cite{Dreiner:2005rd} and is known as proton hexality $\Z6^\mathcal{P}$. Note that from the charge assignment in table~\ref{tab:MSSM_discrete_symmetries} it is not obvious that $\Z6^\mathcal{P} = \Z2^\mathcal{M} \times \Z3^\mathcal{B}$ because the freedom of shifting the \Z{N} charges by hypercharge is used (see appendix~\ref{app:anomaly_mixing}). Proton hexality only allows for the $\mu$-term, the Yukawa couplings and the Weinberg operator $\ell\,h_u \,\ell\, h_u$, while forbidding all other operators in \Eqref{eq:W_MSSM}. It is interesting that $\Z6^\mathcal{P}$ is the unique, anomaly-free, abelian, discrete symmetry with these features.

However, all symmetries in table~\ref{tab:MSSM_discrete_symmetries} have drawbacks. Matter parity $\Z2^\mathcal{M}$ does not forbid the dimension five operators $q\,q\,q\,\ell$ and $\bar u \,\bar u \,\bar d \, \bar e$ which have to be highly suppressed. Both other symmetries $\Z3^\mathcal{B}$ and $\Z6^\mathcal{P}$ are not compatible with the idea of grand unification because matter fields have different charges\footnote{Note that this does not follow trivially from the charges in table~\ref{tab:MSSM_discrete_symmetries}. Again, there is still the freedom of shifting the discrete charges by hypercharge (see appendix~\ref{app:anomaly_mixing}).}. In addition, all symmetries in table~\ref{tab:MSSM_discrete_symmetries} allow the $\mu$-term and are therefore not able to make any statement about the size of $\mu$. 

\section[Discrete Symmetries Commuting with $\SU5$]{Discrete Symmetries Commuting with $\boldsymbol{\SU5}$}
\label{sec:sym_commuting_with_SU5}
In this section we will discuss abelian, discrete symmetries in the MSSM which are reconcilable with the idea of grand unification. We will treat a \Z{N} symmetry whose charge assignment is compatible with the grand unified group \SU5~\cite{Georgi:1974sy} which is the smallest, simple group which contains the SM gauge group. Our notation for the field content and their charges is defined in table~\ref{tab:MSSM_GUT_charges}. This also includes the case of an \SO{10} GUT~\cite{Fritzsch:1974nn} in which case $q_{\rep{10}}=q_{\crep{5}}$.
\begin{table}[htb]
\centering
\begin{tabular}{cccccccccc}
\toprule[1.3pt]
  & $q_i$ & $\bar u_i$ & $\bar d_i$  & $\ell_i$  & $\bar e_i$ & $h_u$ & $h_d$ & $\theta$ \\ 
$\Z{N}$ & $q_{\rep{10}}$ & $q_{\rep{10}}$ & $q_{\crep{5}}$ & $q_{\crep{5}}$ & $q_{\rep{10}}$ & $q_{h_u}$ & $q_{h_d}$ & $q_\theta$ \\
\bottomrule[1.3pt]
\end{tabular}
\caption{The charge assignment of a \Z{N} symmetry which commutes with \SU{5}. The subscripts $\rep{10}$ and $\crep{5}$ indicate the \SU5 representations. $\theta$ is the superspace coordinate.}
\label{tab:MSSM_GUT_charges}
\end{table}
At this stage, the \Z{N} can be an $R$-symmetry or not. Therefore, we introduce the superspace coordinate $\theta$ (cf.\ section~\ref{sec:anomaly_R_symmetries}). For the case of an $R$-symmetry we use the convention $q_\theta=1$, i.e.\ the superpotential has charge two, and $q_\theta=0$ for \Z{N} symmetries which commute with supersymmetry.

A close inspection of table~\ref{tab:MSSM_GUT_charges} reveals that the Higgs fields $h_u$ and $h_d$ do not constitute full \SU5 representations. This means, we assume the usual doublet-triplet splitting problem to be solved. Some solutions to this problem are known~\cite{Dimopoulos:1981xm,Masiero:1982fe} although probably the most attractive mechanism is to consider orbifold GUTs in higher dimensions~\cite{Kawamura:1999nj,Asaka:2001eh}.

\subsection{Anomaly Coefficients}
Let us first evaluate the anomaly constraints as given in section~\ref{sec:anomaly_conditions_abelian}. We will use the following abbreviations for the anomaly coefficients:
\begin{itemize}
 \item $A_3:=$ $\SU3_C$ - $\SU3_C$ - \Z{N} ,
 \item $A_2:=$ $\SU2_L$ - $\SU2_L$ - \Z{N} ,
 \item $A_1:=$ $\U1_Y$  - $\U1_Y$ - \Z{N} ,
 \item $A_0^{\text{MSSM}}:=$  \text{grav} - \text{grav} - \Z{N} .
\end{itemize}
The coefficient $A_0^{\text{MSSM}}$ includes only the MSSM fields and the gravitino, and does not incorporate other states outside the MSSM. These coefficients can be calculated with the aid of \Eqref{eq:anomaly_R_symmetries} to be
\begin{subequations}
\label{eq:MSSM_anomaly_coefficients}
\begin{align}
A_3 & = 3 \cdot \frac{1}{2} \big( 2\underbrace{(q_{\rep{10}}-q_\theta)}_{q} + \underbrace{(q_{\rep{10}}-q_\theta)}_{\bar u} + \underbrace{(q_{\crep{5}}-q_\theta)}_{\bar d} \big) + \underbrace{3q_\theta}_{\text{gauginos}} \nonumber\\
 & =  \frac{9}{2}q_{\rep{10}} + \frac{3}{2}q_{\crep{5}} - 3q_\theta \: , \displaybreak[0]\\
A_2 & = 3 \cdot \frac{1}{2} \big( 3 \underbrace{(q_{\rep{10}}-q_\theta)}_{q} + \underbrace{(q_{\crep{5}}-q_\theta)}_{\ell}) + \frac{1}{2} ((q_{h_u}-q_\theta)+(q_{h_d}-q_\theta) \big) + \underbrace{2q_\theta}_{\text{gauginos}} \nonumber \\
 & = \frac{9}{2}q_{\rep{10}} + \frac{3}{2}q_{\crep{5}} + \frac{1}{2} (q_{h_u}+q_{h_d}) - 5q_\theta \: , \displaybreak[0]\\
A_1 &= \frac{3}{5} \Bigg\{ 3 \Big( \ 6\cdot \underbrace{\left(\frac{1}{6}\right)^2 \cdot (q_{\rep{10}}-q_\theta)}_{q} + 3\cdot \underbrace{\left(-\frac{2}{3}\right)^2 \cdot (q_{\rep{10}}-q_\theta)}_{\bar u} + 3\cdot \underbrace{\left(\frac{1}{3}\right)^2 \cdot (q_{\crep{5}}-q_\theta)}_{\bar d}  \nonumber  \\
 & \quad + 2\cdot \underbrace{\left(-\frac{1}{2}\right)^2 \cdot (q_{\crep{5}}-q_\theta)}_{\ell} + \underbrace{1^2 \cdot (q_{\rep{10}}-q_\theta)}_{\bar e}  \Big) + 2\cdot \left( \frac{1}{2} \right)^2  (q_{h_u}-q_\theta) \nonumber \\
 & \quad + 2\cdot \left( -\frac{1}{2} \right)^2  (q_{h_d}-q_\theta) \Bigg\}  =   \frac{9}{2}q_{\rep{10}} + \frac{3}{2}q_{\crep{5}}  + \frac{3}{5} \left( \frac{1}{2} (q_{h_u}+q_{h_d}) - 11 q_\theta \right)  \: , \displaybreak[0]\\
A_0^{\text{MSSM}} & = 3\cdot \big(6\underbrace{(q_{\rep{10}}-q_\theta)}_{q} + 3\underbrace{(q_{\rep{10}}-q_\theta)}_{\bar u}) + 3 \underbrace{(q_{\crep{5}}-q_\theta)}_{\bar d} + 2 \underbrace{(q_{\crep{5}}-q_\theta)}_{\ell} + \underbrace{(q_{\rep{10}}-q_\theta)}_{\bar e} \big) \nonumber \\
 &\quad + 2(q_{h_u}-q_\theta)+ 2 (q_{h_d}-q_\theta) - \underbrace{21 q_\theta}_{\text{gravitino}} + \underbrace{8 q_\theta + 3 q_\theta + q_\theta}_{\text{gauginos}} \nonumber \\
 & = 30 q_{\rep{10}} + 15 q_{\crep{5}} + 2(q_{h_u}+  q_{h_d}) - 58 q_\theta \: . \label{eq:MSSM_anomaly_coefficients_A0}
\end{align}
\end{subequations}
In the hypercharge anomaly $A_1$, the $\U1_Y$ charges have been rescaled by a factor $\sqrt{5/3}$ (see also the discussion in section~\ref{sec:anomaly_conditions_abelian}) because we assume a GUT and hence gauge coupling unification. Note that the full gravitational anomaly coefficient receives contributions from all states which are charged under \Z{N}.

If we allow for anomaly cancellation via the Green-Schwarz mechanism (cf.\ section~\ref{sec:discrete_GS}), the anomaly constraints amount to 
\begin{subequations}
\label{eq:MSSM_anomaly_conditions}
\begin{alignat}{2}
  A_2 = A_3 &= \rho &\mod \eta \: , \label{eq:MSSM_anomaly_conditions_A2A3}\\
  5 A_1 &= 5 \rho &\mod \eta \: ,\label{eq:MSSM_anomaly_conditions_A1}\\
 A_0^\text{full} &= 24 \rho &\mod \eta \: , \label{eq:MSSM_anomaly_conditions_A0}
\end{alignat}
\end{subequations}
with $\rho\neq 0$ meaning that the Green-Schwarz mechanism is at work. By $A_0^\text{full}$ we mean the full gravitational anomaly coefficient, possibly including contributions from MSSM singlets. As in chapter~\ref{chap:anomalies} we have defined
\begin{equation}
 \eta = 
\left\lbrace 
\begin{array}{ll} 
N & , \quad N\: \text{odd} \\
\frac{N}{2} & ,\quad N\: \text{even}
\end{array}
\right. \:.
\end{equation}

\subsection[Non-$R$ symmetries]{Non-$\boldsymbol{R}$ Symmetries}
In this section we will show that anomaly-free, non-$R$ symmetries compatible with grand unification cannot address the $\mu$-problem. To do so, observe that from \Eqref{eq:MSSM_anomaly_conditions} follows
\begin{equation}
 A_2 - A_3 = 0 \mod \eta \:. 
\end{equation}
Inserting the values for the anomaly coefficients given in \Eqref{eq:MSSM_anomaly_coefficients} leads to (remember that $q_\theta=0$ because we consider non-$R$ symmetries)
\begin{equation}
 \frac{1}{2} (q_{h_u}+q_{h_d})= 0 \mod \eta \:.  \label{eq:MSSM_nonR_anomaly}
\end{equation}
On the other hand, the $\mu$-term is allowed if
\begin{equation}
 q_{h_u}+q_{h_d}= 0 \mod N \:.\label{eq:MSSM_nonR_mu_term}
\end{equation}
Since \Eqref{eq:MSSM_nonR_anomaly} implies \Eqref{eq:MSSM_nonR_mu_term}, a non-$R$ symmetry, which is anomaly-free, cannot forbid the $\mu$-term (cf.\ the similar discussion in~\cite{Hall:2002up}) even if we allow for anomaly cancellation via the Green-Schwarz mechanism. 

\subsection{Order Constraints}
Having seen that non-$R$ symmetries cannot be used to address the $\mu$ problem, we focus on $R$-symmetries. We will derive constraints in the order $N$ of an anomaly-free $\Z{N}^R$ symmetry in the MSSM. 

Again, we use \Eqref{eq:MSSM_anomaly_coefficients} and \Eqref{eq:MSSM_anomaly_conditions} to obtain (now with $q_\theta=1$)
\begin{subequations}
\begin{align}
A_2 - A_3 &= \frac{1}{2}(q_{h_u}+q_{h_d})=2 \mod \eta\;, \label{eq:MSSM_order_constraints1} \\
5(A_1 - A_3) &= 3 \left( \frac{1}{2}(q_{h_u}+q_{h_d}) -6\right) = 0 \mod \eta \:.\label{eq:MSSM_order_constraints2}
\end{align}
\end{subequations}
By inserting \Eqref{eq:MSSM_order_constraints1} into \Eqref{eq:MSSM_order_constraints2} we arrive at
\begin{equation}
 12=0\mod \eta\:. \label{eq:MSSM_order_constraint}
\end{equation}
Equation~\eqref{eq:MSSM_order_constraint} means that $N$ has to be a divisor of 24, i.e.\ $N\in\{1,2,3,4,6,8,12,24\}$. Of course, the case $N=1$ implies that the symmetry is trivial and we will therefore not consider it any further. Additionally, we do not consider $\Z2^R$ symmetries. The superpotential transforms trivially under a $\Z2^R$ symmetry, and hence a $\Z2^R$ symmetry is not meaningful (see the discussion in~\cite{Dine:2009swa}).

\subsection{Classification}
\label{sec:MSSM_classification}
Now we can treat all possible $\Z{N}^R$ symmetries in the MSSM which meet the following criteria:
\begin{itemize}
\item[\ding{172}] All mixed gauge$ \,-\,\Z{N}^R$ anomalies cancel with or without Green-Schwarz mechanism,
\item[\ding{173}] The Yukawa couplings as well as the Weinberg operator $\ell h_u \ell h_u$ are allowed.
\end{itemize}
At this point, we will not take mixed gravitational-$\Z{N}^R$ anomaly into account because it relies on the field content outside the SM. Hence, we end up with the following set of equations in addition to the anomaly freedom conditions in \Eqref{eq:MSSM_anomaly_conditions_A2A3} and \Eqref{eq:MSSM_anomaly_conditions_A1}
\begin{subequations}
\label{eq:MSSM_classification}
\begin{alignat}{2}
\text{up-type Yukawa} \,:\quad &  2 q_{\rep{10}} + q_{h_u}& =2  \mod N \: , \label{eq:MSSM_classification_up_Yukawa}\\
\text{down-type Yukawa} \,:\quad &  q_{\rep{10}} + q_{\crep{5}}+ q_{h_d}& =2 \mod N \: , \label{eq:MSSM_classification_down_Yukawa} \\
\text{Weinberg operator} \,:\quad & 2( q_{\crep{5}}+ q_{h_u})& =2  \mod N \: .
\end{alignat}
\end{subequations}
Before presenting the solutions, let us comment on a general implication of these equations. By adding \Eqref{eq:MSSM_classification_up_Yukawa} and \Eqref{eq:MSSM_classification_down_Yukawa} we arrive at
\begin{equation}
3 q_{\rep{10}} + q_{\crep{5}} + q_{h_u} + q_{h_d} = 4 \mod N \:.
\end{equation}
Anomaly freedom (cf.\ \Eqref{eq:MSSM_order_constraints1}) then gives us
\begin{equation}
 3 q_{\rep{10}} + q_{\crep{5}} = 0 \mod N \: .
\end{equation}
However, this means that the dimension five nucleon decay operators $qqq\ell$ and $\bar u \bar u \bar d \bar e$, which in \SU{5} language read $\rep{10}\,\rep{10}\,\rep{10}\,\crep{5}$, are automatically forbidden under the above assumptions. Hence, we only have to be careful with dimension four operators.

The symmetries, which fulfill criteria \ding{172} and \ding{173} are listed in table~\ref{tab:MSSM_all_symmetries} for all possible values of $N$. We also give the parameter $\rho$ (cf.\ \Eqref{eq:MSSM_anomaly_conditions}) which measures whether the Green-Schwarz mechanism is at work ($\rho\neq 0$) or not ($\rho=0$).
\begin{table}[htb]
\centering
\subfloat[Phenomenologically attractive charge assignments.\label{tab:MSSM_all_symmetries_good}]{
\begin{tabular}{cccccc}
\toprule[1.3pt]
 $N$ & $q_{\rep{10}}$ & $q_{\crep{5}}$ & $q_{h_u}$  & $q_{h_d}$ & $\rho$  \\ \midrule
4 & 1 & 1 & 0 & 0 & 1\\
6 & 5 & 3 & 4 & 0 & 0\\
8 & 5 & 1 & 0 & 4 & 1\\
12 & 5 & 9 & 4 & 0 & 3\\
24 & 5 & 9 & 16 & 12 & 9\\
\bottomrule[1.3pt]
\end{tabular}
}
\qquad \qquad
\subfloat[Charge assignments which do not forbid nucleon decay.\label{tab:MSSM_all_symmetries_bad}]{
\begin{tabular}{cccccc}
\toprule[1.3pt]
 $N$ & $q_{\rep{10}}$ & $q_{\crep{5}}$ & $q_{h_u}$  & $q_{h_d}$ & $\rho$  \\ \midrule
3 & 2 & 0 & 1 & 0 & 0\\
6 & 2 & 0 & 4 & 0 & 0\\
\bottomrule[1.3pt]
\end{tabular}
}
\caption{Anomaly-free $R$-symmetries in the MSSM which allow the Yukawa couplings and the Weinberg operator.}
\label{tab:MSSM_all_symmetries}
\end{table}

The symmetries in table~\ref{tab:MSSM_all_symmetries_bad} do not forbid the dimension four matter parity violating operators in \Eqref{eq:MSSM_b+l_dim4} which in \SU{5} language read $\rep{10}\,\crep{5}\,\crep{5}$. For this reason, we will not consider them any further. In contrast, the symmetries in table~\ref{tab:MSSM_all_symmetries_good} have matter parity $\Z2^\mathcal{M}$ as a subgroup\footnote{This is because all symmetries in table~\ref{tab:MSSM_all_symmetries_good} have a \Z2 subgroup and quarks and leptons have odd charges while the Higgses have even charges.} and therefore the dimension four baryon and lepton number violating operators are absent.

Note that all phenomenologically attractive symmetries in table~\ref{tab:MSSM_all_symmetries_good} are subgroups of the $\Z{24}^R$ symmetry.

\subsection{Comment on the Gravitational Anomaly}
So far, we have neglected the mixed gravitational-$\Z{N}^R$ anomaly because it can receive contributions from fields which are not charged under the SM gauge group. Nevertheless, some general conclusions can be drawn by considering the gravitational anomaly.

From \Eqref{eq:MSSM_anomaly_conditions_A0} we deduce that for anomaly freedom we need
\begin{equation}
  A_0^\text{full}=0\mod\eta
\end{equation}
because $24=0\mod\eta$ for all our symmetries.

Since all symmetries (cf.\ table~\ref{tab:MSSM_all_symmetries_good}) are subgroups of the $\Z{24}^R$ symmetry, we can calculate the anomaly coefficient for $\Z{24}^R$ and it will apply to all other symmetries. The anomaly coefficient, as defined in \Eqref{eq:MSSM_anomaly_coefficients_A0}, is
\begin{equation}
 A_0^\text{MSSM} = 7 \mod 12 \: .
\end{equation}
One subtlety comes from the fact that we possibly rely on anomaly cancellation via the Green-Schwarz mechanism. For the Green-Schwarz mechanism to work we need a light axion $a$ (see section~\ref{sec:discrete_GS}) which is the imaginary part of the scalar component of the dilaton supermultiplet
\begin{equation}
 S = s + \I a + \sqrt{2}\theta\psi + \theta\theta F_S\:
\end{equation}
where $\psi$ is the axino/dilatino. The Green-Schwarz mechanism requires a coupling $\int \D^2 \theta \, S \, W^\alpha W_{\alpha}$ to the gauge fields from which one can infer that the axino carries $R$-charge -1 which has to be added to $A_0^\text{MSSM}$.

In summary, with the MSSM field content and possibly an axino, the case $N=6$ is anomaly free without the Green-Schwarz mechanism, the cases $N=4,12$ are anomaly free with the Green-Schwarz mechanism, and in the cases $N=8,24$ additional fields are needed to cancel the gravitational anomaly. Note that additional fields are required to break supersymmetry unless supersymmetry is broken by the dilaton.

\subsection[Compatibility with $\SO{10}$]{Compatibility with $\boldsymbol{\SO{10}}$}
By looking at the symmetries in table~\ref{tab:MSSM_all_symmetries_good} we observe that only the $\Z4^R$ symmetry is compatible with a complete unification of quarks and leptons ($q_{\rep{10}}=q_{\crep{5}}$). We will now show that also the other cases can potentially be in accordance with \SO{10}. The $\Z{N}^R$ can be a mixture of the $\U1_X$ subgroup of \SO{10} and an additional $\widetilde{\mathbbm{Z}}_N^R$ symmetry which commutes with \SO{10}.

The crucial point is to realize that \SO{10} has rank five, and \SU{5} and the SM gauge group have rank four. Hence, there is an extra $\U1_X$ factor. We will denote the $\U1_X$ charge by $Q_X$. The branching rules are
\begin{subequations}
\begin{align*}
\SO{10}  & \supset \SU{5} \times \U1_X \\
\rep{10} & \rightarrow \rep{5}_{2} + \crep{5}_{-2} \\
\rep{16} & \rightarrow \rep{10}_{-1} + \crep{5}_3 +\rep{1}_{-5} \: .
\end{align*}
\end{subequations}
Consider an \SO{10} GUT with an additional $\widetilde{\mathbbm{Z}}_N^R$ symmetry as given in table~\ref{tab:Z_N_From_SO10_general}.
\begin{table}[htb]
\centering
\subfloat[MSSM field content]{
\begin{tabular}{ccc}
\toprule[1.3pt]
Label & \SO{10} & $\widetilde{\mathbbm{Z}}_N^R$\\
$M_i$ & $\rep{16}$ & $r_M$\\
$H$ & $\rep{10}$ & $r_H$\\
\bottomrule[1.3pt]
\end{tabular}
}
\qquad
\subfloat[Higgs sector]{
\begin{tabular}{ccc}
\toprule[1.3pt]
Label & \SO{10} & $\widetilde{\mathbbm{Z}}_N^R$\\
$\psi_H$ & $\rep{16}$ & $r_{\psi_H}$ \\
$\psi_{\bar H}$ & $\crep{16}$ & $r_{\psi_{\bar H}}$ \\
\bottomrule[1.3pt]
\end{tabular}
}
\\ \vspace{5mm}
\subfloat[$\widetilde{\mathbbm{Z}}_N^R$-charges for different values of $N$, the resulting $\Z{N}^R$-charge as a linear combination of the $\U1_X$ charge $Q_X$ and the $\widetilde{\mathbbm{Z}}_N^R$ charge $r$. In the last column we list the value of the \SO{10}-\SO{10}-$\widetilde{\mathbbm{Z}}_N^R$ anomaly $\rho$.]{
\begin{tabular}{ccccccc}
\toprule[1.3pt]
$N$ & $r_M$ & $r_H$ & $r_{\psi_H}$ & $r_{\psi_{\bar H}}$ & $\Z{N}^R$ & $\rho$\\
6 & 3 & 2 & 2 & 4 & $5(-4Q_X+5r)$ & 2\\
8 & 4 & 2 & 7 & 1 & $3(Q_X-5r)$ & 3\\
12 & 9 & 8 & 8 & 4 & $5(-4Q_X+5r)$ & 5\\
24 & 12 & 2 & 23 & 1 & $7(Q_X-5r)$ & 11\\ 
\bottomrule[1.3pt]
\end{tabular}
}
\caption{Field content of an $\SO{10}\times\widetilde{\mathbbm{Z}}_N^R$ model which can produce the symmetries in table~\ref{tab:MSSM_all_symmetries_good}.}
\label{tab:Z_N_From_SO10_general}
\end{table}
If the \SU{5} singlets contained in $\psi_H$ and $\psi_{\bar H}$ attain VEVs, which have $\U1_X$ charge $Q_X=\pm 5$, we can use the methods developed in chapter~\ref{chap:abelian} to arrive at the following breaking pattern
\begin{equation}
 \SO{10} \times \widetilde{\mathbbm{Z}}_N^R \rightarrow \SU{5} \times \Z{N}^R\:.
\end{equation}
In summary, we can obtain our $\Z{N}^R$ symmetries from an \SO{10} GUT. However, the scenarios presented here are only toy models. First of all, further Higgs fields are needed to break \SU5 down to the SM. In addition, to obtain doublet-triplet splitting and get rid of dimension five operators larger Higgs representations are needed. Also, anomaly matching (cf.\ section~\ref{sec:anomaly_matching}) forces us to introduce extra representations because the value of $\rho$ does not equal the one given in table~\ref{tab:MSSM_all_symmetries_good}.

These consideration also show that the \SU{5} relations are mandatory. Since \SU{5} and the standard model gauge group have the same rank, there is no \U1 with which our $\Z{N}^R$s could mix upon breaking \SU5.

\section[A unique $\Z4^R$ for the MSSM]{A unique $\boldsymbol{\Z4^R}$ for the MSSM}
\label{sec:Z4R}
In section~\ref{sec:MSSM_classification} we found $\Z{N}^R$ symmetries which allow Yukawa couplings and the neutrino mass operator while forbidding dimension four and five proton decay operators and the $\mu$-term. The only such symmetry that respects \SO{10} GUT relations, which is also the simplest such symmetry, is the $\Z4^R$ in table~\ref{tab:MSSM_all_symmetries_good}. Its charge assignment is given in table~\ref{tab:Z4R_charges}.
\begin{table}[htb]
\centering
\begin{tabular}{ccccccccc}
\toprule[1.3pt]
  & $q_i$ & $\bar u_i$ & $\bar d_i$  & $\ell_i$  & $\bar e_i$ & $h_u$ & $h_d$\\ 
$\Z{4}^R$ & 1 & 1 & 1 & 1 & 1 & 0 & 0 \\
\bottomrule[1.3pt]
\end{tabular}
\caption{A unique $\Z4^R$ for the MSSM. The index $i=1,2,3$ labels the generations.}
\label{tab:Z4R_charges}
\end{table}

The $\Z4^R$ has partially been discussed in the literature. In~\cite{Kurosawa:2001iq} this symmetry was discovered. However, Kurosawa, Maru and Yanagida did not allow for anomaly cancellation via the Green-Schwarz mechanism. Therefore, they have to introduce additional fields, which are charged under the SM gauge group, in order to get rid of the mixed gauge-$\Z4^R$ anomalies. The new matter fields cause severe problems with experiment, e.g.\ they contribute to flavor changing neutral currents.

The $\Z4^R$ has also been presented by Babu et. al. in~\cite{Babu:2002tx}. However, they did not mention the uniqueness of the $\Z4^R$ if one requires \SO{10} relations. Also they did not discuss the absence of dimension five proton decay operators as well as the hypercharge and gravitational anomaly. 

\subsection[Phenomenology of Unbroken $\Z4^R$]{Phenomenology of Unbroken $\boldsymbol{\Z4^R}$}
Recall that the perturbative superpotential of the MSSM with the $\Z4^R$ up to dimension five symmetry reads
\begin{equation}
 \mathscr{W}  =  (Y_u)^{ij}\, q_i h_u \bar u_j + (Y_d)^{ij}\, q_i h_d \bar d_j + (Y_e)^{ij}\, \ell_i h_d \bar e_j + \kappa^{(5)}_{ij} \ell_i \ell_j h_u h_u \:. \label{eq:Z4R_superpotential}
\end{equation} 
At the renormalizable level, this is identical to the superpotential of the MSSM with matter parity except for the absence of the $\mu$-term. Hence, at this level the proton is stable as long as the $\Z4^R$ is unbroken. The dimension five Weinberg operator $\ell_i \ell_j h_u h_u$ generates masses for the neutrinos via the See-Saw mechanism~\cite{Minkowski:1977sc,Yanagida:1979as}.

\subsection[Breaking $\Z4^R$]{Breaking $\boldsymbol{\Z4^R}$}
\label{sec:Z4R_breaking}
The $\Z4^R$ needs to be broken for several reasons. First of all, upon supersymmetry breaking the superpotential $\mathscr{W}$ obtains a VEV, and since $\mathscr{W}$ carries $R$-charge two this also breaks the $\Z4^R$. Additionally, we have to generate a $\mu$-term to allow for electroweak symmetry breaking. 
We also expect the $\Z4^R$ to be broken by non-perturbative effects because the symmetry is anomalous without the Green-Schwarz mechanism. For all these reasons the $\Z4^R$ is broken down to its \Z2 subgroup which is equivalent to matter parity.

We can parameterize the $\Z4^R$ breaking with aid of the dilaton superfield $S$ whose existence is required by the Green-Schwarz mechanism (cf.\ section~\ref{sec:discrete_GS}). According to \Eqref{eq:discrete_GS_condition} the dilaton shifts under a $\Z4^R$ transformation 
\begin{equation}
 S \to S + \frac{\I}{2}\Delta_\text{GS} \qquad \text{with} \qquad \Delta_\text{GS}=\frac{1}{4\pi}(\rho\mod\eta)=\frac{1+2\nu}{4\pi}\, ; \quad  (\nu\in\Z{})
\end{equation}
since $\rho=1$ (see table~\ref{tab:MSSM_all_symmetries_good}) and $\eta=2$. It follows that 
\begin{equation} \label{eq:Z4R_trafo_dilaton}
 \exp\left( -8\pi^2 \frac{1+2n}{1+2\nu} S\right) 
\end{equation}
has $\Z4^R$-charge two with $n\in\Z{}$. Hence, we can build superpotential terms by multiplying \Eqref{eq:Z4R_trafo_dilaton} with terms that have $\Z4^R$-charge zero. Thus, we get a non-perturbative superpotential
\begin{equation} \label{eq:W_np}
 \mathscr{W}_\text{np}= \exp\left( -8\pi^2 \frac{1+2n}{1+2\nu}\, S\right) \left(  h_u h_d + qqq\ell +\bar u \bar u \bar d \bar e \right) 
\end{equation}
where we suppressed coupling constants. Once the dilaton obtains a VEV, which is related to the gauge coupling $\left\langle s \right\rangle =g^{-2}$, these terms are proportional to $\exp\left( -\tfrac{8\pi^2}{g^2}\tfrac{1+2n}{1+2\nu}\right) $ which in the case $n=\nu=0$ reproduce the t'Hooft instanton factor $\exp \left( -\tfrac{8\pi^2}{g^2} \right)$ (cf.\ \Eqref{eq:effective_instanton_operator}). 

In summary, the $\Z4^R$ will be broken by non-perturbative effects and $\Z4^R$ violating couplings are exponentially suppressed. However, the matter parity subgroup of the $\Z4^R$ remains exact, e.g.\ dimension four baryon and lepton number violating operators are forbidden also non-perturbatively.

\subsubsection{Gaugino Condensation}
The crucial question now concerns the size and origin of the $e^{-aS}$ term. However, the answer depends on the specific mechanism which provides a superpotential VEV. Let us comment on a prominent way to generate a superpotential VEV and also break supersymmetry, namely gaugino condensation~\cite{Ferrara:1982qs,Nilles:1982ik,Derendinger:1985kk,Dine:1985rz}. In this scheme, a small dimensionful parameter is generated via dimensional transmutation~\cite{Witten:1981nf} by non-perturbative effects. As in other scenarios of supersymmetry breaking, gaugino condensation takes place in a hidden sector. The hidden sector gauge group becomes strongly coupled at a scale
\begin{equation}
 \Lambda = M_\text{Planck} \, \exp \left( -\frac{8\pi^2}{b}\frac{1}{g^2(M_\text{Planck})} \right) 
\end{equation}
where $b$ is the beta function coefficient of the condensing gauge group which depends on the hidden gauge group and the hidden field content and $g$ is the hidden gauge coupling. For example, for a hidden \SU{N_c} gauge theory with $N_f$ superfields in the $\rep{N_c}+\crep{N_c}$ representation $b=3N_c-N_f$. Note that we could have replaced $M_\text{Planck}$ with any scale $\mu$ because $\Lambda$ is actually scale invariant, i.e.\ $\frac{\partial}{\partial\mu}\Lambda=0$. At strong coupling, the gauginos $\lambda$ condense~\cite{Raby:1979my}
\begin{equation}
 \left\langle \lambda \lambda \right\rangle =\Lambda^3
\end{equation}
which provides a VEV for the superpotential
\begin{equation}
 \left\langle W \right\rangle  = \Lambda^3 = M_\text{Planck}^3 \exp \left( -3\frac{8\pi^2}{b}\frac{1}{g^2(M_\text{Planck})} \right)\:.
\end{equation}
If supersymmetry breaking is mediated by gravity to the standard model, the gravitino mass $m_{3/2}$ is of the order of the superpartner masses and is given by the superpotential VEV, i.e.\ $\left\langle W  \right\rangle / M_\text{Planck}^2\sim m_{3/2}$. Since we want supersymmetry to solve the hierarchy problem, $\left\langle W  \right\rangle$ should be around the TeV scale or $10^{-15}$ in Planck units. Combining this observation with \Eqref{eq:W_np}, we conclude that we generate a $\mu$-term of order of the gravitino mass and hence of the order of the other supersymmetry breaking terms in the visible sector. The dimension five baryon and lepton number violating terms are suppressed by a factor $m_{3/2}/M_\text{Planck}^4\sim 10^{-15}/M_\text{Planck}$ which is experimentally safe~\cite{Hinchliffe:1992ad}.

\subsubsection{Domain Walls}
When dealing with broken discrete symmetries, a general problem is the appearance of domain walls (cf.\ \ref{sec:domain_walls}). Domain walls are produced in the early universe and could dominate the energy density today in contrast to observation. They must disappear during inflation or rapidly annihilate~\cite{Vilenkin:1981zs}. Also $R$-symmetries have a potential domain wall problem. For a recent discussion see~\cite{Dine:2010eb}. 

However, a detailed treatment of domain walls requires exact knowledge of how the $\Z4^R$ is broken. An exemplary solution would be that the inflaton carries $R$-charge zero such that the $\Z4^R$ is broken by the $F$-term of the inflaton (remember that the $F$-term carries $R$-charge two). In this case, the domain walls will be inflated away.

\subsection[Phenomenology after $\Z4^R$ Breaking]{Phenomenology after $\boldsymbol{\Z4^R}$ Breaking}
After the $\Z4^R$ is broken we have the following situation. Since matter parity stays exact, the renormalizable terms are identical to the usual MSSM, i.e.\ dimension four proton decay is absent and there is a stable dark matter candidate. Additionally, we have a $\mu$-term of order of the gravitino mass and dimension five proton decay operators are highly suppressed.

\section{Extension to the NMSSM}
The next-to minimal supersymmetric standard model (NMSSM)~\cite{Nilles:1982dy,Frere:1983ag} (see~\cite{Ellwanger:2009dp} for a recent review) adds a standard model singlet $N$ to the field content of the MSSM in order to solve the $\mu$-problem and raise the Higgs mass w.r.t.\ the MSSM. We will now apply our analysis of section~\ref{sec:sym_commuting_with_SU5} to the NMSSM. 

The renormalizable superpotential of the MSSM reads
\begin{equation}\label{eq:W_NMSSM}
 \mathscr{W}_\text{NMSSM}= \mathscr{W}_\text{MSSM} + \lambda N \, h_u h_d + \kappa N^3
\end{equation}
where $\lambda$ and $\kappa$ are coupling constants, and $\mathscr{W}_\text{MSSM}$ the MSSM superpotential given in \Eqref{eq:MSSM_Yukawa_couplings}. A VEV of electroweak size for $N$ will generate a $\mu$-term of the right order. Usually, the NMSSM is defined with a $\Z3$ symmetry under which all chiral superfields carry charge one. However, a VEV for $N$ near the electroweak scale causes severe cosmological problems with domain walls~\cite{Abel:1995wk}. We will not assume the existence of this $\Z3$ symmetry.

Since the SM singlet $N$ does not contribute to the mixed gauge-$\Z{N}^R$ anomalies, our analysis from section~\ref{sec:sym_commuting_with_SU5} still applies. That is, in order for the NMSSM to be compatible with \SU5 and to forbid the $\mu$-term, the only candidate symmetries are based on the ones in table~\ref{tab:MSSM_all_symmetries_good}. We just have to figure out which symmetries and charges $q_N$ for $N$ yield the superpotential in \Eqref{eq:W_NMSSM}. To do so, observe that the two terms in \Eqref{eq:W_NMSSM} require
\begin{subequations}
\begin{align}
q_N + q_{h_u} + q_{h_d} = 2 \mod N \: , \label{eq:NMSSM_order_constraint1}\\
3 q_N = 2 \mod N \:. \label{eq:NMSSM_order_constraint2}
\end{align}
\end{subequations}
From \Eqref{eq:MSSM_order_constraints1} we know that $q_{h_u} + q_{h_d} = 4 \mod N$ because $N$ is even for all symmetries in table~\ref{tab:MSSM_all_symmetries_good}. Inserting this in \Eqref{eq:NMSSM_order_constraint1} yields $q_N = -2 \mod N$. Then \Eqref{eq:NMSSM_order_constraint2} teaches that 
\begin{equation}
 8=0\mod N
\end{equation}
which has the solutions $N=4,8$ (remember that we do not consider $\Z2^R$ symmetries since they do not forbid the $\mu$-term). However, for $N=4$ we have $q_N=2$ and a linear term in $N$ will be allowed which is absent in the NMSSM. Hence, we end up with the $\Z8^R$ symmetry in table~\ref{tab:NMSSM_Z8R-symmetry}
\begin{table}[htb]
\centering
\begin{tabular}{cccccc}
\toprule[1.3pt]
  & $q_{\rep{10}}$ & $q_{\crep{5}}$ & $q_{h_u}$  & $q_{h_d}$ & $q_N$  \\
$\Z8^R$ & 5 & 1 & 0 & 4 &  6\\
\bottomrule[1.3pt]
\end{tabular}
\caption{A $\Z8^R$ symmetry for the NMSSM with \SU5 GUT relations.}
\label{tab:NMSSM_Z8R-symmetry}
\end{table}

\subsection{Phenomenology}
The $\Z8^R$ symmetry in table~\ref{tab:NMSSM_Z8R-symmetry} yields the renormalizable superpotential of the NMSSM as given in \Eqref{eq:W_NMSSM}. Hence, dimension four baryon and lepton number violating couplings are absent, and the $\mu$-term is forbidden. In addition, the Weinberg operator is allowed.

Since, the $\Z8^R$ is anomalous without the Green-Schwarz mechanism, it will be broken non-perturbatively to its matter parity subgroup. Of course, the $\Z8^R$ will also be broken to matter parity if $N$ obtains a VEV. In analogy to \Eqref{eq:W_np}, the non-perturbative superpotential looks like
\begin{equation}
 \mathscr{W}_\text{np} = e^{-aS} \left( qqq\ell +\bar u \bar u \bar d \bar e  \right) + e^{-2aS} N + e^{-3aS} \left(  N^2 + h_u h_d \right) 
\end{equation}
where we have suppressed coupling constants and $a$ is a constant s.t. $e^{-aS}$ has $\Z8^R$ charge two.

In summary, the $\Z8^R$ symmetry reproduces the renormalizable NMSSM superpotential, and suppresses dimension five baryon and lepton number operators. In contrast to the usual \Z3 symmetry in the NMSSM, the domain wall problem with the $\Z8^R$ is less dramatic because the $\Z8^R$ is broken non-pertubatively and not only by a VEV of the singlet $N$.

\clearpage
\thispagestyle{empty}
\chapter{Minimal Flavor Violation}
\label{chap:MFV}
Despite the successes of the MSSM, e.g.\ solving the hierarchy problem or gauge coupling unification, it also introduces some new problems. Issues with baryon and lepton number violation and the $\mu$-problem have been discussed in chapter~\ref{chap:application_MSSM}. In this chapter, we will discuss some aspects of the problems associated with new interactions which contribute to flavor transitions, i.e.\ the flavor problem. See \cite{Isidori:2008qp,Buras:2009if} for recent reviews.

Supersymmetry should be broken at low energies in order to be able to address the hierarchy problem. Since we do not know how supersymmetry is broken, we parameterize the breaking by soft-terms in the Lagrangian~\cite{Martin:1997ns}
\begin{subequations}\label{eq:soft_terms}
\begin{align}
 -\mathscr{L}_{\text{soft}}&=\frac{1}{2} \left(M_3 \tilde{g}^a \tilde{g}^a + M_2 \widetilde{W}^a \widetilde{W}^a + M_1 \tilde{B}\tilde{B}+ \text{h.c.} \right) \label{eq:soft_terms_gauginos} \\
 & \quad + (\boldsymbol{m}^2_{Q})_{ij} \tilde{q}^{\dagger i}  \tilde{q}^j + (\boldsymbol{m}^2_{\tilde{\ell}})_{ij} \tilde{\ell}^{\dagger i} \tilde{\ell}^j + (\boldsymbol{m}^2_{\tilde{u}})_{ij} \tilde{u}^{\dagger j}  \tilde{u}^j + (\boldsymbol{m}^2_{\tilde{d}})_{ij} \tilde{d}^{\dagger i}  \tilde{d}^j + (\boldsymbol{m}^2_{\tilde{e}})_{ij} \tilde{e}^{\dagger i}  \tilde{e}^j \label{eq:soft_terms_sfermions}\\
 &\quad + m_{h_d}^2 h_d^* h_d + m_{h_u}^2 h_u^* h_u + (b\, h_d h_u+ \text{h.c.}) \label{eq:soft_terms_higgsinos}\\
 &\quad + \left( (\boldsymbol{A}_u)_{ij} \tilde{u}^i \tilde{q}^j h_u + (\boldsymbol{A}_d)_{ij} \tilde{d}^i \tilde{q}^j h_d + (\boldsymbol{A}_e)_{ij} \tilde{e}^i \tilde{\ell}^j h_d + \text{h.c.} \right). \label{eq:soft_terms_A}
\end{align}
\end{subequations}
The terms in \Eqref{eq:soft_terms_gauginos} give masses to the gauginos while \Eqref{eq:soft_terms_sfermions} provides mass terms for the sfermions. There are also mass terms for the Higgsinos in \Eqref{eq:soft_terms_higgsinos} and the so-called $\boldsymbol{A}$-terms in \Eqref{eq:soft_terms_A}.

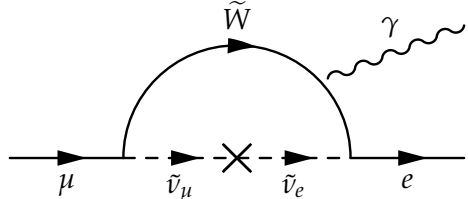
\begin{figure}[htb]
\centering
\begin{fmffile}{feynman}
\begin{fmfgraph*}(170,100)
          \fmfstraight
          \fmfleft{i1,i2,i3} \fmfright{o1,o2,o3}
          \fmf{fermion,label=$\mu$,tension=1}{i2,v1}
          \fmf{fermion,label=$\widetilde W$,left=1,tension=0}{v1,v3}
          \fmf{fermion,label=$e$,tension=1}{v3,o2}
          \fmf{scalar,label=$\tilde \nu_\mu$}{v1,v2}
          \fmfv{decoration.shape=cross}{v2}
          \fmf{scalar,label=$\tilde \nu_e$}{v2,v3}
          \fmffreeze
          \fmf{phantom,tension=2}{v1,v4}
          \fmf{photon,label=$\gamma$,tension=3}{v4,o3}
\end{fmfgraph*} 
\end{fmffile}
\vspace{-15mm}
\caption{An example of a flavor violating process.}
\label{fig:mu_to_e_gamma}
\end{figure}

The soft-masses $\boldsymbol{m}^2$ and the $\boldsymbol{A}$s are $3\times 3$ matrices in generation space. The off-diagonal entries in these matrices contribute to flavor violating processes. For example, a non-zero entry $(\boldsymbol{m}^2_{\tilde{\ell}})_{21}$ could account for the lepton flavor violating decay $\mu\to e\gamma$ (cf.\ figure~\ref{fig:mu_to_e_gamma}). The fact that new contributions to flavor violating processes are tightly constrained by experiment, is known as the supersymmetric flavor problem.

A common way to ameliorate the flavor problem, is to assume that the soft masses $\boldsymbol{m}^2$ are close to the unit matrix and the $\boldsymbol{A}$-terms are proportional to the corresponding Yukawa coupling. In this case, the super-GIM mechanism~\cite{Dimopoulos:1981zb} is at work and unobserved flavor transitions are strongly suppressed. However, this ansatz seems to lack a fundamental motivation. A less restrictive assumption is called minimal flavor violation (MFV)~\cite{Chivukula:1987py,Buras:2000dm,D'Ambrosio:2002ex}. We shall shortly review the MFV ansatz in the next section.

\section{Minimal Flavor Violation: Ansatz}
\label{sec:MFV_ansatz}
In the limit of vanishing Yukawa couplings, the SM enjoys a large flavor symmetry~\cite{Chivukula:1987py} which, apart from some \U1 factors, is given by
\begin{equation}\label{eq:Gflavor}
 G_\text{flavor}= \SU3_q \times \SU3_u \times \SU3_d \times \SU3_\ell \times \SU3_e\; ,
\end{equation}
meaning that the matter field $f$ with $f\in\{q,u,d,\ell,e\}$ transforms as a triplet under $\SU3_f$ and as a singlet under all other groups in \Eqref{eq:Gflavor}. One then views the Yukawas as spurion fields whose VEVs break $G_\text{flavor}$, and give the observed values of the Yukawa couplings. Thus, the spurion fields have to transform under $G_\text{flavor}$ as
\begin{equation}
\Yu\sim(\mathbf{\bar 3},\mathbf{3},0,0,0)\, , \quad \Yd \sim (\mathbf{\bar 3},0,\mathbf{3},0,0) \, , \quad \Ye \sim (0,0,0,\mathbf{\bar 3},\mathbf{3}) \:.
\end{equation}
This leads to the following expansion of the soft supersymmetry breaking operators~\cite{D'Ambrosio:2002ex}
\begin{subequations}\label{eq:softmassdecomp}
\begin{align}
  \boldsymbol{m}_Q^2 &=  \alpha_1\,\mathds{1} + \beta_1 \,\Yu^\dagger \Yu + \beta_2\, \Yd^\dagger \Yd 
 + \beta_3\, \Yd^\dagger \Yd\, \Yu^\dagger \Yu + \beta_3 \,\Yu^\dagger \Yu\, 
 \Yd^\dagger \Yd\;, \\
 \boldsymbol{m}_u^2 &=  \alpha_2\,\mathds{1} + \beta_5\, \Yu \Yu^\dagger  \;,\\
 \boldsymbol{m}_d^2 &= \alpha_3\,\mathds{1} + \beta_6\, \Yd \Yd^\dagger  \;,\\
 \boldsymbol{A}_u &=   \alpha_4\, \Yu + \beta_7\, \Yu\, \Yd^\dagger \Yd  \;,\\
 \boldsymbol{A}_d &=  \alpha_5\, \Yd + \beta_8\, \Yd \,\Yu^\dagger \Yu  \;,\\
 \boldsymbol{A}_e &= \alpha_e\, \Ye \;.
\end{align}
\end{subequations}
As discussed in \cite{D'Ambrosio:2002ex}, higher order terms, i.e.\ terms which
include higher powers of Yukawas, can be neglected due to the Yukawa
hierarchies. In a limit where one only considers top and bottom mass
equation~\eqref{eq:softmassdecomp} is already exact. Note that our notation is
slightly different from the one used in \cite{D'Ambrosio:2002ex} in that our
coefficients $\alpha_i$ and $\beta_i$ carry mass dimension. These
coefficients are related to the original $a_i$ and $b_i$ in
\cite{D'Ambrosio:2002ex} by $\alpha_1=m_0^2\,a_1$ and $\alpha_4=A\, a_4$ etc.,
where $m_0$, $A$ are soft mass scales. This is done in order to simplify the expressions to be
presented below. (The original MFV decomposition involves one parameter
more, e.g.\ $m_0^2$, $a_2$ and $b_5$ instead of  $\alpha_2$ and
$\beta_5$ in the case of $\boldsymbol{m}_u^2$.) In our notation of Yukawa
couplings and scalar soft mass squareds we follow \cite{Martin:1993zk}.

\subsection{Anomalies}
\label{sec:MFV_anomalies}
The flavor symmetry $G_\text{flavor}$ in \Eqref{eq:Gflavor} is anomalous. The anomalous combinations are (cf.\ section~\ref{sec:anomaly_conditions_global})
\begin{subequations}\label{eq:MFV_anomalies}
\begin{alignat}{2}
 &\SU3_f - \SU3_f - \U1_Y  \quad & \text{for} \quad  & f=q,u,d,\ell,e \: ,\\
 &\SU3_f - \SU3_f - \SU3_f  \quad & \text{for} \quad &  f=q,\ell \:.
\end{alignat}
\end{subequations}
The anomalies in the first line appear because the spurions do not carry hypercharge. The second  case comes from the fact that the reducible representation, under which the fields transform, is neither real nor pseudo-real.

\subsection{Discrete Minimal Flavor Violation}
\label{sec:discrete_MFV}
Having seen that the usual justification of the MFV ansatz rests on an anomalous symmetry, we can contemplate replacing the \SU3s in \Eqref{eq:Gflavor} by anomaly-free, discrete, non-abelian subgroups of \SU3. 

The finite subgroups of \SU3 were first listed in a physical context in \cite{fairbairn:1038}. In order to preserve the MFV structure, we look for an \SU3 subgroup which has a three-dimensional complex representation. The smallest choice is the group $\Sigma(168)$ in \cite{fairbairn:1038}. In the newer literature this group is called $\mathcal{PSL}_2(7)$, the projective linear group of $2\times 2$-matrices over $\mathbb{F}_7$, the field with seven elements. This means $\mathcal{PSL}_2(7)\cong \mathrm{SL}(2,\mathbb{F}_7)/\mathbb{Z}_2$. $\Sigma(168)$ is simple and in fact the second-smallest nonabelian simple group after the alternating group $A_5$. Hence, we consider replacing \Eqref{eq:Gflavor} by
\begin{equation}
  G'_\text{flavor}= \Sigma(168)_q \times \Sigma(168)_u \times \Sigma(168)_d \times \Sigma(168)_\ell \times \Sigma(168)_e\; .
\end{equation}
$\Sigma(168)$ has six irreducible representations $\rep{1},\rep{3},\crep{3},\rep{6},\rep{7},\rep{8}$. The lowest dimensional representations of \SU3 decompose into $\Sigma(168)$ representations as~\cite{Luhn:2007yr}
\begin{align}
  \SU3 & \supset \Sigma(168) \nonumber \\
  \rep{3} & \to \rep{3} \nonumber\\
 \crep{3} & \to \crep{3}\nonumber\\
  \crep{6} & \to \rep{6}\nonumber\\
 \rep{6} & \to \rep{6}\nonumber\\
 \rep{8} & \to \rep{8} \label{eq:branching_rules_SU3_PSL7}
\end{align}
So the structure of the two groups is quite similar despite the fact that the $\rep{6}$ of $\Sigma(168)$ is real. Also the Kronecker product of irreducible representations decomposes in a similar manner. Especially, we have~\cite{Luhn:2007yr} for $\Sigma(168)$
\begin{align}
 \rep{3} \otimes \crep{3} &= \rep{1} \oplus \rep{8} \nonumber \\
 \rep{8} \otimes \rep{8} &= \rep{1} \oplus \ldots \:.
\end{align}
Hence, all arguments in \cite{D'Ambrosio:2002ex}, leading to the expansion of the soft parameters in terms of the Yukawas in the MSSM as in \Eqref{eq:softmassdecomp}, are still valid. In~\cite{Zwicky:2009vt} a similar analysis, which found $\Sigma(168)$ and other groups, was performed. 

Of course, there will be corrections to \Eqref{eq:softmassdecomp}, if we use $\Sigma(168)$ instead of \SU3. The corrections come from the fact that there are invariant contractions of $\Sigma(168)$ representations which are not invariant under \SU3. However, these corrections are of higher order in the Yukawa couplings.

We can now check that this finite symmetry is indeed anomaly free. Anomaly constraints for non-abelian, discrete groups have been discussed in section~\ref{sec:anomaly_non_abelian_discrete}. There we concluded that anomalies are absent if the determinant of the generators of the representation matrices equals one. In our case, this happens because of the correspondence between the representations of \SU3 and $\Sigma(168)$ as described in \Eqref{eq:branching_rules_SU3_PSL7}. This can also be checked explicitly by computing the determinate of the generators as given in \cite{fairbairn:1038}.

Let us finally comment on the possible origin of the $\Sigma(168)$. If one wants to obtain $\Sigma(168)$ from an \SU3 by spontaneous symmetry breaking, note that the smallest representation which contains a $\Sigma(168)$ singlet has dimension 15~\cite{King:1982sj}. This is in accordance with the findings in~\cite{Adulpravitchai:2009kd} that large irreducible representations are needed to obtain non-abelian, discrete symmetries via spontaneous symmetry breaking from a continuous group. Of course, the $\Sigma(168)$ could also have a different origin, e.g.\ as arising from compactifying higher dimensions~\cite{Altarelli:2006kg,Kobayashi:2006wq,Adulpravitchai:2009id} on suitably chosen spaces.

\section{Running Minimal Flavor Violation}

\subsection{Renormalization Group Equations for the MFV Parameters}
\label{sec:RGE_MFV}
The MFV ansatz is usually imposed at the electroweak scale. However, from a top-down perspective we expect the flavor structure to arise by some mechanism at the GUT scale, e.g.\ by discrete symmetries as explained in section~\ref{sec:discrete_MFV}. This raises the question, whether renormalization effects destroy the MFV ansatz in \Eqref{eq:softmassdecomp}. Since the spurion argument does not imply a
preferred scale, one can infer that, if the MFV decomposition applies at one
renormalization scale, it must apply at any other scale as well. 
That is, renormalization effects will modify
the values of the coefficients, $\alpha_i$ and $\beta_i$, but not the validity
of the ansatz. 

Since the MFV ansatz is based on an anomalous symmetry (cf.\ \ref{sec:MFV_anomalies}), a independent numerical confirmation of the scale independence seems in order. We have checked explicitly that it holds: we start with soft terms
complying with the decomposition \eqref{eq:softmassdecomp} at the GUT scale
and run them down to the SUSY breaking scale, i.e.\ solve the corresponding
renormalization group equations (RGEs). Then we successfully fit the low energy
soft masses by the decomposition \eqref{eq:softmassdecomp}, i.e.\ when inserting the
Yukawa matrices at the low scale we find values of the MFV parameters $\alpha_i$
and $\beta_i$ such that the mass matrices are reproduced with high
accuracy. The details of our numerical studies are deferred to
appendix~\ref{app:MFVcheck}.

Now that we have seen that the running of the soft masses can be described in terms of scale
dependent MFV coefficients $\alpha_i$ and $\beta_i$ we now study the behavior of
these coefficients under the renormalization group.
We calculate the RGEs for the $\alpha_i$ and $\beta_i$ by inserting
\eqref{eq:softmassdecomp} in the one-loop RGEs for the soft-masses and the
trilinear couplings (cf.\ \cite{Martin:1993zk}). Note that there are two sources
for the running of the MFV coefficients: first, the soft terms run, and second,
the Yukawa matrices, to which we match the soft terms, also depend on the
renormalization scale. Neglecting the Yukawa couplings of the first and second
generation, the results read~\cite{Paradisi:2008qh}
\begin{subequations}\label{eq:MFVRGEs}
\begin{align}
 16\pi^2 \frac{\D \alpha_1}{\D t} 
 &= -\frac{32}{3} g_3^2 |M_3|^2 - 6g_2^2|M_2|^2 - \frac{2}{15} g_1^2|M_1|^2 +
 \frac{1}{5}g_1^2 S\;, \displaybreak[0]\\
 16\pi^2 \frac{\D \alpha_2}{\D t} 
 &= -\frac{32}{3} g_3^2 |M_3|^2 - \frac{32}{15} g_1^2|M_1|^2 - \frac{4}{5}g_1^2
 S\;, \displaybreak[0]\\
 16\pi^2 \frac{\D \alpha_3}{\D t} 
 &= -\frac{32}{3} g_3^2 |M_3|^2 - \frac{8}{15} g_1^2|M_1|^2 + \frac{2}{5}g_1^2
 S\;, \displaybreak[0]\\
 16\pi^2 \frac{\D \alpha_4}{\D t} 
 &= 12 \alpha_4 y_t^2 + 10 \beta_7y_t^2 y_b^2 + 2 \beta_8
 y_t^2y_b^2 +  \frac{32}{3}g_3^2 M_3 + 6 g_2^2 M_2 + \frac{26}{15}
 g_1^2 M_1\;,  \displaybreak[0]\\
 16\pi^2 \frac{\D \alpha_5}{\D t} 
 &= 12 \alpha_5 y_b^2 +
 10 \beta_8y_t^2 y_b^2 + 2\beta_7 y_t^2y_b^2 + 
 \frac{32}{3}g_3^2 M_3 + 6 g_2^2 M_2 + \frac{14}{15} g_1^2 M_1 + 2 \alpha_e
 y_\tau^2\;, \\
 16\pi^2 \frac{\D \beta_1}{\D t} 
 &= 2 m_{H_u}^2 + 2 \alpha_4^2 + 2\beta_8^2y_t^2 y_b^2 +  2\alpha_1
 + 2\alpha_2 \nonumber\\
 &  \quad- 10\beta_1y_t^2 +2 \beta_5y_t^2 + \beta_1 \left(
 \frac{32}{3}g_3^2 + 6 g_2^2 + \frac{26}{15}g_1^2 \right)\;, \displaybreak[0]\\
 16\pi^2 \frac{\D \beta_2}{\D t} 
 &= 2 m_{H_d}^2 + 2\alpha_5^2 + 2
 \beta_7^2y_t^2 y_b^2 +  2\alpha_1 + 2\alpha_3 -
 10\beta_2y_b^2 - 2\beta_2y_\tau^2 + 2\beta_6y_b^2
 \nonumber\\
 &\quad+ \beta_2 \left( \frac{32}{3}g_3^2 + 6 g_2^2 + \frac{14}{15}g_1^2
 \right)\;, \displaybreak[0]\\
 16\pi^2 \frac{\D \beta_3}{\D t} 
 &= 2 \alpha_4\beta_7 +2 \alpha_5\beta_8 -12\beta_3y_t^2 -
 12\beta_3y_b^2 - 2\beta_3y_\tau^2  + \beta_3
 \left(\frac{64}{3}g_3^2 + 12g_2^2 + \frac{8}{3}g_1^2 \right)\;,  \displaybreak[0] \\
 16\pi^2 \frac{\D \beta_5}{\D t} 
 &= 4 m_{H_u}^2 +4 \left( \alpha_4 + \beta_7y_b^2 \right)^2 +  4\alpha_1
 + 4\alpha_2 + 4 \beta_1y_t^2 + 4\beta_2y_b^2 +
 8\beta_3y_t^2y_b^2 \nonumber\\
 & \quad   + \beta_5 \left( -8y_t^2-2y_b^2 +\frac{32}{3}g_3^2 + 6 g_2^2 + \frac{26}{15}g_1^2 \right)\;,  
 \displaybreak[0]\\
 16\pi^2 \frac{\D \beta_6}{\D t} 
 &= 4 m_{H_d}^2 + 4\left( \alpha_5 + \beta_8y_t^2\right)^2 +  4\alpha_1 + 4\alpha_3 + 4\beta_1y_t^2 + 4\beta_2y_b^2 + 8\beta_3y_t^2y_b^2 
 \nonumber\\
 &  \quad + \beta_6\left( - 2y_t^2 - 8y_b^2-2y_\tau^2 +
 \frac{32}{3}g_3^2 + 6g_2^2 + \frac{14}{15}g_1^2  \right)\;, \displaybreak[0]\\
 16\pi^2   \frac{\D \beta_7}{\D t} 
 &= 2\alpha_5 + \beta_7\left( -12y_b^2 - 2y_\tau^2 +
 \frac{32}{3}g_3^2 + 6g_2^2 + \frac{14}{15}g_1^2 \right)\;, \displaybreak[0]\\
 16\pi^2   \frac{\D \beta_8}{\D t} 
 &= 2\alpha_4 + \beta_8\left( -12y_t^2 +
 \frac{32}{3}g_3^2 + 6g_2^2 + \frac{26}{15}g_1^2 \right)\;.
 \end{align}
\end{subequations}
Here $\D/\D t$ denotes the logarithmic derivative w.r.t.\ the
renormalization scale, $g_1$, $g_2$, $g_3$ are the gauge couplings,
$M_1$, $M_2$, $M_3$ the gaugino masses,  $y_t$,
$y_b$, $y_\tau$ the third family Yukawa couplings and $m_{H_u}$, $m_{H_d}$
the Higgs soft mass terms. We have further defined
\begin{align}
 S &=
 m_{H_u}^2 - m_{H_d}^2 + \mathrm{Tr}\bigg[ \alpha_1\mathds{1} + \beta_1 \Yu^\dagger \Yu + \beta_2 \Yd^\dagger \Yd + 2 \beta_3 \Yd^\dagger \Yd \Yu^\dagger \Yu \nonumber\\
& \quad - 2 \alpha_2\mathds{1} - 2 \beta_5 \Yu \Yu^\dagger  
  + \alpha_3\mathds{1} + \beta_6 \Yd \Yd^\dagger - \boldsymbol{m}_L^2 + \boldsymbol{m}_e^2 \bigg] 
\end{align}
with $\boldsymbol{m}_L^2$ and $\boldsymbol{m}_e^2$ denoting the $3\times3$ mass
matrices for the charged lepton doublets and singlets, respectively.

\subsection{Approximations of low-energy MFV Coefficients}

We derive approximate relations between the values at the GUT scale and the low scale.
Here we assume mSUGRA inspired initial conditions~(for details see
equation~\eqref{eq:universal})
and allow for one non-zero $\beta_i$ while the others are set to zero. 
The value of $\tan\beta$ is fixed to 10. The formulae are obtained by varying the initial
values of $m_{\nicefrac{1}{2}},m_0,A,\beta_i$, running them down to the low
scale and fitting a linear combination of the parameters to the obtained points
in parameter-space. Details and the results are shown in appendix~\ref{app:Approximations}. From these formulas one can already infer that the low energy values of the $\beta_i$ are rather insensitive to their high-energy values. This leads to a fixed-point behavior as discussed in the next section.

\subsection{Fixed Points}

Let us now come to the discussion of the relation between the boundary values
for the MFV coefficients at the high scale and the values they attain at the low
scale. 
A crucial feature of the low-energy values of the MFV coefficients $\beta_i$ is
that they are rather insensitive to their GUT boundary values. 
It is, of course,
well known that the soft masses tend to get aligned due to the renormalization
group evolution
\cite{Dine:1990jd,Brignole:1993dj,Choudhury:1994pn,Brax:1995up,Chankowski:2005jh}. Our
results make this statement more precise. There is an on-going competition
between the alignment process, triggered mainly by the positive gluino
contributions, and misalignment process, driven by negative effects proportional
to the Yukawa matrices. 
These effects are so strong that the memory to the initial conditions gets
almost wiped out, at least as long as the ratio between scalar and gaugino
masses at the high scale is not too large.

\begin{figure}
\centering
\psfrag{lbX}{\small $\log \frac{Q}{\text{GeV}}$}
\psfrag{lbY}{$\frac{\beta_1}{\alpha_1}$}
\psfrag{2}{\footnotesize 2}
\psfrag{3}{\footnotesize 3}
\psfrag{4}{\footnotesize 4}
\psfrag{5}{\footnotesize 5}
\psfrag{6}{\footnotesize 6}
\psfrag{7}{\footnotesize 7}
\psfrag{8}{\footnotesize 8}
\psfrag{9}{\footnotesize 9}
\psfrag{10}{\footnotesize 10}
\psfrag{11}{\footnotesize 11}
\psfrag{12}{\footnotesize 12}
\psfrag{13}{\footnotesize 13}
\psfrag{14}{\footnotesize 14}
\psfrag{15}{\footnotesize 15}
\psfrag{16}{\footnotesize 16}
\psfrag{1.0}{\footnotesize 1.0}
\psfrag{0.5}{\footnotesize 0.5}
\psfrag{0.0}{\footnotesize 0.0}
\psfrag{-0.5}{\footnotesize -0.5}
\psfrag{-1.0}{\footnotesize -1.0}
\subfloat[SPS1a]{\includegraphics{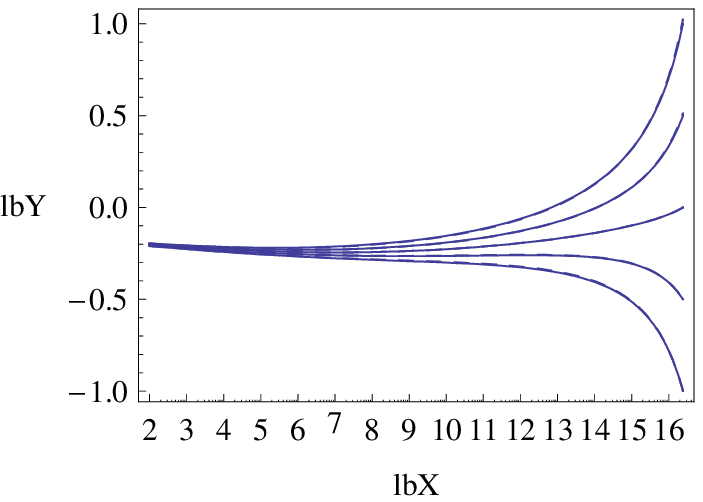}}
\qquad 
\subfloat[SPS1b]{\includegraphics{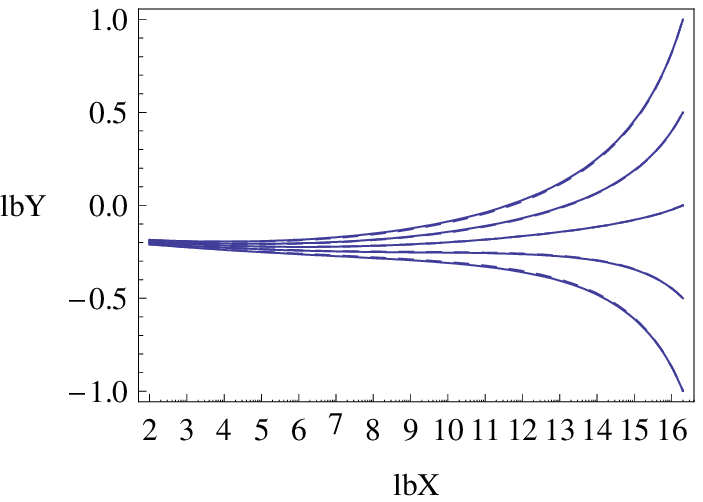}}
\\[0.3cm]
\subfloat[SPS2]{\includegraphics{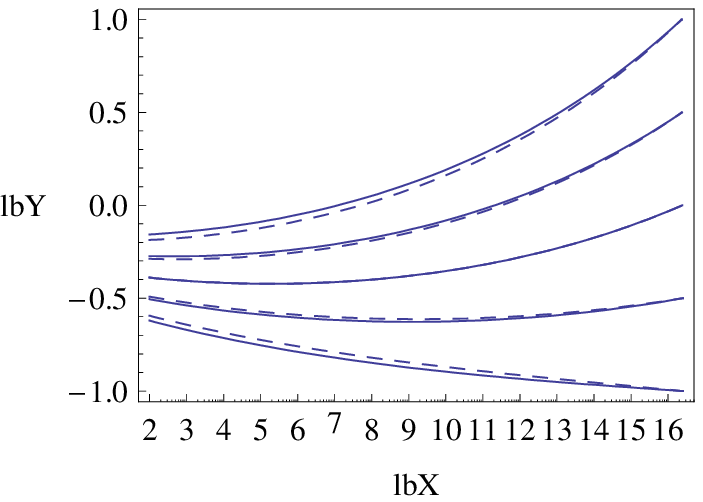}}
\qquad 
\subfloat[SPS3]{\includegraphics{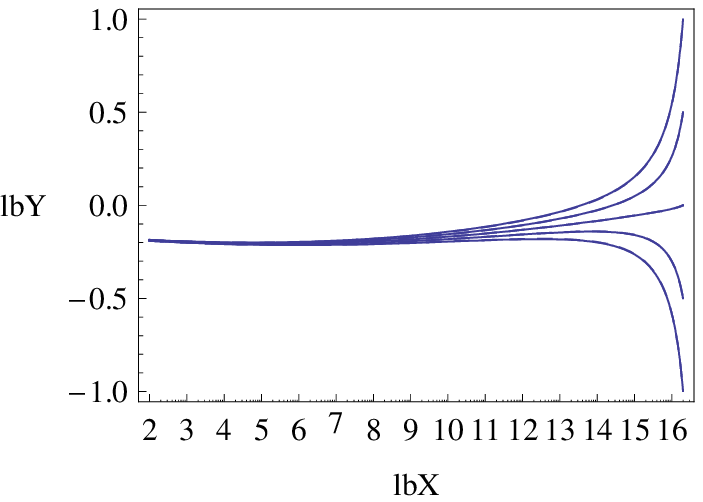}}
\\[0.3cm]
\subfloat[SPS4]{\includegraphics{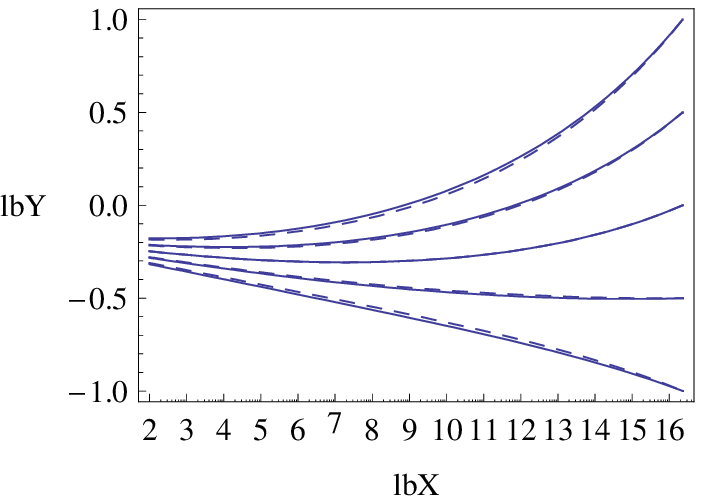}}
\qquad 
\subfloat[SPS5]{\includegraphics{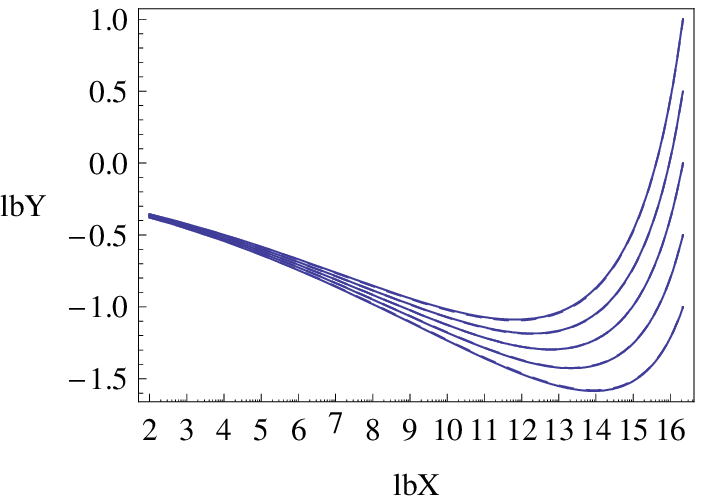}}
\caption{The running of $\frac{\beta_1}{\alpha_1}$. For the solid curve
only $\beta_1$ is non-zero at the high scale while for the dashed curve all
$\beta_i$ are switched on. }
\label{fig:b_1/a_1}
\end{figure}

\begin{figure}
\centering
\psfrag{lbX}{\small $\log \frac{Q}{\text{GeV}}$}
\psfrag{lbY}{$\frac{\beta_6}{\alpha_3}$}
\psfrag{2}{\footnotesize 2}
\psfrag{3}{\footnotesize 3}
\psfrag{4}{\footnotesize 4}
\psfrag{5}{\footnotesize 5}
\psfrag{6}{\footnotesize 6}
\psfrag{7}{\footnotesize 7}
\psfrag{8}{\footnotesize 8}
\psfrag{9}{\footnotesize 9}
\psfrag{10}{\footnotesize 10}
\psfrag{11}{\footnotesize 11}
\psfrag{12}{\footnotesize 12}
\psfrag{13}{\footnotesize 13}
\psfrag{14}{\footnotesize 14}
\psfrag{15}{\footnotesize 15}
\psfrag{16}{\footnotesize 16}
\psfrag{1.0}{\footnotesize 1.0}
\psfrag{0.5}{\footnotesize 0.5}
\psfrag{0.0}{\footnotesize 0.0}
\psfrag{-0.5}{\footnotesize -0.5}
\psfrag{-1.0}{\footnotesize -1.0}
\subfloat[SPS1a]{\includegraphics{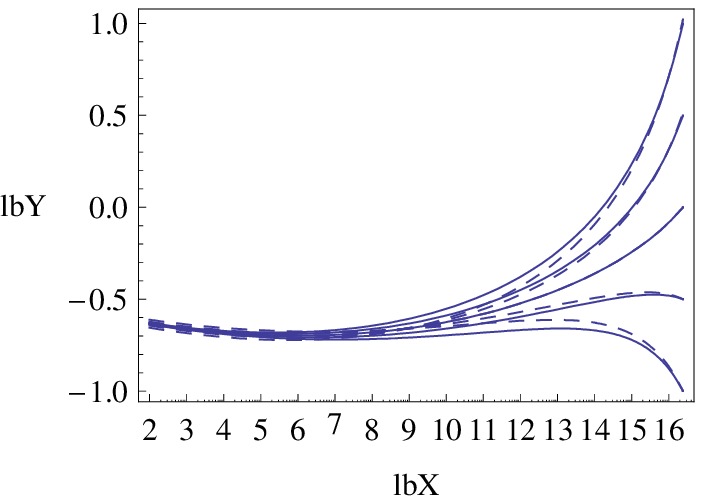}} 
\qquad
\subfloat[SPS1b]{\includegraphics{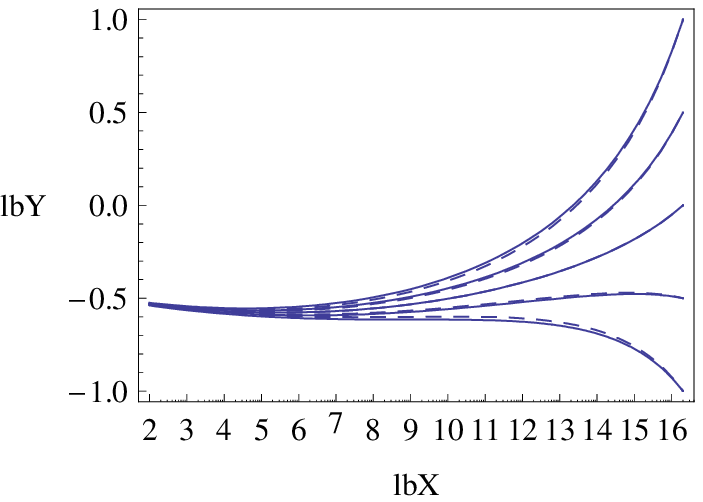}}
\\[0.3cm]
\subfloat[SPS2]{\includegraphics{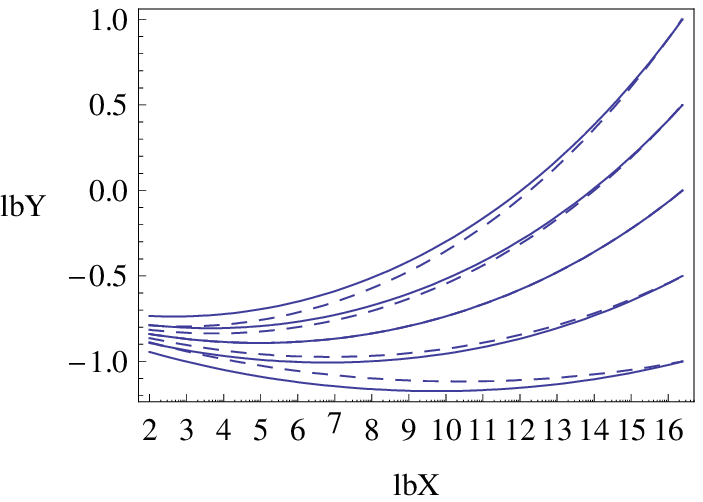}}
\qquad
\subfloat[SPS3]{\includegraphics{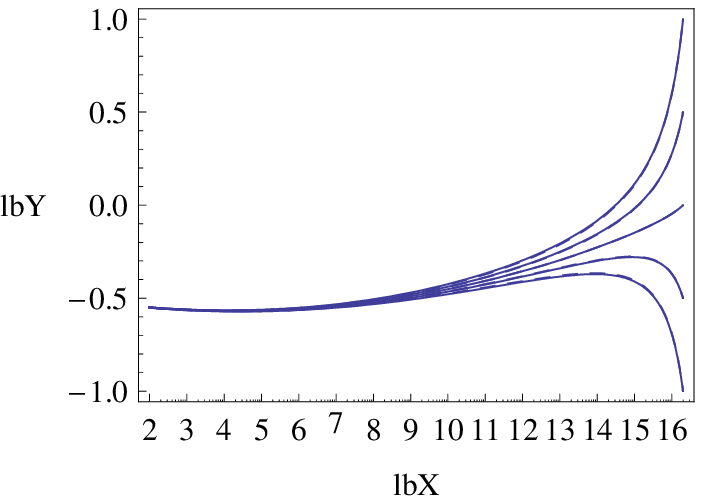}}
\\[0.3cm]
\subfloat[SPS4]{\includegraphics{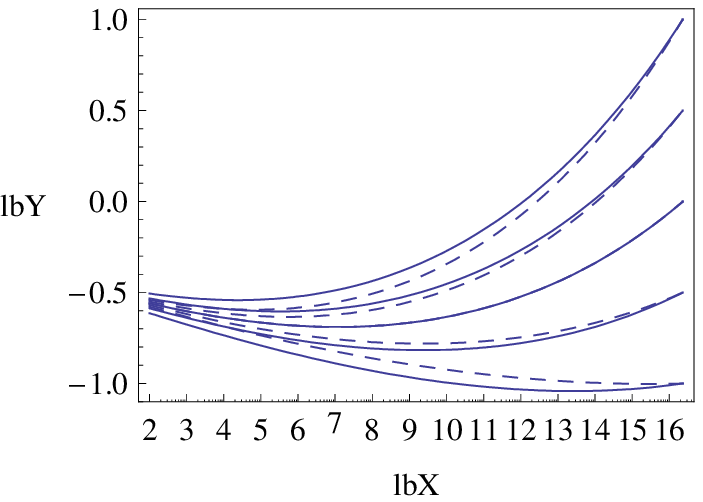}}
\qquad
\subfloat[SPS5]{\includegraphics{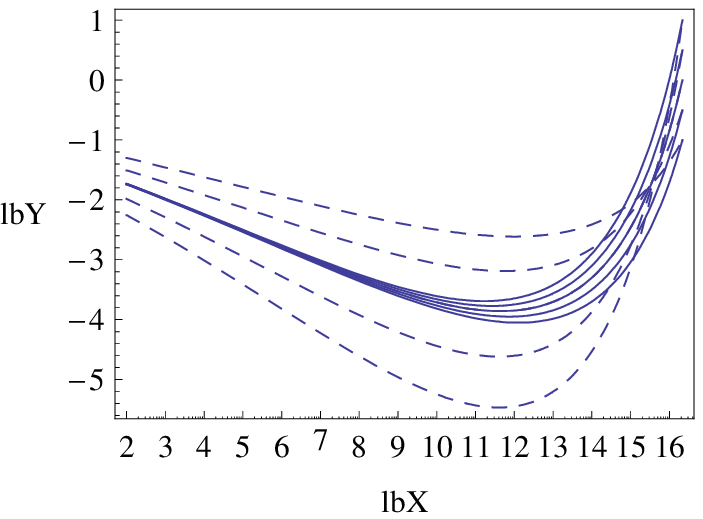}}
\caption{The running of $\frac{\beta_6}{\alpha_3}$. For the solid curve only $\beta_6$ is non-zero at the high scale while for the dashed curve all $\beta_i$ are switched on.}
 \label{fig:b_5/a_2}
\end{figure}

To illustrate the behavior under the renormalization group, we analyze the
situation at several benchmark points. 
These points were chosen to be the so-called SPS points \cite{Allanach:2002nj}
(cf.\ table~\ref{tab:SPS}) amended by corrections in the MFV form. 
\begin{table}[htb]
\centering
\begin{tabular}{ccccc}
\toprule[1.3pt]
  Point & $m_0$ & $m_{\nicefrac{1}{2}}$ & $A$ & $\tan\beta$ \\ 
1a & \unit[100]{GeV} & \unit[250]{GeV} & \unit[-100]{GeV} & 10 \\ 
1b & \unit[200]{GeV} & \unit[400]{GeV} & 0 & 30 \\
2 & \unit[1450]{GeV} & \unit[300]{GeV} & 0 & 10\\
3 & \unit[90]{GeV} & \unit[400]{GeV} & 0 & 10\\
4 & \unit[400]{GeV} & \unit[300]{GeV} & 0 & 10 \\
5 & \unit[150]{GeV} & \unit[300]{GeV} & \unit[-1000]{GeV} & 5 \\ 
\bottomrule[1.3pt]
\end{tabular}
\caption{Survey of SPS points.}
\label{tab:SPS}
\end{table}
Examples for the RG behavior are displayed in figures~\ref{fig:b_1/a_1}
and \ref{fig:b_5/a_2} (see pages~\pageref{fig:b_1/a_1} and \pageref{fig:b_5/a_2}). We show the ratio $\beta_1/\alpha_1$ and
$\beta_6/\alpha_3$, respectively. Note that these ratios coincide with $b_1/a_1$
and $b_6/a_3$ in the original MFV decomposition \cite{D'Ambrosio:2002ex}. These
ratios parameterize the deviations of the soft terms from unit
matrices.
In our illustrations, we use two different initial conditions for the
$\beta_i$. For the solid curve only the shown parameter is set non-zero at the
high scale, i.e.\ in figure~\ref{fig:b_1/a_1} only $\beta_1$ has a non-zero initial
value, while the dashed curves correspond to universal initial conditions for the
$\beta_i$. That is, we choose input values of the soft terms of the form
\eqref{eq:softmassdecomp} with 
\begin{equation}
 \alpha_i=m_0^2
 \quad\text{and}\quad \left( 
 \beta_j=0 \:\forall j\ne k\quad\text{or}\quad
 \beta_i = \text{universal} \right) \;.
\end{equation}
We observe that the ratios get driven to non-trivial, i.e.\ non-zero, fixed
points. The corresponding low-energy fixed point values can be inferred from our
numerical approximations in appendix~\ref{app:Approximations}. 
These fixed points emerge from the competition from alignment and misalignment
processes, as discussed above.

The observed fixed point behavior has been confirmed by an independent study~\cite{Colangelo:2008qp}. The authors of~\cite{Colangelo:2008qp} included the running of complex phases in the couplings in \Eqref{eq:soft_terms}. The CP-violating phases are found to be driven to a fixed point at zero. This relaxes the CP problem associated with these phases.

\subsection{Beyond MFV}

The MFV ansatz is usually justified by the spurion argument. However, there are
some drawbacks to this reasoning. First of all, $G_\text{flavor}$ is anomalous (cf.\ \Eqref{eq:MFV_anomalies}). Secondly, it is
hardly conceivable that one (spurion) field VEV can give rise to a rank three
Yukawa coupling with hierarchical eigenvalues. From these considerations we
infer that the flavor symmetry will likely be broken by more than one field,
such that Yukawa couplings and corrections to soft parameters are proportional
to linear combinations of such fields with, in general, different coefficients.
In this picture one would expect corrections to the MFV scheme. 

Assuming that both new physics interactions and flavor models operate at the
high scale, it is worthwhile to  understand the implications of the presence of
non-MFV terms $\boldsymbol{\Delta  m}^2_f, \boldsymbol{\Delta  A}_f$, which we
add to the ansatz \eqref{eq:softmassdecomp}. That is, we decompose the soft terms
according to
\begin{subequations}\label{eq:nonMFV}
\begin{align}
 \boldsymbol{m}^2_f & = (\boldsymbol{m}^2_f)_\mathrm{MFV}  + \boldsymbol{\Delta m}^2_f\;,\\
 \boldsymbol{A}_f  & = (\boldsymbol{A}_f)_\mathrm{MFV} + \boldsymbol{\Delta A}_f\;.
 \end{align}
\end{subequations}
We require the norm (cf.\ equation~\eqref{eq:norm}) of the non-MFV terms  to be
minimal, which makes the decomposition unambiguous. In other words, we demand
orthogonality between the MFV and the non-MFV terms, for instance
\begin{equation} 
\tr(\boldsymbol{\Delta m}^2_f)=0\;,\quad \tr(\Yu \Yu^\dagger\boldsymbol{\Delta m}^2_u)=0\;,\quad\text{etc.}
\end{equation}
As we know from our considerations in section~\ref{sec:RGE_MFV}, the non-MFV
terms will -- to high accuracy -- not be generated by the running of the MFV
terms. This implies that, in a decomposition of the soft terms in MFV and
non-MFV terms, the evolution of non-MFV terms is always proportional to non-MFV
terms.  To make this statement more precise, let us spell out the RGEs for the
non-MFV terms. With the additional requirement of orthogonality the derivation
is analogous to the one that led us to \Eqref{eq:MFVRGEs}. One obtains
\begin{subequations}
\begin{align}
16\pi^2\,\frac{\D}{\D t} \boldsymbol{\Delta m}_u^2
 &  =  
2\boldsymbol{\Delta m}_u^2\,\Yu \Yu^\dagger
+4\,\Yu\,\boldsymbol{\Delta m}_Q^2\,\Yu^\dagger
+2\,\Yu \Yu^\dagger\,\boldsymbol{\Delta m}_u^2
+4\,\boldsymbol{\Delta}(\boldsymbol{A}_u\boldsymbol{A}_u^\dagger)\;,
\\
16\pi^2\,\frac{\D}{\D t} \boldsymbol{\Delta m}_d^2
& =  
2\boldsymbol{\Delta m}_d^2\,\Yd \Yd^\dagger
+4\,\Yd\,\boldsymbol{\Delta m}_Q^2\,\Yd^\dagger
+2\,\Yd \Yd^\dagger\,\boldsymbol{\Delta m}_d^2
+4\,\boldsymbol{\Delta}(\boldsymbol{A}_d\boldsymbol{A}_d^\dagger)\;,
\\
16\pi^2\,\frac{\D}{\D t} \boldsymbol{\Delta m}_Q^2
 &=  
\boldsymbol{\Delta m}_Q^2\,\left(
\Yu^\dagger \Yu+
\Yd^\dagger \Yd
\right)
+ \left(
\Yu^\dagger \Yu+
\Yd^\dagger \Yd
\right)\,\boldsymbol{\Delta m}_Q^2
\nonumber\\
&  \quad
+2\,\Yu^\dagger\,\boldsymbol{\Delta m}_u^2\,\Yu
+2\,\Yd^\dagger\,\boldsymbol{\Delta m}_d^2\,\Yd
+2\,\boldsymbol{\Delta}(\boldsymbol{A}_u^\dagger\boldsymbol{A}_u)
+2\,\boldsymbol{\Delta}(\boldsymbol{A}_d^\dagger\boldsymbol{A}_d)\;,
\\
 16\pi^2\,
 \frac{\D }{\D t}\boldsymbol{\Delta A}_u
 & =  
 \boldsymbol{\Delta A}_u\,
 \left[3\tr(\Yu\Yu^\dagger)
 +5\,\Yu^\dagger \Yu
 +\Yd^\dagger \Yd
 -\frac{16}{3}g_3^2-3g_2^2-\frac{13}{15}g_1^2 \right]
 \nonumber\\
  &  \quad+\Yu\,
 \left[
  4\,\Yu^\dagger\,\boldsymbol{\Delta A}_u
 + 2\,\Yd^\dagger\,\boldsymbol{\Delta A}_d\right]\;,\\
 16\pi^2\,
 \frac{\D }{\D t}\boldsymbol{\Delta A}_d
 & =  
 \boldsymbol{\Delta A}_d\,
 \left[\tr(3\,\Yd\Yd^\dagger
 	+\Ye^\dagger\Ye)
 +5\,\Yd^\dagger \Yd
 +\Yu^\dagger \Yu
 -\frac{16}{3}g_3^2-3g_2^2-\frac{7}{15}g_1^2 \right]
 \nonumber\\
 &  \quad+\Yd\,
 \left[
  4\,\Yd^\dagger\,\boldsymbol{\Delta A}_d
 + 2\,\Yu^\dagger\,\boldsymbol{\Delta A}_u\right]\;,
 \label{eq:RGE4DelataAu}
\end{align}
\end{subequations}
where
$\boldsymbol{\Delta}(\boldsymbol{A}_f\boldsymbol{A}_f^\dagger)=\boldsymbol{\Delta
A}_f\,\boldsymbol{A}_f^\dagger+\boldsymbol{A}_f\,\boldsymbol{\Delta
A}_f^\dagger+\boldsymbol{\Delta A}_f\,\boldsymbol{\Delta A}_f^\dagger$. An
important point to notice is that the $\boldsymbol{\Delta m}^2_f$ terms get only
contributions from the Yukawas but not from the gauge couplings. This is not
true for the $\boldsymbol{\Delta A}_f$ terms, where the running is substantial.
However, the $\boldsymbol{\Delta A}_f$ terms cannot be too large since they are
constrained by FCNC processes and the requirement of avoiding charge and color
breaking minima \cite{Gabbiani:1996hi,Casas:1996de}. We have also checked that,
due to the hierarchical structure of the Yukawas, non-MFV terms will to a good
accuracy  stay non-MFV, i.e.\ orthogonal to the MFV terms, under the RGE. This
statement applies as long as the corrections $\boldsymbol{\Delta A}_f$ are not
too large, which we assume, as discussed. We would like to close by summarizing
the following observations:
\begin{enumerate}
\item
For vanishing $\boldsymbol{\Delta A}_f$ the $\beta$-functions of the $(\boldsymbol{ \Delta m}^2_f)_{\mathrm{off-diagonal}}$ are only proportional to the Yukawa
couplings. Hence the $\boldsymbol{(\Delta m}^2_f)_{\mathrm{off-diagonal}}$ stay almost constant.

\item
By contrast, the $\boldsymbol{\Delta A}_f$ do change due to the running. The
dominant contributions are a scaling effect proportional to the gauge couplings
and a lowering proportional to the top Yukawa. The net evolution can 
be approximated by $\boldsymbol{\Delta A}_f|_\mathrm{low-scale} \approx
(1\!-\!3)\cdot \boldsymbol{\Delta A}_f|_\mathrm{high-scale}$.
\end{enumerate}

\section{Discussion}
We have seen that the MFV parameters get driven to non-trivial fixed points when running from the GUT scale to the electroweak scale, at least if the ratio between scalar and gaugino masses is order unity. This means that the phenomenology is rather insensitive to the exact structure at the GUT scale. Hence, the flavor problem is relaxed in this context because we can obtain an acceptable phenomenology without extensively fine-tuning the flavor structure at the GUT scale.

Additionally, we stressed that the usual MFV ansatz is based on an anomalous flavor symmetry $G_\text{flavor}$. This raises the question whether $G_\text{flavor}$ can be of fundamental nature. We proposed to use a non-abelian, anomaly-free, discrete flavor symmetry instead which gives higher order corrections to the MFV ansatz. Because of the described properties of the renormalization group equations, these corrections are expected not to cause significant deviations. 

\clearpage
\thispagestyle{empty}

\chapter{Application to String Theory}
\label{chap:application_string}

In this section we will show how the methods developed in the previous chapters can be used in string theory. The importance of discrete symmetries in string theory has recently been stressed~\cite{Nilles:2009yd}. Specifically, we will discuss a model which exhibits the $\Z4^R$ symmetry of section~\ref{sec:Z4R} such that proton stability and the $\mu$-problem are under control, and has a flavor structure similar to MFV.

\section{Orbifold Compactifications}
String theory~\cite{Polchinski:1998rr,Polchinski:1998rq,Green:1987sp,Green:1987mn,Becker:2007zj} is a candidate for a theory of quantum gravity and a potential theory of all known interactions. String theory with fermions is only consistent in ten dimensions. We will only discuss heterotic string theory~\cite{Gross:1985fr,Gross:1985rr}. In order to build models of the real world, one compactifies the six additional dimensions on a compact manifold, an idea which goes back to Kaluza and Klein~\cite{Kaluza:1921tu,Klein:1926tv}. The compactification manifold has to be of Calabi-Yau type~\cite{Candelas:1985en} if we want $\mathcal{N}=1$ supersymmetry in four dimensions. Another option is the compactification on orbifolds~\cite{Dixon:1985jw,Dixon:1986jc,Bailin:1999nk} whose geometry is much easier to handle. For a recent review see~\cite{Nilles:2008gq}.

It proved to be a difficult task to reconstruct the standard model from string theory. There are myriads of different vacua in string theory, all with different physical features, known as the 'string landscape'. Recently, there has been some progress in building realistic models from the heterotic string~\cite{Buchmuller:2005jr,Buchmuller:2006ik,Lebedev:2006kn,Lebedev:2007hv}. There, a string compactification with the exact MSSM particle spectrum and other phenomenologically appealing features as matter parity is presented. However, the models were not completely realistic mainly because of uncontrolled proton decay. 

\section{A Concrete Model}
We will treat a model based on a $\Z2\times\Z2$ orbifold compactification of the heterotic string with an additional freely acting \Z2. A similar model was presented in~\cite{Blaszczyk:2009in}. However, in this model all Higgs fields get large masses, i.e. there is no Higgs pair with masses around the eletroweak scale.

The $\E8\times\E8$ gauge group in ten dimensions is broken upon compactification to four dimensions to
\begin{equation} \label{eq:string_gauge_group}
 \SU3_C\times\SU2_L\times\U1_Y\times \left[  \SU3\times \SU2 \times \SU2\times\U1^8 \right]\:,
\end{equation}
that is the standard model gauge group times a hidden sector times eight \U1s. There is also a $\Z4^R\times\Z4^R\times\Z4^R$ $R$-symmetry coming from the Lorenz symmetry in the compactified dimensions. The complete spectrum, including all quantum numbers, is given in appendix~\ref{app:details_model}. It is summarized in table~\ref{tab:massless_spectrum}.
\begin{table}[htb]
\centering
\begin{tabular}{cccccccccccccc}
\toprule[1.3pt]
Label & $q$ & $\bar u$ & $\bar d$ & $\ell$ & $\bar e$ & $\delta$ & $\bar\delta$ & $h$ & $\bar h$ & $x$ & $\bar x$ & $y$ & $z$ \\ \midrule
\# & 3 & 3 & 3 & 3 & 3 & 3 & 3 & 6 & 6 & 5 & 5 & 6 & 6\\ \midrule
$\SU3_C$ & $\rep{3}$ & $\crep{3}$ & $\crep{3}$ & $\rep{1}$ & $\rep{1}$ & $\rep{3}$ & $\crep{3}$ & $\rep{1}$ & $\rep{1}$ & $\rep{1}$ & $\rep{1}$ & $\rep{1}$ & $\rep{1}$\\
 $\SU2_L$ & $\rep{2}$ & $\rep{1}$ & $\rep{1}$ & $\rep{2}$ & $\rep{1}$ & $\rep{1}$ & $\rep{1}$ & $\rep{2}$ & $\rep{2}$ & $\rep{1}$ & $\rep{1}$ & $\rep{1}$ & $\rep{1}$\\
 $\U1_Y$ & $\tfrac{1}{6}$ & -$\tfrac{2}{3}$ & $\tfrac{1}{3}$ & -$\tfrac{1}{2}$ & 1 & -$\tfrac{1}{3}$ & $\tfrac{1}{3}$ & -$\tfrac{1}{2}$ & $\tfrac{1}{2}$ & 0 & 0 & 0 & 0\\ \midrule
 $\SU3$ & $\rep{1}$ & $\rep{1}$ & $\rep{1}$ & $\rep{1}$ & $\rep{1}$ & $\rep{1}$ & $\rep{1}$ & $\rep{1}$ & $\rep{1}$ & $\rep{3}$ & $\crep{3}$ & $\rep{1}$ & $\rep{1}$\\
 $\SU2$ & $\rep{1}$ & $\rep{1}$ & $\rep{1}$ & $\rep{1}$ & $\rep{1}$ & $\rep{1}$ & $\rep{1}$ & $\rep{1}$ & $\rep{1}$ & $\rep{1}$ & $\rep{1}$ & $\rep{2}$ & $\rep{1}$\\
 $\SU2$ & $\rep{1}$ & $\rep{1}$ & $\rep{1}$ & $\rep{1}$  & $\rep{1}$ & $\rep{1}$ & $\rep{1}$ & $\rep{1}$& $\rep{1}$ & $\rep{1}$ & $\rep{1}$ & $\rep{1}$ & $\rep{2}$\\
\bottomrule[1.3pt]
\end{tabular}
\caption{The massless spectrum of the $\Z2\times\Z2$ model. The representations are given w.r.t.\ the gauge group $\SU3_C\times\SU2_L\times\U1_Y \times \left[  \SU3\times \SU2  \times \SU2\right]_\text{hidden}$.}
\label{tab:massless_spectrum}
\end{table}

The spectrum consists of three generations of quarks ($q$, $\bar{u}$, $\bar{d}$) and leptons ($\ell$, $\bar{e}$). There are also three vector-like pairs of exotic $\SU3_C$ triplets ($\delta$, $\bar\delta$) and six pairs of Higgs candidates ($h$, $\bar h$). Note that the states $\bar d$ and $\bar \delta$ have the same quantum numbers. We will justify this notation in retrospect because in the vacuum we will choose below, a symmetry, namely the matter parity subgroup of the $\Z4^R$, will distinguish between quarks and exotics. The same is true for $\ell$ and $h$.

Finally, there are 22 fields which are charged under the hidden sector ($x$, $\bar x$, $y$, $z$). In addition, we have 37 singlets under all non-abelian gauge groups.

\subsection{VEV Assignment}
In a particular vacuum, some of the massless fields will obtain a VEV. One reason for this is the so-called Fayet-Iliopoulos (FI) term~\cite{Fayet:1974jb}. 

In general, global supersymmetry the scalar potential, which needs to vanish for supersymmetry to be unbroken, is given by
\begin{equation} 
 V = \frac{1}{2}\sum_a D_a^2 + \sum_i |F_i|^2
\end{equation}
where $a$ labels the gauge groups and the index $i$ runs over the chiral superfields in the theory. The auxiliary fields $D_a$ and $F_i$ are given by
\begin{subequations}
\begin{align}
D_a &= \sum_i \phi_i^\dagger\,\mathbf{t}_a \, \phi_i\:, \\
F_i &= \frac{\partial \mathscr{W}}{\partial \phi_i}
\end{align}
\end{subequations}
with the superpotential $\mathscr{W}$ and $\phi_i$ is the scalar component of a chiral superfield. $\mathbf{t}_a$ label the generators of the gauge group.

It is a common feature of heterotic orbifold compactifications that the low energy gauge group contains an anomalous $\U1_{\text{anom}}$~\cite{Kobayashi:1996pb}. The anomaly is cancelled by the Green-Schwarz mechanism~\cite{Green:1984sg} (cf.\ section~\ref{sec:discrete_GS}). The crucial point is now that the $D$-term associated with $\U1_{\text{anom}}$ gets an additional contribution~\cite{Dine:1987xk,Atick:1987gy} at one loop
\begin{equation}
 D_\text{anom} = \sum_i Q_\text{anom}^{(i)}\, |\phi_i|^2 + \frac{g\, M^2_\text{Planck}}{192\pi^2} \tr Q_\text{anom}
\end{equation}
where $g$ is the four-dimensional gauge coupling. Since we do not want to break supersymmetry near the Planck scale $M_\text{Planck}$, some fields, with negative charges under $\U1_{\text{anom}}$, need to obtain VEVs in order to cancel the FI-term.

With the methods developed in chapter~\ref{chap:abelian} we search for a set of fields which lead to an acceptable phenomenology when getting a VEV. Especially, we want to obtain the $\Z4^R$ of section~\ref{sec:Z4R} at low energies. We will consider a vacuum configuration in which the fields
\begin{align}
 \widetilde{\phi}^{(i)} &=
  \{\phi _1, \phi _2, \phi _3, \phi _4, \phi _5, \phi _6, \phi _7, \phi _8, 
 \phi _9, \phi _{10}, \phi _{11}, \phi _{12}, \phi _{13}, \phi _{14}, 
 \phi _{15}, \phi _{16}, \phi _{17}, \phi _{18}, \nonumber \\
 & \qquad x_1,x_2,x_3,\bar x_1,\bar x_2,\bar x_3,\bar x_4 ,z_1,z_2,z_3,z_4,z_5,z_6\} \;.
\label{eq:VEVfields}
\end{align}
will obtain VEVs.

\subsection{Remnant discrete symmetries}
Giving VEVs to the fields in \Eqref{eq:VEVfields} breaks the hidden sector \SU3 and one of the hidden \SU2s. Additionally, we can use the methods of chapter~\ref{chap:abelian}, especially the \texttt{Mathematica} package described in appendix~\ref{app:package}, to conclude that all \U1s but hypercharge in \Eqref{eq:string_gauge_group} and the $R$-symmetries are broken via the scheme
\begin{equation}
 \U1^8\times\Z4^R\times\Z4^R\times\Z4^R\quad\to\quad\Z4^R \: .
\end{equation}
The $\Z4^R$ charges of the fields which are charged under the standard model gauge group is given in table~\ref{tab:matter_charges}. The complete charge assignment is tabulated in appendix~\ref{app:details_model}.

\begin{table}[htb]
\centering
\subfloat[Quarks and leptons.]{
\begin{tabular}{cccccc}
\toprule[1.3pt]
 & $q_i$ & $\bar u_i$ & $\bar d_i$ & $\ell_i$ & $\bar e_i$ \\
$\Z4^R$ & 1 & 1 & 1 & 1 & 1\\
\bottomrule[1.3pt]
\end{tabular}
}
\\
\subfloat[Higgs and exotics.]{
\begin{tabular}{ccccccccccccccc}
\toprule[1.3pt]
 & $h_1$ & $h_2$& $h_3$& $h_4$& $h_5$& $h_6$ & $\bar h_1$ & $\bar h_2$ & $\bar h_3$ & $\bar h_4$ & $\bar h_5$ &  $\bar h_6$ & $\delta_i$ & $\bar\delta_i$ \\
$\Z4^R$ & 2 & 0 & 0 & 0 & 0 & 0 & 2 & 0 & 0 & 2 & 2 & 2 & 2 & 0 \\
\bottomrule[1.3pt]
\end{tabular}
}
\caption{The $\Z4^R$ charges of the fields which are charges under the standard model. The index $i$ takes values $i=1,2,3$.}
\label{tab:matter_charges}
\end{table}
In the matter sector, this is exactly the $\Z4^R$ symmetry we discussed in section~\ref{sec:Z4R} (cf. table~\ref{tab:Z4R_charges}). We can now justify our labeling of states. All Higgses and exotics have even charges under the $\Z4^R$ while quarks and leptons have odd charges.

\subsection{Decoupling of exotics}
In string compactification one usually encounters exotics, i.e. fields with standard model quantum numbers which are not part of the SM field content. In our concrete case, these are the exotic triplets $\delta$, $\bar\delta$ and the additional Higgses $h$, $\bar h$. The exotics have to be decoupled from the low-energy spectrum by giving large masses to them. 

The usual way of checking decoupling of some exotics $X_i$ and $\bar X_j$ is to check the existence of a coupling of the form~\cite{Nilles:2008gq}
\begin{equation}
 \widetilde{\phi}^n \, X_i \bar X_j \: , \label{eq:exotics_decoupling}
\end{equation}
making use of string selection rules. $\widetilde{\phi}^n$ is a monomial of order $n$ in the VEV fields in \Eqref{eq:VEVfields}. When the $\widetilde{\phi}$s obtain VEVs, the couplings in \Eqref{eq:exotics_decoupling} will generate a mass matrix for the exotics. 

The discrete symmetries allow for checking the decoupling. With the charges in table~\ref{tab:matter_charges} we can readily analyze the structure of the mass matrices. As a crosscheck, we computed the couplings directly with the string selection rules in appendix~\ref{app:string_selection_rules}. Note that it is a highly non-trivial check of our methods that both calculations, based on the one hand on discrete symmetries and on the other hand on string selection rules, give the same result. The $\bar h-h $ Higgs mass matrix turns out to be
\begin{equation}
 \mathcal{M}_h  =
\begin{pmatrix} 
0 & \phi_{7} & \phi_{5} & \widetilde{\phi}^{7} & \phi_{9} & \phi_{10}\\ 
\phi_{6} & 0 & 0 & 0 & 0 & 0 \\ 
\phi_{4} & 0 & 0 & 0 & 0 & 0 \\ 
0 & \phi_{2}\phi_{4}\phi_{7} & \phi_{2} &  \widetilde{\phi}^{7} & \phi_{15} & \phi_{16} \\
0 & \phi_{17} & \widetilde{\phi}^{3} & \phi_{13} & \phi_{1}& \widetilde{\phi}^{3}\\
0 & \phi_{18} & \widetilde{\phi}^{3} & \phi_{14} & \widetilde{\phi}^{3} & \phi_{1}\\
\end{pmatrix}\;.
\end{equation}
Clearly, this mass matrix has rank five. The massless Higgs pair, which can be identified with the MSSM Higgs fields, is given by the linear combination
\begin{subequations}
\begin{eqnarray}
 h_u & = & a_2\, \bar h_2 +a_3\, \bar h_3\;,\\
 h_d & = & b_3\,h_3 +b_4\,h_4 +b_5\,h_5 +b_6\,h_6 \;,
\end{eqnarray}
\end{subequations}
with $a_i$ and $b_j$ denoting coefficients. The $\bar \delta-\delta$ mass matrix is
\begin{equation}
 \mathcal{M}_\delta =
\begin{pmatrix}
 \widetilde \phi ^9 & \phi_{15}  &  \phi_{16}  \\
 \phi_{13}  & \phi_{1}  &  \phi_{6}(\phi_{10}\phi_{17}+\phi_{9}\phi_{18}) \\
 \phi_{14}  &  \phi_{6}(\phi_{10}\phi_{17}+\phi_{9}\phi_{18}) & \phi_{1} 
\end{pmatrix} \:.
\end{equation}
Hence, the matrix has full rank and all exotics decouple at linear level in the $\widetilde{\phi}$ fields, i.e.\ they get masses near the Planck scale. Note that the block structure of $\mathcal{M}_\delta$ is not a coincidence but a consequence of the fact that $\delta_2/\delta_3$ and $\bar \delta_2/\bar \delta_3$ form $D_4$ doublets. 

\subsection{Yukawa couplings}
The effective Yukawa couplings are defined by 
\begin{equation}
 \mathscr{L}_Y~=~
 \sum_{i=2,3}\left[ (Y_u^{(i)})^{fg}\,q_f\,\bar u_g\,\bar h_i \right] 
 +\sum_{i=3}^6
 \left[(Y_d^{(i)})^{fg}\,q_f\,\bar d_g\,h_i
 +(Y_e^{(i)})^{fg}\,\ell_f\,\bar e_g\,h_i\right]\;.
\label{eq:definition_yukawas}
\end{equation}
The Yukawa coupling structures are
\begin{subequations}\label{eq:YukawaMatricesZ2xZ2_1-1}
\begin{eqnarray}
Y_u^{(2)} & = & 
\left(
\begin{array}{ccc}
 1 &  \widetilde{\phi} ^4 &  \widetilde{\phi} ^4 \\
 \widetilde{\phi} ^4 & 1 &  \widetilde{\phi} ^4 \\
  \widetilde{\phi} ^4 &  \widetilde{\phi} ^4 &  \widetilde{\phi} ^2
\end{array}
\right) \;, 
\quad
Y_u^{(3)}  ~=~  
\left(
\begin{array}{ccc}
  \phi_{4}\phi_{7} &   \widetilde{\phi} ^6 &  \widetilde{\phi} ^6 \\
   \widetilde{\phi} ^6 & \phi_{4}\phi_{7} &  \widetilde{\phi} ^6 \\
  \widetilde{\phi} ^6 &  \widetilde{\phi} ^6 & 1
\end{array}
\right)\;,
\label{eq:Yu}\\
Y_e^{(5)} =  (Y_d^{(5)})^T & = & 
\left(
\begin{array}{ccc}
  \widetilde{\phi} ^6 &  \widetilde{\phi} ^6 &  \widetilde{\phi} ^4 \\
  \widetilde{\phi} ^6 &  \widetilde{\phi} ^6 & 1 \\
   \widetilde{\phi} ^4 & 1 &  \widetilde{\phi} ^4
\end{array}
\right) \;,
\\
Y_e^{(6)} =  (Y_d^{(6)})^T & = & 
\left(
\begin{array}{ccc}
  \widetilde{\phi} ^6 &  \widetilde{\phi} ^6 & 1 \\
  \widetilde{\phi} ^6 &  \widetilde{\phi} ^6 &  \widetilde{\phi} ^4 \\
 1 &  \widetilde{\phi} ^4 &  \widetilde{\phi} ^4
\end{array}
\right) \;.
\label{eq:YdYe}
\end{eqnarray}
\end{subequations}
$Y_d$ and $Y_e$ coincide at tree-level, i.e.\ they exhibit \SU5
GUT relations. There are additional contributions to $Y_e$/$Y_d$
from couplings to $h_{3,4}$ which appear at order
$\widetilde\phi^{10}$ and can be neglected if the VEVs of the
$\widetilde\phi$-fields are small.

Because of the localization of the matter fields, we expect the large diagonal entries in $Y_u^{(2)}$ and the large  (1,3) and (3,1) entries in $Y_e^{(6)}$ to be exponentially suppressed. 

\subsection[$D_4$ flavor symmetry]{$\boldsymbol{D_4}$ flavor symmetry}
\label{sec:D4_symmetry}
The block structure of the
Yukawa matrices is not a coincidence but a consequence of a $D_4$ flavor
symmetry \cite{Grimus:2003kq}, related to the vanishing Wilson line in the $e_1$
direction, $W_1 = 0$ (cf.\ e.g.\ \cite{Kobayashi:2006wq,Ko:2007dz}).  The first two
generations transform as a $D_4$ doublet, while the third generation is a $D_4$
singlet. The two generators of the doublet representation are
\begin{equation}
\sigma_3=\begin{pmatrix}1&0\\0&-1\end{pmatrix}\: ,
\qquad
\sigma_2=\begin{pmatrix}0&1\\1&0\end{pmatrix}\: .
\end{equation}
Below we will write formulas with explicit $D_4$ doublets contracted
with the standard scalar product. Note that there are invariants with more than
two $D_4$ charged fields which cannot be written in terms of a scalar product.

A consequence of the $D_4$ symmetry is that the soft masses (see \Eqref{eq:soft_terms}) are of the form~\cite{Ko:2007dz}
\begin{equation}
 \boldsymbol{m}^2 = 
\begin{pmatrix}
a&0&0 \\
0&a&0 \\
0&0&b
\end{pmatrix}
+
D_4\,\text{breaking VEVs} \:.
\end{equation}
This structure is very similar to the MFV ansatz in section~\ref{sec:MFV_ansatz}. Hence, we expect the flavor problem to be relaxed by the fixed-point behavior of the MFV parameters described in chapter~\ref{chap:MFV}.

\subsection{Neutrino masses}
In our model we have 14 neutrinos, i.e.\ SM singlets whose charges are odd under $\Z4^R$ meaning that
they have odd $\Z2^\mathcal{M}$ charge, where $\Z2^\mathcal{M}$ is the matter parity subgroup of $\Z4^R$. In our vacuum the neutrinos are
\begin{equation}
 N= \{ n_1 , n_2 , n_3 , \bar n_1 , \bar n_2 , \bar n_3 , \bar n_4 , \bar n_5 , \bar n_6 , \bar n_7 , \bar n_8 , x_4 , x_5 , \bar x_5 \} \: .
\end{equation}
The see-saw mechanism~\cite{Minkowski:1977sc,Yanagida:1979as} is an attractive way to account for the smallness of neutrino masses. In supersymmetry the see-saw mechanism is described by the superpotential terms
\begin{equation}
 \mathscr{W}\supset (Y_\nu)^{ij}\, h_u \, \ell_i \, N_j +\frac{1}{2} M_{ij}\, N_i \, N_j \:.
\end{equation}
When $h_u$ obtains a VEV during the electroweak phase transition, the neutrinos get small masses of $\mathcal{O}(\unit[10^{-3}]{eV})$ if $M$ is around the GUT scale \unit[$10^{16}$]{GeV}.

The neutrino Yukawa coupling $Y_\nu$ is a $3\times 14$ matrix and has full
rank in our vacuum. $M$ is a $14\times 14$ mass matrix and  has rank 14 at the perturbative level. Hence the neutrino see-saw with many neutrinos \cite{Buchmuller:2007zd} is at work.

\subsection{Proton decay operators} The $\Z4^R$ symmetry forbids
all dimension four and five proton decay operators at the
perturbative level (see section~\ref{sec:Z4R}). In addition, the
non-anomalous matter parity subgroup $\Z2^\mathcal{M}$ forbids all
dimension four operators also non-perturbatively. The dimension
five operators like $q\,q\,q\,\ell$ are generated
non-pertubatively, as we will discuss below.

\subsection[Non-perturbative violation of $\Z4^R$]{Non-perturbative violation of $\boldsymbol{\Z4^R}$}
As discussed in section~\ref{sec:Z4R_breaking} the $\Z4^R$ is broken if the superpotential attains a VEV. In the present model, we expect the hidden sector \SU2 to generate a superpotential VEV and break supersymmetry via gaugino condensation. This can effectively be described by the term $e^{-aS}$ (see section~\ref{sec:Z4R_breaking}). 

Once we include the terms that are only forbidden by the $\Z4^R$ symmetry, we obtain further couplings. 
An example for such an additional term is the dimension five proton decay operator, 
\begin{equation}
 \mathscr{W}_{\text{np}}
 \supset
 q_1\,q_1\,q_2\,\ell_1\,
 \mathrm{e}^{-a\,S}\, (z_1z_2)^2 
\left[\begin{pmatrix}\phi_{17}\\\phi_{18}\end{pmatrix} \cdot \begin{pmatrix}\phi_{17}\\\phi_{18}\end{pmatrix}\right]^2
\phi_2\,\phi_4^2\,\phi_7^2
\left[\begin{pmatrix}\phi_{12}\\\phi_{11}\end{pmatrix} \cdot \begin{pmatrix}\phi_{17}\\\phi_{18}\end{pmatrix}\right]
\end{equation}
where we suppressed coefficients. This is a concrete realization of the discussion in section~\ref{sec:Z4R_breaking}. Especially, $S$ is the dilaton, and $a$ is chosen s.t.\ $e^{-aS}$ has charge 15 under the anomalous $\U1_\text{anom}$ and is odd under the space group selection rule $\Z{2}^{n_5}$. The bracket structure between the $\phi_{17}$/$\phi_{18}$ and $\phi_{11}$/$\phi_{12}$ is a consequence of the non-Abelian $D_4$ symmetry, where these fields transform as doublets.

\subsection[Solution to the $\mu$-Problem]{Solution to the $\boldsymbol{\mu}$ problem}
In this model any allowed superpotential term can serve as an effective $\mu$ term
(cf.\ the discussion in \cite{Kappl:2008ie}). This fact can be seen from
higher-dimensional gauge invariance \cite{Brummer:2010fr}.
Therefore, the (non-perturbative) $\mu$ term is of the order of the gravitino mass,
\begin{equation}
 \mu\sim\langle\mathscr{W}_\mathrm{np}\rangle\sim m_{3/2}
\end{equation}
in Planck units. If hidden sector strong dynamics induces a
non-trivial $\langle\mathscr{W}_\mathrm{np}\rangle$, the $\mu$ problem is
solved as discussed in section~\ref{sec:Z4R_breaking}.

\subsection{Anomaly Mixing}
In our model we have two anomalous symmetries. On the one hand, we have the anomalous $\U1_\text{anom}$. On the other hand, the space group selection rule $\Z2^{n_5}$ for $n_5$, corresponding to the anomalous space group element in \Eqref{eq:anomalous_space_group_element}, is anomalous. We want to answer the question, if it is possible to rotate the $\Z2^{n_5}$ anomaly completely into the anomalous $\U1_\text{anom}$. This question has generally been discussed in section~\ref{sec:anomaly_conditions_abelian}. 

Hence, we have to consider the two anomaly coefficients
\begin{subequations}
\begin{alignat}{2}
 G-G-\U1_\text{anom}:&\quad A & = 15  \: , \\
 G-G-\Z2^{n_5}:&\quad B & = \frac{1}{2} \: ,
\end{alignat} 
\end{subequations}
with $G=\{\SU3_C,\,\SU2_L\}$. Note that we have scaled to $\U1_\text{anom}$ charges to integers. The two anomalies are not independent if (cf. \Eqref{eq:anomaly_mixing})
\begin{equation}
 B+nA = \frac{1}{2}+15 n =0\mod 1
\end{equation}
has a solution for $n\in\Z{}$ which is obviously not the case, and therefore the anomaly of $\Z2^{n_5}$ cannot be removed. Thus, we have two independent sources of anomalies in our model.

\clearpage
\thispagestyle{empty}
\chapter{Summary}

In this thesis, discrete symmetries in the MSSM were discussed. In order to understand the origin of these symmetries, a method was developed which allows to calculate the remnant discrete symmetry of a spontaneously broken $\U1^N$ symmetry. This method was then generalized to other cases such as the breaking of $\Z{N}^R$ symmetries. Additionally, a way to visualize the breaking was highlighted.

Another important ingredient, when studying discrete symmetries, are anomalies. Anomalies were discussed in the path integral. Special attention was paid to anomaly cancelation via the Green-Schwarz mechanism. With these methods a unique $\Z4^R$ symmetry in the MSSM was found which
\begin{itemize}
\item forbids dimension four and five proton decay operators at the perturbative level,
\item contains matter parity as a subgroup, and therefore dimension four proton decay operators remain forbidden also non-perturbatively,
\item allows the Weinberg operator for neutrino masses,
\item solves the $\mu$-problem,
\item is anomaly-free with the Green-Schwarz mechanism,
\item is compatible with grand unification.
\end{itemize}
In addition, the flavor problem of the MSSM in the context of MFV was addressed. An anomaly-free, discrete symmetry was presented that could replace the continuous flavor symmetry which is usually used in the formulation of MFV. It was shown that the renormalization group running of the MFV coefficients exhibits non-trivial fixed points, thus relaxing the flavor problem.

Finally, a string theory model with realistic features such as the exact MSSM spectrum  was described which illustrates the discussed methods and symmetries. The \texttt{Mathematica} package of appendix~\ref{app:package} was, for example, used in finding this model. The model also possesses the presented $\Z4^R$ symmetry and a flavor structure similar to MFV.

\clearpage
\thispagestyle{empty}
\chapter*{Acknowledgments}
\addcontentsline{toc}{chapter}{Acknowledgments} 

At this point, I want to give credit to all the people who contributed to this thesis in whichever way. Especially, I would like to thank
\begin{itemize}
 \item Michael Ratz for his support as a supervisor. I appreciate his guidance, all the discussion time, and the possibility to visit conferences and summer schools all over the world.
 \item Hyun Min Lee, Paride Paradisi, Björn Petersen, Stuart Raby, Graham Ross, Kai Schmidt-Hoberg, Cristoforo Simonetto, and Patrick Vaudrevange for fruitful collaborations.
 \item the whole T30e team, in particular Rolf Kappl, Björn Petersen and Martin Winkler, for creating a very pleasant working environment, and for spending their time with me also outside the physics department.
 \item Cristoforo Simonetto for being my favorite room mate, for all discussions, and all the tea he thankfully accepted. 
 \item Rolf Kappl, Kai Schmidt-Hoberg, and Cristoforo Simonetto for proof reading the manuscript.
 \item Karin Ramm for all the bureaucracy she took care of.
\end{itemize}
Last but not least I want to thank my parents for their constant support, and D\'{e}sir\'{e}e for bearing me.
\begin{appendix}

\addcontentsline{toc}{part}{Appendix} 

\makeatletter 
\renewcommand{\chapter}{\@startsection{chapter}{1}{0cm}{-\baselineskip}{1.5\baselineskip}{\LARGE\slshape\bfseries}}
\renewcommand{\section}{\@startsection{section}{2}{0cm}{-\baselineskip}{0,5\baselineskip}{\normalsize\slshape\bfseries}}
\renewcommand{\subsection}{\@startsection{subsection}{3}{0cm}{-\baselineskip}{0,5\baselineskip}{\normalsize\slshape\bfseries}}
\makeatother 

\chapter{Finite abelian groups}
\label{app:finite_abelian_groups}
In this section we will present some mathematical background on finite abelian groups and fix our notation. We will not given an extensive treatment of the subject, but rather collect important results which are used in this thesis. For an overview on the subject see the classic reference~\cite{WA03} or a modern presentation e.g.\ \cite{Summit04}.

\section{Cyclic groups}
As we will see below, the building blocks of finite, abelian groups are finite, cyclic groups. A cyclic group $H$ is abstractly defined as a group that is generated by a single element
\begin{equation}
 H=\{ a^n \, | \, n\in\Z{} \}\:.
\end{equation}
Every finite, cyclic group is isomorphic to a cyclic subgroup \Z{N} of \Z{}. The elements of \Z{N} are $\{0,\ldots,N-1\}$ and the group operation is addition modulo $N$.

\subsection*{Representations}
Abelian groups only have one-dimensional, irreducible representations. \Z{N} has $N$ irreducible representations which are labeled by their charge $\{0,\ldots,N-1\}$. A field $\phi$ with charge $q$ transforms under a \Z{N} transformation as
\begin{equation}
 \phi\to e^{2\pi\I\, q \frac{n}{N}} \phi\:, \label{eq:general_ZN_trafo}
\end{equation}
where $n\in\{0,\ldots,N-1\}$ parameterizes the transformation.

\section{Structure of finite abelian groups}
\label{sec:structure_of_finite_abelian_groups}
Finite abelian groups have been fully classified. At the heart of the mathematical treatment lies the fundamental theorem of finite abelian groups which can be found in any algebra textbook (see for example~\cite{WA03}).
\paragraph{Fundamental theorem of finite abelian groups}
Every finite abelian group $G$ is isomorphic to a direct product of cyclic groups.
\begin{equation}
G\cong \Z{a_1} \times \cdots \times \Z{a_n} 
\end{equation}
The numbers $a_i$ are not unique. There are two forms in which they are conveniently presented.
\begin{description}
	\item[invariant factors] In this form the $a_i$ fulfil the condition that $a_i$ divides $a_{i+1}$.
	\item[elementary divisors] In this form the $a_i$ are prime powers, i.e.\ $a_i=p_i^{e_i}$ where $p_i$ is prime. 
\end{description}
Of course, one can switch from one form to the other. The method is given in~\cite{Summit04}. \comment{noch genauer erklären wie man hin und her wechselt!!} In table~\ref{tab:fundamental_theorem_abelian_groups} some examples of the two forms are presented.
\begin{table}[htbp]
\centering
\begin{tabular}{cccc}
\toprule[1.3pt]
invariant factors & \Z6 & \Z8 & $\Z2\times\Z6\times\Z{12}$ \\
elementary divisors & $\Z2\times\Z3$ & \Z8 & $\Z2\times\Z2\times\Z3\times\Z3\times\Z4$\\
\bottomrule[1.3pt]
\end{tabular}
\caption{Some examples of different representations of finite abelian groups.}
\label{tab:fundamental_theorem_abelian_groups}
\end{table}

\section{Automorphisms}
\label{sec:automorphisms_abelian_groups}
An automorphism $\phi$ of a group $G$ is an one-to-one, onto map
\begin{equation}
 \phi: \: G \to G \quad\text{with}\quad \phi(ab)=\phi(a)\phi(b)\:.
\end{equation}
Hence, an automorphism gives merely another way to write the group. From a more physical point of view, an automorphism of an abelian group connects two equivalent charge assignments. The set of automorphisms form a group which we denote by $\Aut{G}$.

A special, but important case, is $G=\Z{N}$. In this case, all automorphisms are given by multiplication with a number which is coprime to $N$. The number of coprimes of an integer $N$ is given by the Euler phi function $\varphi(N)$. Hence, $\Aut{\Z{N}}=\Z{\varphi(N)}$.

The automorphisms of a general, finite, abelian groups are described in~\cite{Hillar:2007}. If we define $H_p=\Z{p^{e_1}}\times\ldots\times\Z{p^{e_n}}$, then Section~\ref{sec:structure_of_finite_abelian_groups} teaches that we can write a finite, abelian group in the form $G\cong H_{p_1}\times\ldots\times H_{p_m}$ with pairwise distinct primes $p_i$. In this elementary divisor decomposition the automorphisms factorize, i.e.\
\begin{equation}
\Aut{G}\cong\Aut{H_{p_1}\times\ldots\times H_{p_m}}\cong \Aut{H_{p_1}}\times\ldots\times \Aut{H_{p_m}}\:.
\end{equation}
Thus, it is sufficient to study \Aut{H_p}.

\subsection*{The automorphisms of $H_p$}
We will denote the elements, more physically speaking the charges, of $H_p=\Z{p^{e_1}}\times\ldots\times\Z{p^{e_n}}$ with a column vector $q=(q_1,\ldots,q_n)^T$. Remember that $p$ is prime. As before $q_i$ is defined modulo $p^{e_i}$. Consider the following set of matrices
\begin{equation}
 R(H_p)=\{ (a_{ij})\in\Z{}^{n\times n}\, |\, p^{e_i-e_j}  \text{ divides } a_{ij}  \text{ for all } 1\leq j \leq i \leq n\}\:.
\end{equation}
As a simple example, take $n=3$, $e_1=1$, $e_2=2$ and $e_3=5$, i.e.\ $H_p=\Z{p}\times\Z{p^2}\times\Z{p^5}$. Then
\begin{equation}
R(H_p)=\left\lbrace
\left(
\begin{array}{ccc}
b_{11} & b_{12} & b_{13} \\ 
b_{21} p & b_{22} & b_{23} \\ 
b_{31} p^4 & b_{32}p^3 & b_{33}
\end{array}
\right) 
\, | \,
b_{ij}\in\Z{}
\right\rbrace \:. 
\end{equation}
With these matrices, we can define a map $\Psi_A$ from $H_p$ to $H_p$ by 
\begin{equation}
 \Psi_A(q) = q' = A\, q \qquad\text{with}\qquad A\in R(H_p)
\end{equation}
where, again, each $q'_i$ is defined modulo $p^{e_i}$. These maps are the endomorphisms of $H_p$. The automorphisms are given by those elements of $R(H_p)$ which are invertible modulo $p$. That is
\begin{equation}
 \Aut{H_p}=\{\Psi_A\,|\, A \:(\text{mod}\, p) \in \text{GL}_n(\Z{p}) \}\:,
\end{equation}
where $A \:(\text{mod}\, p)$ means that every entry in $A$ is taken modulo $p$. In case $H_p=\Z{p}\times\ldots\times\Z{p}$, i.e.\ $e_i=1$ for all $i$, we have $\Aut{H_p}=\text{GL}_n(\Z{p})$. 

If we define $d_k=\max\{ l\,|\, e_l=e_k \}$ and $c_k=\min\{ l\,|\, e_l=e_k \}$, the number of automorphisms is given by~\cite{Hillar:2007}
\begin{equation}
 |\Aut{H_p}|=\prod_{k=1}^n (p^{d_k}-p^{k-1})\: \prod_{j=1}^n (p^{e_j})^{n-d_j}\:\prod_{i=1}^n (p^{e_i-1})^{n-c_i+1}\:.
\label{eq:number_automorphisms}
\end{equation}

\paragraph{Example}
Let us consider the automorphism of $H_2=\Z{2}\times\Z{4}$ that is $p=2$, $n=2$, $e_1=1$ and $e_2=2$. From \Eqref{eq:number_automorphisms} we know that $|\Aut{H_p}|=8$, since $c_1=d_1=1$ and $c_2=d_2=2$. The automorphisms are given by the matrices
\begin{align}
\left(\begin{array}{cc} 1 & 0 \\ 0 & 1 \end{array}\right)\:,
\left(\begin{array}{cc} 1 & 1 \\ 0 & 1 \end{array}\right)\:,
\left(\begin{array}{cc} 1 & 0 \\ 2 & 1 \end{array}\right)\:,
\left(\begin{array}{cc} 1 & 1 \\ 2 & 1 \end{array}\right)\:,\label{eq:automorphisms_Z2xZ4} \\ \nonumber
\left(\begin{array}{cc} 1 & 0 \\ 0 & 3 \end{array}\right)\:,
\left(\begin{array}{cc} 1 & 1 \\ 0 & 3 \end{array}\right)\:,
\left(\begin{array}{cc} 1 & 0 \\ 2 & 3 \end{array}\right)\:,
\underbrace{\left(\begin{array}{cc} 1 & 1 \\ 2 & 3 \end{array}\right)}_A\:.
\end{align}
Let us call the last matrix in \Eqref{eq:automorphisms_Z2xZ4} $A$. The automorphism $\Psi_A\in\Aut{H_p}$ is the mapping
\begin{align}
\begin{pmatrix} 0 \\ 0 \end{pmatrix} \to \begin{pmatrix} 0 \\ 0 \end{pmatrix}\:,
\begin{pmatrix} 0 \\ 1 \end{pmatrix} \to \begin{pmatrix} 1 \\ 3 \end{pmatrix}\:,
\begin{pmatrix} 0 \\ 2 \end{pmatrix} \to \begin{pmatrix} 0 \\ 2 \end{pmatrix}\:,
\begin{pmatrix} 0 \\ 3 \end{pmatrix} \to \begin{pmatrix} 1 \\ 1 \end{pmatrix}\:,\\\nonumber
\begin{pmatrix} 1 \\ 0 \end{pmatrix} \to \begin{pmatrix} 1 \\ 2 \end{pmatrix}\:,
\begin{pmatrix} 1 \\ 1 \end{pmatrix} \to \begin{pmatrix} 0 \\ 1 \end{pmatrix}\:,
\begin{pmatrix} 1 \\ 2 \end{pmatrix} \to \begin{pmatrix} 1 \\ 0 \end{pmatrix}\:,
\begin{pmatrix} 1 \\ 3 \end{pmatrix} \to \begin{pmatrix} 0 \\ 3 \end{pmatrix}\:.
\end{align}

\section[Mixing with a \U1]{Mixing with a $\boldsymbol{\U1}$}
\label{app:anomaly_mixing}
In the previous section we have shown how the automorphisms of a finite, abelian group can be used to redefine the charges while leaving the physics invariant. In this section, we will discuss a way to obtain equivalent charge assignments by mixing a \Z{N} symmetry with a \U1.

Consider a theory with a $\U1\times\Z{N}$ symmetry. We will take the \U1 to be gauged but this is not necessary for our purposes. A symmetry transformation acts on the fields in the theory as
\begin{subequations}
\begin{alignat}{2}
&\Z{N} \, &: \quad\phi^{(i)} & \to e^{\frac{2\pi\I}{N}q^{(i)}n} \phi^{(i)} \:, \\
&\U1 \, &: \quad\phi^{(i)} & \to e^{\I\, Q^{(i)}\alpha(x)} \phi^{(i)} \:,
\end{alignat}
\end{subequations}
where $n=0,\ldots, N-1$ parameterizes the $\Z{N}$ transformation, and $\alpha(x)$ the \U1 transformation. Every time we do a \Z{N} transformation, we can also perform an \U1 transformation with $\alpha=\frac{2\pi}{N}n$. In this case, the fields behave as
\begin{equation}
 \phi^{(i)} \to \exp \left( \frac{2\pi\I}{N}(q^{(i)}+Q^{(i)})n \right) \: \phi^{(i)}\:,
\end{equation}
which is a \Z{N} transformation with redefined charges. In general, we can add integer multiples of the \U1 charges to the \Z{N} charges that is we are free to define new $\Z{N}$ charges $q'^{(i)}$ by
\begin{equation}
 q'^{(i)} = q^{(i)}+ k Q^{(i)} \quad \text{with} \quad k \in \Z{} \: .
\end{equation}

\newpage

\chapter{\textbf{\texttt{Mathematica}} Package for Abelian Discrete Symmetries}
\label{app:package}
A \texttt{Mathematica} package which automatizes most of the methods developed in chapter~\ref{chap:abelian} dealing with abelian, discrete symmetries can be obtained from the webpage
\begin{center}
 \href{http://einrichtungen.physik.tu-muenchen.de/T30e/codes/DiscreteBreaking/}{\texttt{http://einrichtungen.physik.tu-muenchen.de/T30e/codes/DiscreteBreaking/}} .
\end{center}
This appendix is meant to serve as an user guide to this package. We will shortly describe the usage of each routine illustrated with an example.

\section{\texttt{GetRemnantDiscreteSymmetry}}
The routine \texttt{GetRemnantDiscreteSymmetry} determines the remnant discrete symmetries that arise from $\U1^N$ 
theories by spontaneous breaking. The methods incorporated in this procedure are explained in section~\ref{sec:obtaining_abelian_discrete_symmetries}. The routine is called with
\begin{center}
 \texttt{ GetRemnantDiscreteSymmetries[VEVcharges,MatterCharges] }
\end{center}
where \texttt{VEVcharges} is a list of \U1 charges of fields that attain a 
vacuum expectation value and \texttt{MatterCharges} contains a list of \U1
charges of the fields in whose discrete symmetries one is interested. 

The output is a list with two elements. The first element is a list of integers $\mathtt{\{d_1,\ldots,d_N\}}$ indicating that the subgroup of the initial $\U1^N$ is broken down to
\begin{equation} \label{eq:app_general_symmetry}
 \Z{d_1} \times \ldots \times \Z{d_N}\: .
\end{equation}
A value $d_i=0$ indicates an unbroken \U1 factor. The second list in the output are the corresponding charges of the \texttt{MatterCharges}.

\paragraph{Example}
Let us show how the example of section~\ref{sec:example_abelian_breaking} is calculated with the aid of the presented routine. The charges are given in table~\ref{tab:example_abelian_breaking_charges}. The input is 
\begin{center}
 \texttt{GetRemnantDiscreteSymmetry[\{\{8,-2\},\{4,2\},\{2,4\}\}, \{\{1, 3\},\{1, 5\}\}]} .
\end{center}
The output
\begin{center}
 \texttt{ \{\{6\}, \{\{5\}, \{3\}\}\} }
\end{center}
means that we are left with a \Z6 symmetry. Note that this not the outcome as given in table~\ref{tab:example_abelian_breaking_charges} but the charges as given in and below \Eqref{eq:example_redundacies_2}.

\section{\texttt{SimplifyZSymmetries}}
\texttt{SimplifyZSymmetries} brings a given discrete symmetry in the canonical form and eliminates possible redundancies as explained in section~\ref{sec:redundandies}. The routine is called with
\begin{center}
 \texttt{ SimplifyZSymmetries[Symmetry,Charges] }
\end{center}
The input \texttt{Symmetry} contains a list of integers corresponding to a abelian, discrete symmetry (as in \Eqref{eq:app_general_symmetry}) and \texttt{Charges} contains lists of charges under this symmetry. The output is formated in the same way as for the procedure \texttt{GetRemnantDiscreteSymmetry}.

\paragraph{Example}
We will treat the example of a $\Z4\times\Z8$ symmetry in table~\ref{tab:example_redundancies1}. The input is
\begin{center}
 \texttt{SimplifyZSymmetries[\{4,8\}, \{\{2,4\},\{3,3\}\}]} ,
\end{center}
which produces the output
\begin{center}
 \texttt{\{\{8,2\}, \{\{4,1\},\{3,1\}\}\}} .
\end{center}
This tells us that the symmetry is actually a $\Z8\times\Z2$ symmetry.

\section{\texttt{EquivalentChargeAssignments}}
\texttt{EquivalentChargeAssignments} calculates all equivalent charge assignments for a given symmetry. The methods are developed in appendix~\ref{sec:automorphisms_abelian_groups}. The routine is called with
\begin{center}
 \texttt{EquivalentChargeAssignments[Symmetry,Charges] }
\end{center}
The input \texttt{Symmetry} contains a list of integers corresponding to a abelian, discrete symmetry (as in \Eqref{eq:app_general_symmetry}) and \texttt{Charges} contains lists of charges under this symmetry. The output is a list of equivalent charge assignments.

\paragraph{Example}
We will illustrate the usage with the example of a $\Z2\times\Z4$ symmetry at the end of appendix~\ref{sec:automorphisms_abelian_groups} on page~\pageref{eq:automorphisms_Z2xZ4}. The input is
\begin{center}
 \texttt{EquivalentChargeAssignments[\{2,4\}, \{\{1,0\}, \{0,1\}\}]}
\end{center}
which gives the output
\begin{center}
 \texttt{\{\{\{1, 0\}, \{0, 1\}\}, \{\{1, 0\}, \{1, 1\}\}, \{\{1, 2\}, \{0, 1\}\}, \{\{1, 2\}, \{1, 
   1\}\},} \\
\texttt{ \{\{1, 0\}, \{0, 3\}\}, \{\{1, 0\}, \{1, 3\}\}, \{\{1, 2\}, \{0, 3\}\}, \{\{1, 
   2\}, \{1, 3\}\}\}} .
\end{center}
There are eight equivalent charge assignments corresponding to the eight automorphisms of $\Z2\times\Z4$.

\section{\texttt{FindMatterParity}}
\texttt{FindMatterParity} searches for a \Z2 subgroup of a given abelian, discrete symmetry under which all fields are odd. The general procedure is described in section~\ref{sec:identifying_matter_parity}. The routine is called with
\begin{center}
 \texttt{FindMatterParity[Symmetry,Charges]} .
\end{center}
The input \texttt{Symmetry} contains a list of integers corresponding to a abelian, discrete symmetry (as in \Eqref{eq:app_general_symmetry}) and \texttt{Charges} contains lists of charges under this symmetry. The output is a symmetry written with explicit \Z2 factors and a 
list of replacement rules which takes the original charges to the 
explicit \Z2 representation rendering the matter parity obvious.

\paragraph{Example}
We will treat the example of a $\Z2\times\Z4$ in table~\ref{tab:example_matter_parity}. The input is
\begin{center}
 \texttt{FindMatterParity[\{2,4\}, \{\{1,2\},\{0,3\},\{0,1\}\}]}
\end{center}
which gives the output
\begin{center}
 \texttt{ \{\{2,4\},\{\{\{0,0\} $\to$ \{0,0\},\{0,1\} $\to$ \{1,3\},\{0,2\} $\to$
   \{0,2\},\{0,3\} $\to$ \{1,1\}, } \\
\texttt{\{1,0\} $\to$ \{1,2\},\{1,1\} $\to$
   \{0,1\},\{1,2\} $\to$ \{1,0\},\{1,3\} $\to$ \{0,3\}\}\}\} } .
\end{center}
The non-empty output means that there is a matter parity subgroup contained in the $\Z2\times\Z4$. Note that there are more replacement rules which make the matter parity obvious but we do not list them here. We rather give the replacements which lead to the charges in table~\ref{tab:example_matter_parity2}.


\newpage

\chapter{Details of the MFV Analysis}

\section{Numerical Checks}
\label{app:MFVcheck}

In this appendix we describe how we numerically check the scale-independent
validity of the MFV decomposition.
For our numerical calculations we use SOFTSUSY Version 2.0.14
\cite{Allanach:2001kg}. We restrict ourselves to real matrices only (partially
because of the corresponding limitation of SOFTSUSY). Apart from the usual
GUT-relations
\begin{eqnarray}
 M_1 &= & M_2 ~=~ M_3 ~=:~ m_{\nicefrac{1}{2}} \;, \nonumber\\
 \alpha_1 &=& \alpha_2 ~=~ \alpha_3 ~=~ m^2_{H_u} ~=~ m^2_{H_d} ~=:~ m_0^2,
 \quad \boldsymbol{m}^2_e ~=~ \boldsymbol{m}^2_L ~=~ m_0^2\,\mathds{1}\;,\nonumber\\
 \alpha_4 &=& \alpha_5 ~=:~ A\;,
\label{eq:universal}
\end{eqnarray}
we consider here only universal $\beta_i$:
\begin{equation}
 \beta_1~=~\dots~=~\beta_6~=:~b\,m_0^2\;,\quad
 \beta_7~=~\beta_8~=~b\, A\;.
\end{equation}  
Restricting ourselves to $\tan\beta=10$, we perform a scan over the following
region in parameter space:
\begin{gather*}
 \unit[-1000]{GeV}~<~ A~ <~ \unit[1000]{GeV}\;, \qquad |A| ~\leq~ m_0\;, \quad
 \unit[200]{GeV}~ <~ m_{\nicefrac{1}{2}} ~<~ \unit[500]{GeV}\;, \\
 \unit[100]{GeV} ~<~ m_0 ~<~ \unit[1500]{GeV}\;,\quad
 |\beta_{1,2,3,4,5,6}| ~\leq ~m_0^2\;, \qquad |\beta_{7,8}|~ \leq ~|A|\;.
\end{gather*}

At the low scale, defined by SOFTSUSY as $\sqrt{m_{\tilde t_1}\, m_{\tilde
t_2}}$, a best fit decomposition is used to minimize the absolute difference to
the form of equation~\eqref{eq:softmassdecomp}. To this end, we use the matrix
norm
\begin{equation}
 |M_{ij}|~=~\sqrt{ \sum_{ij} |M_{ij}|^2}\;.
 \label{eq:norm}
\end{equation}
Define now $\boldsymbol{\Delta m}_f$ $\left(\boldsymbol{\Delta A}_f\right)$ as the
part of $\boldsymbol{m}_f$ ($\boldsymbol{A}_f$) which is orthogonal to the MFV
decomposition at the low scale (cf.\ equation~\eqref{eq:nonMFV}). Then for all the points in our scan the ratio
$\frac{|\boldsymbol{\Delta m}_f|}{|\boldsymbol{m}_f|}$ 
($\frac{|\boldsymbol{\Delta A}_f|}{m_{\nicefrac{1}{2}}}$) lies below the indicated number (table~\ref{tab:Deviations}). 
We normalize $\boldsymbol{\Delta A}_f$ to $m_{\nicefrac{1}{2}}$ rather than
$|\boldsymbol{A}_f|$ since the later can approach zero at the low scale. Notice
also that we truncate the MFV decomposition as specified in 
\eqref{eq:softmassdecomp}. The deviations (table~\ref{tab:Deviations}) are of
the order of higher-order MFV terms, i.e.\ we expect that the MFV approximation
gets practically perfect when higher order terms are included.

\begin{table}[htb]
\centering
\begin{tabular}{lccccc}
\toprule[1.3pt]
 quantity &
 $\boldsymbol{m}^2_Q$ & $\boldsymbol{m}^2_u$ & $\boldsymbol{m}^2_d$ &
 $\boldsymbol{A}_u$ & $\boldsymbol{A}_d$ \\ 
 deviation &
 $10^{-7}$ & $10^{-6}$ & $10^{-5}$ & $10^{-2}$ &  $10^{-3}$ \\
\bottomrule[1.3pt]
\end{tabular}
\caption{Deviations from scale independence of the MFV ansatz.}
\label{tab:Deviations}
\end{table}

We observe that for the bilinear soft masses the MFV-decomposition holds with
great accuracy. In the case of the trilinears the error is of the order
$\frac{m_c}{m_t}\approx 1\%$, as one might have expected.

In summary, for $|\beta_i| \leq \alpha_j$ (comparing only coefficients of same
mass dimension), soft terms which are in the MFV form at the GUT scale will be
in this form at the low scale with good precision.

\begin{landscape}
\section{Approximations on low-energy MFV Coefficients}
\label{app:Approximations}

We numerically solve the RGEs as described in appendix~\ref{app:MFVcheck}, but
with only one $\beta_i$ set different from zero. A test with random initial
conditions for $\beta_i$ confirms our results.

The following formulae reproduce the exact SOFTSUSY results up to an error of
\begin{equation*}
 \frac{|\alpha_{i,\textrm{fit}}-\alpha_{i,\textrm{SOFTSUSY}}|}{  |\alpha_{i,\textrm{SOFTSUSY}}|} < 0.1 \qquad \text{and} \qquad
 \frac{|\beta_{1,2,3;4;5,\textrm{fit}}-\beta_{1,2,3;4;5,\textrm{SOFTSUSY}}|}{|\alpha_{1;2;3,\textrm{SOFTSUSY}}|},
\frac{|\beta_{7,8,\textrm{fit}}-\beta_{7,8,\textrm{SOFTSUSY}}|}{m_{\nicefrac{1}{2}}} < 0.02\;.
\end{equation*}
In the following formulae, the variables on the left hand side denote the values
at the low scale, while those on the right hand side are the quantities at the
high scale. 
\tabcolsep=0.5pt 
\begin{center}
\begin{tabular}{rclllllll}
$\alpha_1$ &~=~& $+0.94\,m_0^2$&$+5.04\,m^2_{\nicefrac{1}{2}}$\\ 
$\alpha_2$ & = & $+0.95\,m_0^2$&$+4.72\,m^2_{\nicefrac{1}{2}}$\\ 
$\alpha_3$ & = & $+0.95\,m_0^2$&$+4.61\,m^2_{\nicefrac{1}{2}}$\\
$\alpha_4$ & = && $-2.00\,m_{\nicefrac{1}{2}}$&& $+0.32\,A$ \\ 
$\alpha_5$ & = && $-3.23\,m_{\nicefrac{1}{2}}$&& $+0.98\,A$   \\
$\beta_1$ & = & $-0.41\,m_0^2$   & $-0.96\,m^2_{\nicefrac{1}{2}}$ & $+0.16\,A\,m_{\nicefrac{1}{2}}$ & $-0.04\,A^2$ &  $+0.27\,\beta_1-0.03\,\beta_5$\\ 
$\beta_2$ & = & $-0.43\,m_0^2$   & $-1.38\,m^2_{\nicefrac{1}{2}}$  & $+0.57\,A\,m_{\nicefrac{1}{2}}$ & $-0.15\,A^2$  &  $-0.02\,\beta_1+0.1\,\beta_2+0.01\,\beta_5$\\
$\beta_3$ & = && $+0.13\,m^2_{\nicefrac{1}{2}}$  & $-0.13\,A\,m_{\nicefrac{1}{2}}$&  $+0.04\,A^2$ & $+0.02\,\beta_1+0.03\,\beta_3-0.01\beta_5-(0.01\,\beta_7+0.04\,\beta_8)\,A+(0.03\,\beta_7+0.08\,\,\beta_8)\,m_{\nicefrac{1}{2}}$  \\
$\beta_5$ & = & $-0.83\,m_0^2$  &  $-1.96\,m^2_{\nicefrac{1}{2}}$  & $+0.32\,A\,m_{\nicefrac{1}{2}}$ & $-0.09\,A^2$ &  $-0.07\,\beta_1+0.24\,\beta_5$ \\
$\beta_6$ & = & $-0.86\,m_0^2$  & $-2.57\,m^2_{\nicefrac{1}{2}}$  & $+0.94\,A\,m_{\nicefrac{1}{2}}$ & $-0.25\,A^2$  &   $-0.07\,\beta_1+0.01\beta_5+0.12\beta_6-0.14\,A\,\beta_8+0.25\,m_{\nicefrac{1}{2}}\,\beta_8 $ \\
$\beta_7$ & = &&$+0.51\,m_{\nicefrac{1}{2}}$&& $-0.27\,A$ &$+0.10\,\beta_7$ \\
$\beta_8$ & = &&$+0.27\,m_{\nicefrac{1}{2}}$&& $-0.14\,A$ &$+0.30\,\beta_8$ \\
\end{tabular}
\end{center}
\end{landscape}

\newpage

\chapter{Details of the String Model}
\label{app:details_model}

The orbifold model is defined by a torus lattice that is spanned by six orthogonal vectors $e_\alpha$, $\alpha=1,\ldots, 6$, the $\Z2\times\Z2$ twist vectors $v_1=(0,1/2,-1/2)$ and $v_2=(-1/2,0,1/2)$, and the associated shifts
\begin{subequations}
\begin{align}
V_1 & =  \left( \frac{1}{2}, -\frac{1}{2},  0,  0,  0,  0,  0,  0 \right) \left(0,  0,  0,  0,  0,  0,  0,  0\right) \: , \\
V_2 & = \left( 0,\frac{1}{2}, -\frac{1}{2},  0,  0,  0,  0,  0 \right) \left(0,  0,  0,  0,  0,  0,  0,  0\right) \: , 
\end{align}
\end{subequations}
and the six discrete Wilson lines
\begin{subequations}
\begin{align}
W_1 & =  \left(0^8 \right) \left(0^8 \right)\: , \\
W_2 & =  \left(  -\frac{5}{4},  \frac{3}{4},  \frac{3}{4},  \frac{3}{4}, -\frac{5}{4}, -\frac{1}{4}, -\frac{1}{4}, -\frac{1}{4} \right) \left(    1, -\frac{1}{2}, -\frac{1}{2}, -\frac{1}{2},  0,  0,  \frac{1}{2},  1 \right)\: , \\
W_3 & =  \left( -\frac{3}{4}, -\frac{3}{4},  \frac{5}{4}, -\frac{3}{4},  \frac{1}{4},  \frac{1}{4},  \frac{1}{4},  \frac{1}{4}\right) \left(    -\frac{3}{4}, -\frac{3}{4},  \frac{1}{4},  \frac{1}{4},  \frac{1}{4},  \frac{1}{4},  \frac{1}{4},  \frac{1}{4}\right)\: , \\
W_5 & = \left(  \frac{1}{4},  \frac{1}{4}, -\frac{7}{4},  \frac{5}{4}, -\frac{3}{4},  \frac{1}{4},  \frac{1}{4},  \frac{5}{4}  \right) \left( -1,  0, -1, -1,  0,  0,  0,  0\right)\: , \\
W_4 &= W_6 ~ = ~ W_2\; ,
\end{align}
\end{subequations}
corresponding to the six torus directions $e_\alpha$. Additionally, we divide out the \Z2 symmetry corresponding to
\begin{equation}
 \tau = \frac{1}{2} (e_2 + e_4 + e_6)
\end{equation}
with a gauge embedding denoted by $W$ where
\begin{equation}
 W= \frac{1}{2} (W_2 + W_4 + W_6) = \frac{3}{2} W_2 \;.
\end{equation}
The anomalous space group element reads
\begin{equation}
 g_\text{anom} = (k,\ell;n_1,n_2,n_3,n_4,n_5,n_6) 
 = (0, 0;0, 1, 0, 0, 1, 0)\:, \label{eq:anomalous_space_group_element}
\end{equation}
where the boundary conditions of twisted string are
\begin{equation} \label{eq:string_boundary_conditions}
 X(\tau,\sigma+2\pi)
 =
 \vartheta^k\,\omega^\ell\,X(\tau,\sigma)+ n_\alpha e_\alpha
\end{equation}
with $\vartheta$ and $\omega$ denoting the rotations corresponding to $v_1$ and $v_2$. 

\section{Selection Rules}
\label{app:string_selection_rules}
A superpotential term $\Pi_i \Phi^{(i)}$ is allowed if it gauge invariant w.r.t.\ the gauge group in \Eqref{eq:string_gauge_group} and fulfils the following conditions~\cite{Blaszczyk:2009in}
\begin{subequations} \label{eq:string_selection_rules}
\begin{alignat}{2}
R-\text{invariance} :\quad &\sum_i R^{(i)}_j &= -1  & \mod 2  \quad \text{where} \quad j=1,2,3 \: , \\
\text{point group rule} :\quad &\sum_i k^{(i)}&= 0 & \mod 2 \: , \\
\quad&\sum_i \ell^{(i)} &= 0  & \mod 2 \: , \\
\text{space group rule} :\quad &\sum_i n^{(i)}_j &= 0 & \mod 2 \quad \text{where} \quad j=1,3,5 \: , \\
\quad &\sum_i n^{(i)}_2 + n^{(i)}_4 + n^{(i)}_6 &= 0  & \mod 2 \: .
\end{alignat}
\end{subequations}

\section{Spectrum}
In the following we list all fields of the string model. In the column 'irreps' we give the irreducible representation w.r.t.\ the non-abelian part of gauge group
\begin{equation}
 \SU3_C\times\SU2_L\times\U1_Y\times \left[  \SU3\times \SU2 \times \SU2 \right]_\text{hidden}\:.
\end{equation}
$R_{i}$ are the discrete $R$-charges. $k,\ell$ and $n_1,n_2,n_3,n_4,n_5,n_6$ defined the boundary conditions of twisted strings as defined in \Eqref{eq:string_boundary_conditions}. Note that we list $\Sigma  n_{\text{even}}= n_2+n_4+n_6$ as only this combination enters the selection rules in \Eqref{eq:string_selection_rules}. $q_Y$ denotes hypercharge and $q_\text{anom}$ is the charge w.r.t.\ the anomalous \U1. $q_1,\ldots,q_7$ label the remaining \U1 charges. If there are two labels in one line, this means that these states form a $D_4$ doublet. For the $D_4$ doublets we have $n_1=0$ for the first component and $n_1=1$ for the second.

\begin{landscape}
 \tabcolsep=5pt 
\begin{longtable}{ccccccccccccccccccccccccc}
\toprule[1.3pt] 
\multicolumn{5}{c}{irrep} & $ R_1 $ & $ R_2 \
$ & $ R_3 $ & $ k $ & $ \ell $ & $ n_1 $ & $ n_3 $ & $ n_5 $ & $ \Sigma  n_{\text{even}} $ & $ q_Y $ & $ \
q_{\text{anom}} $ & $ q_1 $ & $ q_2 $ & $ q_3 $ & $ q_4 $ & $ q_5 $ & $ q_6 $ & $ q_7 $ & $ \Z4^R $ & $ \
\text{label} $ \\ \midrule \endhead 
\bottomrule[1.3pt] \\ \endfoot
$ \bar{3} $ & $ 1 $ & $ 1 $ & $ 1 $ & $ 1 $ & $ 0 $ & -$\tfrac{1}{2} $ & -$\tfrac{1}{2} $ & $ 1 $ & $ 0 $ & $ 0 $ & $ 0 $ & $ 0 $ & $ 0 $ & $ \tfrac{1}{3} $ & $ 0 $ & -$\tfrac{1}{4} $ & $ 0 $ & $ 0 $ & -$\tfrac{55}{4} $ & $ 4 $ & $ 4 $ & $ 
\tfrac{3}{2} $ & $ 1 $ & $ \overline{d}_1 $ \\ $
 \bar{3} $ & $ 1 $ & $ 1 $ & $ 1 $ & $ 1 $ & -$\tfrac{1}{2} $ & -$\tfrac{1}{2} $ & $ 0$ & $ 1 $ & $ 1 $ & $ n_1 $ & $ 0 $ & $ 
0 $ & $ 0 $ & $ \tfrac{1}{3} $ & $ \tfrac{1}{2} $ & -$\tfrac{1}{4} $ & $ \tfrac{1}{2} $ & $ 
-\tfrac{63}{4} $ & -$4 $ & $ 4 $ & $ 4 $ & $ \tfrac{3}{2} $ & $ 1 $ & $ \overline{d}_2, \overline{d}_3 
$ \\ $
 1 $ & $ 1 $ & $ 1 $ & $ 1 $ & $ 1 $ & $ 0 $ & -$\tfrac{1}{2} $ & -$\tfrac{1}{2} $ & $ 1 $ & $ 0 $ & $ 0 $ & $ 0 $ & $ 0 $ & $ 0 
$ & $ 1 $ & $ 1 $ & $ 0 $ & $ 1 $ & $ \tfrac{51}{4} $ & -$2 $ & $ 2 $ & $ 2 $ & $ \tfrac{3}{4} $ & $ 1 $ & $ 
\overline{e}_1 $ \\ $
 1 $ & $ 1 $ & $ 1 $ & $ 1 $ & $ 1 $ & -$\tfrac{1}{2} $ & -$\tfrac{1}{2} $ & $ 0 $ & $ 1 $ & $ 1 $ & $ n_1 $ & $ 0 $ & $ 0 $ & $ 0 
$ & $ 1 $ & $ \tfrac{3}{2} $ & $ 0 $ & $ \tfrac{3}{2} $ & -$3 $ & $ \tfrac{31}{4} $ & $ 2 $ & $ 2 $ & $ 
\tfrac{3}{4} $ & $ 1 $ & $ \overline{e}_2, \overline{e}_3 $ \\ $
 \bar{3} $ & $ 1 $ & $ 1 $ & $ 1 $ & $ 1 $ & $ 0 $ & -$\tfrac{1}{2} $ & -$\tfrac{1}{2} $ & $ 1 $ & $ 0 $ & $ 0 $ & $ 0 $ & $ 
0 $ & $ 0 $ & -$\tfrac{2}{3} $ & $ 1 $ & $ 0 $ & $ 1 $ & $ \tfrac{51}{4} $ & -$2 $ & $ 2 $ & $ 2 $ & $ \tfrac{3}{4} $ & $ 
1 $ & $ \overline{u}_1 $ \\ $
 \bar{3} $ & $ 1 $ & $ 1 $ & $ 1 $ & $ 1 $ & -$\tfrac{1}{2} $ & -$\tfrac{1}{2} $ & $ 0 $ & $ 1 $ & $ 1 $ & $ n_1 $ & $ 0 $ & $ 
0 $ & $ 0 $ & -$\tfrac{2}{3} $ & $ \tfrac{3}{2} $ & $ 0 $ & $ \tfrac{3}{2} $ & -$3 $ & $ \tfrac{31}{4} $ & $ 2 
$ & $ 2 $ & $ \tfrac{3}{4} $ & $ 1 $ & $ \overline{u}_2, \overline{u}_3 $ \\ $
 1 $ & $ 2 $ & $ 1 $ & $ 1 $ & $ 1 $ & $ 0 $ & -$\tfrac{1}{2} $ & -$\tfrac{1}{2} $ & $ 1 $ & $ 0 $ & $ 0 $ & $ 0 $ & $ 0 $ & $ 0 
$ & -$\tfrac{1}{2} $ & $ 0 $ & -$\tfrac{1}{4} $ & $ 0 $ & $ 0 $ & -$\tfrac{55}{4} $ & $ 4 $ & $ 4 $ & $ 
\tfrac{3}{2} $ & $ 1 $ & $ \ell_1 $ \\ $
 1 $ & $ 2 $ & $ 1 $ & $ 1 $ & $ 1 $ & -$\tfrac{1}{2} $ & -$\tfrac{1}{2} $ & $ 0 $ & $ 1 $ & $ 1 $ & $ n_1 $ & $ 0 $ & $ 0 $ & $ 0 
$ & -$\tfrac{1}{2} $ & $ \tfrac{1}{2} $ & -$\tfrac{1}{4} $ & $ \tfrac{1}{2} $ & -$\tfrac{63}{4} $ & $ 
-4 $ & $ 4 $ & $ 4 $ & $ \tfrac{3}{2} $ & $ 1 $ & $ \ell_2, \ell_3 $ \\ $
 3 $ & $ 2 $ & $ 1 $ & $ 1 $ & $ 1 $ & $ 0 $ & -$\tfrac{1}{2} $ & -$\tfrac{1}{2} $ & $ 1 $ & $ 0 $ & $ 0 $ & $ 0 $ & $ 0 $ & $ 0 
$ & $ \tfrac{1}{6} $ & $ 1 $ & $ 0 $ & $ 1 $ & $ \tfrac{51}{4} $ & -$2 $ & $ 2 $ & $ 2 $ & $ \tfrac{3}{4} $ & $ 1 $ & $ 
q_1 $ \\ $
 3 $ & $ 2 $ & $ 1 $ & $ 1 $ & $ 1 $ & -$\tfrac{1}{2} $ & -$\tfrac{1}{2} $ & $ 0 $ & $ 1 $ & $ 1 $ & $ n_1 $ & $ 0 $ & $ 0 $ & $ 0 
$ & $ \tfrac{1}{6} $ & $ \tfrac{3}{2} $ & $ 0 $ & $ \tfrac{3}{2} $ & -$3 $ & $ \tfrac{31}{4} $ & $ 2 $ & $ 2 $ & $ 
\tfrac{3}{4} $ & $ 1 $ & $ q_2, q_3 $ \\ $
 1 $ & $ 2 $ & $ 1 $ & $ 1 $ & $ 1 $ & $ 0 $ & -$1 $ & $ 0 $ & $ 0 $ & $ 0 $ & $ 0 $ & $ 0 $ & $ 0 $ & $ 0 $ & $ \tfrac{1}{2} $ & -$1 $ & $ 
-\tfrac{1}{2} $ & -$1 $ & $ 2 $ & $ 4 $ & -$4 $ & -$4 $ & -$\tfrac{3}{2} $ & $ 2 $ & $ \overline{h}_1 $ \\ $
 1 $ & $ 2 $ & $ 1 $ & $ 1 $ & $ 1 $ & $ 0 $ & $ 0 $ & -$1 $ & $ 0 $ & $ 0 $ & $ 0 $ & $ 0 $ & $ 0 $ & $ 0 $ & $ \tfrac{1}{2} $ & -$3 $ & $ 
0 $ & -$3 $ & $ 6 $ & -$\tfrac{31}{2} $ & -$4 $ & -$4 $ & -$\tfrac{3}{2} $ & $ 0 $ & $ \overline{h}_2 $ \\ $
 1 $ & $ 2 $ & $ 1 $ & $ 1 $ & $ 1 $ & -$1 $ & $ 0 $ & $ 0 $ & $ 0 $ & $ 0 $ & $ 0 $ & $ 0 $ & $ 0 $ & $ 0 $ & $ \tfrac{1}{2} $ & -$2 $ & $ 
0 $ & -$2 $ & -$\tfrac{51}{2} $ & $ 4 $ & -$4 $ & -$4 $ & -$\tfrac{3}{2} $ & $ 0 $ & $ \overline{h}_3 $ \\ $
 1 $ & $ 2 $ & $ 1 $ & $ 1 $ & $ 1 $ & $ 0 $ & -$\tfrac{1}{2} $ & -$\tfrac{1}{2} $ & $ 1 $ & $ 0 $ & $ 0 $ & $ 1 $ & $ 0 $ & $ 0 
$ & $ \tfrac{1}{2} $ & $ 2 $ & $ \tfrac{1}{8} $ & -$\tfrac{11}{2} $ & -$\tfrac{89}{8} $ & $ 
-\tfrac{13}{8} $ & -$\tfrac{13}{4} $ & -$\tfrac{13}{4} $ & -$\tfrac{27}{8} $ & $ 2 $ & $ 
\overline{h}_4 $ \\ $
 1 $ & $ 2 $ & $ 1 $ & $ 1 $ & $ 1 $ & -$\tfrac{1}{2} $ & -$\tfrac{1}{2} $ & $ 0 $ & $ 1 $ & $ 1 $ & $ n_1 $ & $ 1 $ & $ 0 $ & $ 0 
$ & $ \tfrac{1}{2} $ & $ \tfrac{3}{2} $ & $ \tfrac{1}{8} $ & -$6 $ & $ \tfrac{37}{8} $ & $ 
-\tfrac{91}{8} $ & -$\tfrac{13}{4} $ & -$\tfrac{13}{4} $ & -$\tfrac{27}{8} $ & $ 2 $ & $ 
\overline{h}_5, \overline{h}_6 $ \\ $
 1 $ & $ 2 $ & $ 1 $ & $ 1 $ & $ 1 $ & $ 0 $ & -$1 $ & $ 0 $ & $ 0 $ & $ 0 $ & $ 0 $ & $ 0 $ & $ 0 $ & $ 0 $ & -$\tfrac{1}{2} $ & $ 1 $ & $ 
\tfrac{1}{2} $ & $ 1 $ & -$2 $ & -$4 $ & $ 4 $ & $ 4 $ & $ \tfrac{3}{2} $ & $ 2 $ & $ h_1 $ \\ $
 1 $ & $ 2 $ & $ 1 $ & $ 1 $ & $ 1 $ & $ 0 $ & $ 0 $ & -$1 $ & $ 0 $ & $ 0 $ & $ 0 $ & $ 0 $ & $ 0 $ & $ 0 $ & -$\tfrac{1}{2} $ & $ 3 $ & $ 
0 $ & $ 3 $ & -$6 $ & $ \tfrac{31}{2} $ & $ 4 $ & $ 4 $ & $ \tfrac{3}{2} $ & $ 0 $ & $ h_2 $ \\ $
 1 $ & $ 2 $ & $ 1 $ & $ 1 $ & $ 1 $ & -$1 $ & $ 0 $ & $ 0 $ & $ 0 $ & $ 0 $ & $ 0 $ & $ 0 $ & $ 0 $ & $ 0 $ & -$\tfrac{1}{2} $ & $ 2 $ & $ 
0 $ & $ 2 $ & $ \tfrac{51}{2} $ & -$4 $ & $ 4 $ & $ 4 $ & $ \tfrac{3}{2} $ & $ 0 $ & $ h_3 $ \\ $
 1 $ & $ 2 $ & $ 1 $ & $ 1 $ & $ 1 $ & $ 0 $ & -$\tfrac{1}{2} $ & -$\tfrac{1}{2} $ & $ 1 $ & $ 0 $ & $ 0 $ & $ 1 $ & $ 1 $ & $ 1 
$ & -$\tfrac{1}{2} $ & $ 0 $ & -$\tfrac{1}{8} $ & $ \tfrac{15}{2} $ & $ \tfrac{57}{8} $ & $ 
-\tfrac{51}{8} $ & -$\tfrac{33}{4} $ & -$\tfrac{33}{4} $ & $ \tfrac{27}{8} $ & $ 0 $ & $ h_4 $ \\ $
 1 $ & $ 2 $ & $ 1 $ & $ 1 $ & $ 1 $ & -$\tfrac{1}{2} $ & $ 0 $ & -$\tfrac{1}{2} $ & $ 0 $ & $ 1 $ & $ n_1 $ & $ 0 $ & $ 0 $ & $ 0 
$ & -$\tfrac{1}{2} $ & -$\tfrac{3}{2} $ & $ \tfrac{1}{4} $ & -$\tfrac{3}{2} $ & $ 3 $ & $ 6 $ & -$6 $ & -$6 
$ & -$\tfrac{9}{4} $ & $ 0 $ & $ h_5, h_6 $ \\ $
 3 $ & $ 1 $ & $ 1 $ & $ 1 $ & $ 1 $ & $ 0 $ & -$\tfrac{1}{2} $ & -$\tfrac{1}{2} $ & $ 1 $ & $ 0 $ & $ 0 $ & $ 1 $ & $ 0 $ & $ 0 
$ & -$\tfrac{1}{3} $ & $ 2 $ & $ \tfrac{1}{8} $ & -$\tfrac{11}{2} $ & -$\tfrac{89}{8} $ & $ 
-\tfrac{13}{8} $ & -$\tfrac{13}{4} $ & -$\tfrac{13}{4} $ & -$\tfrac{27}{8} $ & $ 2 $ & $ \delta_1 $ \\ $
 3 $ & $ 1 $ & $ 1 $ & $ 1 $ & $ 1 $ & -$\tfrac{1}{2} $ & -$\tfrac{1}{2} $ & $ 0 $ & $ 1 $ & $ 1 $ & $ n_1 $ & $ 1 $ & $ 0 $ & $ 0 $ & -$\tfrac{1}{3} $ & $ \tfrac{3}{2} $ & $ \tfrac{1}{8} $ & -$6 $ & $ \tfrac{37}{8} $ & $ -\tfrac{91}{8} $ & -$\tfrac{13}{4} $ & -$\tfrac{13}{4} $ & -$\tfrac{27}{8} $ & $ 2 $ & $ \delta 
_2, \delta _3 $ \\ $
 \bar{3} $ & $ 1 $ & $ 1 $ & $ 1 $ & $ 1 $ & $ 0 $ & -$\tfrac{1}{2} $ & -$\tfrac{1}{2} $ & $ 1 $ & $ 0 $ & $ 0 $ & $ 1 $ & $ 1 $ & $ 1 $ & $ \tfrac{1}{3} $ & $ 0 $ & -$\tfrac{1}{8} $ & $ \tfrac{15}{2} $ & $ \tfrac{57}{8} $ & $ 
-\tfrac{51}{8} $ & -$\tfrac{33}{4} $ & -$\tfrac{33}{4} $ & $ \tfrac{27}{8} $ & $ 0 $ & $ 
\overline{\delta }_1 $ \\ $
 \bar{3} $ & $ 1 $ & $ 1 $ & $ 1 $ & $ 1 $ & -$\tfrac{1}{2} $ & $ 0 $ & -$\tfrac{1}{2} $ & $ 0 $ & $ 1 $ & $ n_1 $ & $ 0 $ & $ 
0 $ & $ 0 $ & $ \tfrac{1}{3} $ & -$\tfrac{3}{2} $ & $ \tfrac{1}{4} $ & -$\tfrac{3}{2} $ & $ 3 $ & $ 6 $ & -$6 
$ & -$6 $ & -$\tfrac{9}{4} $ & $ 0 $ & $ \overline{\delta }_2, \overline{\delta }_3 $ \\ $
 1 $ & $ 1 $ & $ \bar{3} $ & $ 1 $ & $ 1 $ & $ 0 $ & -$\tfrac{1}{2} $ & -$\tfrac{1}{2} $ & $ 1 $ & $ 0 $ & $ 0 $ & $ 0 $ & $ 
1 $ & $ 0 $ & $ 0 $ & $ 1 $ & $ \tfrac{1}{8} $ & $ 1 $ & -$\tfrac{75}{8} $ & $ \tfrac{15}{8} $ & $ 3 $ & $ 3 $ & $ 
\tfrac{55}{8} $ & $ 2 $ & $ \overline{x}_1 $ \\ $
 1 $ & $ 1 $ & $ \bar{3} $ & $ 1 $ & $ 1 $ & $ 0 $ & -$\tfrac{1}{2} $ & -$\tfrac{1}{2} $ & $ 1 $ & $ 0 $ & $ 0 $ & $ 0 $ & $ 
1 $ & $ 1 $ & $ 0 $ & $ 1 $ & -$\tfrac{1}{4} $ & $ 16 $ & -$\tfrac{5}{2} $ & $ \tfrac{35}{4} $ & $ 1 $ & $ 1 $ & $ 
-\tfrac{5}{2} $ & $ 1 $ & $ \overline{x}_2 $ \\ $
 1 $ & $ 1 $ & $ 3 $ & $ 1 $ & $ 1 $ & $ 0 $ & -$\tfrac{1}{2} $ & -$\tfrac{1}{2} $ & $ 1 $ & $ 0 $ & $ 0 $ & $ 0 $ & $ 1 $ & $ 1 
$ & $ 0 $ & $ 1 $ & -$\tfrac{1}{4} $ & -$14 $ & -$\tfrac{3}{2} $ & $ \tfrac{43}{4} $ & -$1 $ & -$1 $ & $ 
\tfrac{5}{2} $ & $ 1 $ & $ x_2 $ \\ $
 1 $ & $ 1 $ & $ 1 $ & $ 2 $ & $ 1 $ & $ 0 $ & -$\tfrac{1}{2} $ & -$\tfrac{1}{2} $ & $ 1 $ & $ 0 $ & $ 0 $ & $ 1 $ & $ 0 $ & $ 1 
$ & $ 0 $ & $ \tfrac{3}{2} $ & $ \tfrac{1}{4} $ & -$6 $ & -$\tfrac{11}{4} $ & $ \tfrac{33}{4} $ & $ 
-\tfrac{33}{4} $ & $ \tfrac{29}{4} $ & -$\tfrac{15}{4} $ & $ 2 $ & $ y_1 $ \\ $
 1 $ & $ 1 $ & $ 1 $ & $ 1 $ & $ 2 $ & $ 0 $ & -$\tfrac{1}{2} $ & -$\tfrac{1}{2} $ & $ 1 $ & $ 0 $ & $ 0 $ & $ 1 $ & $ 0 $ & $ 1 
$ & $ 0 $ & $ \tfrac{3}{2} $ & $ \tfrac{1}{4} $ & -$6 $ & -$\tfrac{11}{4} $ & $ \tfrac{33}{4} $ & $ 
\tfrac{45}{4} $ & -$\tfrac{17}{4} $ & -$\tfrac{15}{4} $ & $ 2 $ & $ z_1 $ \\ $
 1 $ & $ 1 $ & $ 3 $ & $ 1 $ & $ 1 $ & $ 0 $ & -$\tfrac{1}{2} $ & -$\tfrac{1}{2} $ & $ 1 $ & $ 0 $ & $ 0 $ & $ 1 $ & $ 0 $ & $ 1 
$ & $ 0 $ & $ 2 $ & $ \tfrac{1}{4} $ & $ \tfrac{19}{2} $ & -$\tfrac{17}{4} $ & $ \tfrac{21}{4} $ & $ 
-\tfrac{21}{4} $ & -$\tfrac{21}{4} $ & -$\tfrac{5}{4} $ & $ 3 $ & $ x_3 $ \\ $
 1 $ & $ 1 $ & $ \bar{3} $ & $ 1 $ & $ 1 $ & $ 0 $ & -$\tfrac{1}{2} $ & -$\tfrac{1}{2} $ & $ 1 $ & $ 0 $ & $ 0 $ & $ 1 $ & $ 
1 $ & $ 0 $ & $ 0 $ & $ 0 $ & -$\tfrac{1}{4} $ & -$\tfrac{15}{2} $ & $ \tfrac{1}{4} $ & -$\tfrac{53}{4} $ & $ 
-\tfrac{25}{4} $ & -$\tfrac{25}{4} $ & $ \tfrac{5}{4} $ & $ 3 $ & $ \overline{x}_3 $ \\ $
 1 $ & $ 1 $ & $ 1 $ & $ 2 $ & $ 1 $ & $ 0 $ & -$\tfrac{1}{2} $ & -$\tfrac{1}{2} $ & $ 1 $ & $ 0 $ & $ 0 $ & $ 1 $ & $ 1 $ & $ 0 
$ & $ 0 $ & $ \tfrac{3}{2} $ & $ \tfrac{1}{4} $ & -$6 $ & -$\tfrac{11}{4} $ & $ \tfrac{33}{4} $ & $ 
\tfrac{45}{4} $ & -$\tfrac{17}{4} $ & -$\tfrac{15}{4} $ & $ 2 $ & $ y_2 $ \\ $
 1 $ & $ 1 $ & $ 1 $ & $ 1 $ & $ 2 $ & $ 0 $ & -$\tfrac{1}{2} $ & -$\tfrac{1}{2} $ & $ 1 $ & $ 0 $ & $ 0 $ & $ 1 $ & $ 1 $ & $ 0 
$ & $ 0 $ & $ \tfrac{3}{2} $ & $ \tfrac{1}{4} $ & -$6 $ & -$\tfrac{11}{4} $ & $ \tfrac{33}{4} $ & $ 
-\tfrac{33}{4} $ & $ \tfrac{29}{4} $ & -$\tfrac{15}{4} $ & $ 2 $ & $ z_2 $ \\ $
 1 $ & $ 1 $ & $ 1 $ & $ 2 $ & $ 1 $ & -$\tfrac{1}{2} $ & $ 0 $ & -$\tfrac{1}{2} $ & $ 0 $ & $ 1 $ & $ n_1 $ & $ 0 $ & $ 0 $ & $ 1 $ & $ 0 $ & $ 3 $ & -$\tfrac{1}{8} $ & $ 3 $ & $ \tfrac{11}{8} $ & -$\tfrac{33}{8} $ & $ 9 $ & $ 
-\tfrac{13}{2} $ & $ \tfrac{15}{8} $ & $ 0 $ & $ y_3, y_5 $ \\ $
 1 $ & $ 1 $ & $ 1 $ & $ 1 $ & $ 2 $ & -$\tfrac{1}{2} $ & $ 0 $ & -$\tfrac{1}{2} $ & $ 0 $ & $ 1 $ & $ n_1 $ & $ 0 $ & $ 0 $ & $ 1 $ & $ 0 $ & $ 3 $ & -$\tfrac{1}{8} $ & $ 3 $ & $ \tfrac{11}{8} $ & -$\tfrac{33}{8} $ & -$\tfrac{21}{2} $ & $ 
5 $ & $ \tfrac{15}{8} $ & $ 0 $ & $ z_3, z_5 $ \\ $
 1 $ & $ 1 $ & $ \bar{3} $ & $ 1 $ & $ 1 $ & -$\tfrac{1}{2} $ & $ 0 $ & -$\tfrac{1}{2} $ & $ 0 $ & $ 1 $ & $ n_1 $ & $ 0 $ & $ 
0 $ & $ 1 $ & $ 0 $ & $ \tfrac{5}{2} $ & -$\tfrac{1}{8} $ & -$\tfrac{25}{2} $ & $ \tfrac{23}{8} $ & $ 
-\tfrac{9}{8} $ & $ 6 $ & $ 6 $ & -$\tfrac{5}{8} $ & $ 3 $ & $ \overline{x}_4, \overline{x}_5 $ \\ $
 1 $ & $ 1 $ & $ 1 $ & $ 2 $ & $ 1 $ & -$\tfrac{1}{2} $ & $ 0 $ & -$\tfrac{1}{2} $ & $ 0 $ & $ 1 $ & $ n_1 $ & $ 0 $ & $ 1 $ & $ 0 
$ & $ 0 $ & $ 3 $ & -$\tfrac{1}{8} $ & $ 3 $ & $ \tfrac{11}{8} $ & -$\tfrac{33}{8} $ & -$\tfrac{21}{2} $ & $ 
5 $ & $ \tfrac{15}{8} $ & $ 0 $ & $ y_4, y_6 $ \\ $
 1 $ & $ 1 $ & $ 1 $ & $ 1 $ & $ 2 $ & -$\tfrac{1}{2} $ & $ 0 $ & -$\tfrac{1}{2} $ & $ 0 $ & $ 1 $ & $ n_1 $ & $ 0 $ & $ 1 $ & $ 0 
$ & $ 0 $ & $ 3 $ & -$\tfrac{1}{8} $ & $ 3 $ & $ \tfrac{11}{8} $ & -$\tfrac{33}{8} $ & $ 9 $ & $ 
-\tfrac{13}{2} $ & $ \tfrac{15}{8} $ & $ 0 $ & $ z_4, z_6 $ \\ $
 1 $ & $ 1 $ & $ 3 $ & $ 1 $ & $ 1 $ & -$\tfrac{1}{2} $ & -$\tfrac{1}{2} $ & $ 0 $ & $ 1 $ & $ 1 $ & $ n_1 $ & $ 1 $ & $ 0 $ & $ 1 
$ & $ 0 $ & $ \tfrac{3}{2} $ & $ \tfrac{1}{4} $ & $ 9 $ & $ \tfrac{23}{2} $ & -$\tfrac{9}{2} $ & $ 
-\tfrac{21}{4} $ & -$\tfrac{21}{4} $ & -$\tfrac{5}{4} $ & $ 3 $ & $ x_4, x_5 $ \\ $
 1 $ & $ 1 $ & $ 3 $ & $ 1 $ & $ 1 $ & $ 0 $ & -$\tfrac{1}{2} $ & -$\tfrac{1}{2} $ & $ 1 $ & $ 0 $ & $ 0 $ & $ 0 $ & $ 1 $ & $ 0 
$ & $ 0 $ & $ 2 $ & $ \tfrac{1}{8} $ & $ 2 $ & -$\tfrac{91}{8} $ & -$\tfrac{17}{8} $ & $ 7 $ & $ 7 $ & $ 
-\tfrac{25}{8} $ & $ 0 $ & $ x_1 $ \\ $
 1 $ & $ 1 $ & $ 1 $ & $ 1 $ & $ 1 $ & $ 0 $ & -$\tfrac{1}{2} $ & -$\tfrac{1}{2} $ & $ 1 $ & $ 0 $ & $ 0 $ & $ 1 $ & $ 0 $ & $ 0 
$ & $ 0 $ & $ 0 $ & -$\tfrac{3}{8} $ & $ \tfrac{15}{2} $ & -$\tfrac{61}{8} $ & $ \tfrac{43}{8} $ & $ 
\tfrac{37}{4} $ & $ \tfrac{37}{4} $ & $ \tfrac{45}{8} $ & $ 0 $ & $ \phi _1 $ \\ $
 1 $ & $ 1 $ & $ 1 $ & $ 1 $ & $ 1 $ & $ 0 $ & -$\tfrac{1}{2} $ & -$\tfrac{1}{2} $ & $ 1 $ & $ 0 $ & $ 0 $ & $ 1 $ & $ 0 $ & $ 0 
$ & $ 0 $ & -$4 $ & -$\tfrac{1}{8} $ & $ \tfrac{7}{2} $ & -$\tfrac{115}{8} $ & $ \tfrac{45}{8} $ & $ 
-\tfrac{3}{4} $ & -$\tfrac{3}{4} $ & $ \tfrac{15}{8} $ & $ 0 $ & $ \phi _2 $ \\ $
 1 $ & $ 1 $ & $ 1 $ & $ 1 $ & $ 1 $ & $ 0 $ & -$\tfrac{1}{2} $ & -$\tfrac{1}{2} $ & $ 1 $ & $ 0 $ & $ 0 $ & $ 1 $ & $ 1 $ & $ 1 
$ & $ 0 $ & $ 0 $ & $ \tfrac{1}{8} $ & $ \tfrac{15}{2} $ & -$\tfrac{61}{8} $ & -$\tfrac{177}{8} $ & $ 
-\tfrac{9}{4} $ & -$\tfrac{9}{4} $ & $ \tfrac{45}{8} $ & $ 0 $ & $ \phi _3 $ \\ $
 1 $ & $ 1 $ & $ 1 $ & $ 1 $ & $ 1 $ & $ 0 $ & $ 0 $ & -$1 $ & $ 0 $ & $ 0 $ & $ 0 $ & $ 0 $ & $ 0 $ & $ 0 $ & $ 0 $ & $ 1 $ & $ 
-\tfrac{1}{2} $ & $ 1 $ & $ \tfrac{55}{2} $ & $ 0 $ & $ 0 $ & $ 0 $ & $ 0 $ & $ 0 $ & $ \phi _4 $ \\ $
 1 $ & $ 1 $ & $ 1 $ & $ 1 $ & $ 1 $ & $ 0 $ & $ 0 $ & -$1 $ & $ 0 $ & $ 0 $ & $ 0 $ & $ 0 $ & $ 0 $ & $ 0 $ & $ 0 $ & -$1 $ & $ 
\tfrac{1}{2} $ & -$1 $ & -$\tfrac{55}{2} $ & $ 0 $ & $ 0 $ & $ 0 $ & $ 0 $ & $ 0 $ & $ \phi _5 $ \\ $
 1 $ & $ 1 $ & $ 1 $ & $ 1 $ & $ 1 $ & -$1 $ & $ 0 $ & $ 0 $ & $ 0 $ & $ 0 $ & $ 0 $ & $ 0 $ & $ 0 $ & $ 0 $ & $ 0 $ & $ 2 $ & $ 
-\tfrac{1}{2} $ & $ 2 $ & -$4 $ & $ \tfrac{39}{2} $ & $ 0 $ & $ 0 $ & $ 0 $ & $ 0 $ & $ \phi _6 $ \\ $
 1 $ & $ 1 $ & $ 1 $ & $ 1 $ & $ 1 $ & -$1 $ & $ 0 $ & $ 0 $ & $ 0 $ & $ 0 $ & $ 0 $ & $ 0 $ & $ 0 $ & $ 0 $ & $ 0 $ & -$2 $ & $ \tfrac{1}{2} $ & -$2 $ & $ 4 $ & -$\tfrac{39}{2} $ & $ 0 $ & $ 0 $ & $ 0 $ & $ 0 $ & $ \phi _7 $ \\ $
 1 $ & $ 1 $ & $ 1 $ & $ 1 $ & $ 1 $ & $ 0 $ & -$\tfrac{1}{2} $ & -$\tfrac{1}{2} $ & $ 1 $ & $ 0 $ & $ 0 $ & $ 0 $ & $ 1 $ & $ 0 
$ & $ 0 $ & $ 2 $ & $ \tfrac{1}{8} $ & -$28 $ & -$\tfrac{83}{8} $ & -$\tfrac{1}{8} $ & $ 5 $ & $ 5 $ & $ 
\tfrac{15}{8} $ & $ 0 $ & $ \phi _8 $ \\ $
 1 $ & $ 1 $ & $ 1 $ & $ 1 $ & $ 1 $ & -$\tfrac{1}{2} $ & $ 0 $ & -$\tfrac{1}{2} $ & $ 0 $ & $ 1 $ & $ n_1 $ & $ 0 $ & $ 0 $ & $ 0 
$ & $ 0 $ & $ \tfrac{5}{2} $ & $ \tfrac{1}{4} $ & $ \tfrac{5}{2} $ & -$5 $ & -$10 $ & $ 10 $ & $ 10 $ & $ 
\tfrac{15}{4} $ & $ 0 $ & $ \phi _{10}, \phi _9 $ \\ $
 1 $ & $ 1 $ & $ 1 $ & $ 1 $ & $ 1 $ & -$\tfrac{1}{2} $ & $ 0 $ & -$\tfrac{1}{2} $ & $ 0 $ & $ 1 $ & $ n_1 $ & $ 0 $ & $ 1 $ & $ 1 
$ & $ 0 $ & $ \tfrac{5}{2} $ & $ 0 $ & $ \tfrac{5}{2} $ & -$\tfrac{79}{4} $ & $ \tfrac{7}{4} $ & $ 
-\tfrac{23}{2} $ & -$\tfrac{23}{2} $ & $ 0 $ & $ 0 $ & $ \phi _{11}, \phi _{12} $ \\ $
 1 $ & $ 1 $ & $ 1 $ & $ 1 $ & $ 1 $ & -$\tfrac{1}{2} $ & $ 0 $ & -$\tfrac{1}{2} $ & $ 0 $ & $ 1 $ & $ n_1 $ & $ 0 $ & $ 1 $ & $ 1 
$ & $ 0 $ & -$\tfrac{3}{2} $ & $ 0 $ & -$\tfrac{3}{2} $ & -$\tfrac{47}{4} $ & $ \tfrac{71}{4} $ & $ 
\tfrac{23}{2} $ & $ \tfrac{23}{2} $ & $ 0 $ & $ 0 $ & $ \phi _{13}, \phi _{14} $ \\ $
 1 $ & $ 1 $ & $ 1 $ & $ 1 $ & $ 1 $ & -$\tfrac{1}{2} $ & -$\tfrac{1}{2} $ & $ 0 $ & $ 1 $ & $ 1 $ & $ n_1 $ & $ 1 $ & $ 0 $ & $ 0 
$ & $ 0 $ & -$\tfrac{1}{2} $ & -$\tfrac{3}{8} $ & $ 7 $ & $ \tfrac{65}{8} $ & -$\tfrac{35}{8} $ & $ 
\tfrac{37}{4} $ & $ \tfrac{37}{4} $ & $ \tfrac{45}{8} $ & $ 0 $ & $ \phi _{15}, \phi _{16} $ \\ $
 1 $ & $ 1 $ & $ 1 $ & $ 1 $ & $ 1 $ & -$\tfrac{1}{2} $ & -$\tfrac{1}{2} $ & $ 0 $ & $ 1 $ & $ 1 $ & $ n_1 $ & $ 1 $ & $ 0 $ & $ 0 
$ & $ 0 $ & -$\tfrac{9}{2} $ & -$\tfrac{1}{8} $ & $ 3 $ & $ \tfrac{11}{8} $ & -$\tfrac{33}{8} $ & $ 
-\tfrac{3}{4} $ & -$\tfrac{3}{4} $ & $ \tfrac{15}{8} $ & $ 0 $ & $ \phi _{17}, \phi _{18} $ \\ $
 1 $ & $ 1 $ & $ 1 $ & $ 1 $ & $ 1 $ & -$\tfrac{1}{2} $ & -$\tfrac{1}{2} $ & $ 0 $ & $ 1 $ & $ 1 $ & $ n_1 $ & $ 0 $ & $ 0 $ & $ 0 
$ & $ 0 $ & $ \tfrac{5}{2} $ & $ \tfrac{1}{4} $ & $ \tfrac{5}{2} $ & $ \tfrac{39}{4} $ & $ \tfrac{39}{2} 
$ & $ 0 $ & $ 0 $ & $ 0 $ & $ 1 $ & $ n_1, n_2 $ \\ $
 1 $ & $ 1 $ & $ 1 $ & $ 1 $ & $ 1 $ & $ 0 $ & -$\tfrac{1}{2} $ & -$\tfrac{1}{2} $ & $ 1 $ & $ 0 $ & $ 0 $ & $ 0 $ & $ 0 $ & $ 0 
$ & $ 0 $ & $ 2 $ & $ \tfrac{1}{4} $ & $ 2 $ & $ \tfrac{51}{2} $ & $ \tfrac{39}{4} $ & $ 0 $ & $ 0 $ & $ 0 $ & $ 1 $ & $ n_3 $ \\ $
 1 $ & $ 1 $ & $ 1 $ & $ 1 $ & $ 1 $ & $ 0 $ & -$1 $ & $ 0 $ & $ 0 $ & $ 0 $ & $ 0 $ & $ 0 $ & $ 0 $ & $ 0 $ & $ 0 $ & -$1 $ & $ 0 $ & -$1 $ & $ 
\tfrac{63}{2} $ & -$\tfrac{39}{2} $ & $ 0 $ & $ 0 $ & $ 0 $ & $ 2 $ & $ \overline{\phi }_1 $ \\ $
 1 $ & $ 1 $ & $ 1 $ & $ 1 $ & $ 1 $ & $ 0 $ & -$\tfrac{1}{2} $ & -$\tfrac{1}{2} $ & $ 1 $ & $ 0 $ & $ 0 $ & $ 0 $ & $ 1 $ & $ 0 $ & $ 0 $ & $ 1 $ & $ \tfrac{1}{8} $ & $ 31 $ & -$\tfrac{83}{8} $ & -$\tfrac{1}{8} $ & $ 5 $ & $ 5 $ & $ 
\tfrac{15}{8} $ & $ 2 $ & $ \overline{\phi }_2 $ \\ $
 1 $ & $ 1 $ & $ 1 $ & $ 1 $ & $ 1 $ & $ 0 $ & -$\tfrac{1}{2} $ & -$\tfrac{1}{2} $ & $ 1 $ & $ 0 $ & $ 0 $ & $ 1 $ & $ 0 $ & $ 0 
$ & $ 0 $ & $ 2 $ & -$\tfrac{1}{8} $ & -$\tfrac{11}{2} $ & $ \tfrac{29}{8} $ & $ \tfrac{113}{8} $ & $ 
-\tfrac{37}{4} $ & -$\tfrac{37}{4} $ & -$\tfrac{45}{8} $ & $ 2 $ & $ \overline{\phi }_3 $ \\ $
 1 $ & $ 1 $ & $ 1 $ & $ 1 $ & $ 1 $ & $ 0 $ & -$1 $ & $ 0 $ & $ 0 $ & $ 0 $ & $ 0 $ & $ 0 $ & $ 0 $ & $ 0 $ & $ 0 $ & $ 1 $ & $ 0 $ & $ 1 $ & $ 
-\tfrac{63}{2} $ & $ \tfrac{39}{2} $ & $ 0 $ & $ 0 $ & $ 0 $ & $ 2 $ & $ \overline{\phi }_4 $ \\ $
 1 $ & $ 1 $ & $ 1 $ & $ 1 $ & $ 1 $ & $ 0 $ & -$\tfrac{1}{2} $ & -$\tfrac{1}{2} $ & $ 1 $ & $ 0 $ & $ 0 $ & $ 1 $ & $ 1 $ & $ 1 $ & $ 0 $ & $ 2 $ & $ \tfrac{1}{8} $ & -$\tfrac{11}{2} $ & $ \tfrac{147}{8} $ & $ \tfrac{19}{8} $ & $ 
\tfrac{49}{4} $ & $ \tfrac{49}{4} $ & -$\tfrac{15}{8} $ & $ 2 $ & $ \overline{\phi }_5 $ \\ $
 1 $ & $ 1 $ & $ 1 $ & $ 1 $ & $ 1 $ & $ 0 $ & -$\tfrac{1}{2} $ & -$\tfrac{1}{2} $ & $ 1 $ & $ 0 $ & $ 0 $ & $ 1 $ & $ 1 $ & $ 1 $ & $ 0 $ & -$2 $ & $ \tfrac{3}{8} $ & -$\tfrac{19}{2} $ & $ \tfrac{93}{8} $ & $ \tfrac{21}{8} $ & $ 
\tfrac{9}{4} $ & $ \tfrac{9}{4} $ & -$\tfrac{45}{8} $ & $ 2 $ & $ \overline{\phi }_6 $ \\ $
 1 $ & $ 1 $ & $ 1 $ & $ 1 $ & $ 1 $ & -$\tfrac{1}{2} $ & -$\tfrac{1}{2} $ & $ 0 $ & $ 1 $ & $ 1 $ & $ n_1 $ & $ 1 $ & $ 0 $ & $ 0 $ & $ 0 $ & $ \tfrac{3}{2} $ & -$\tfrac{1}{8} $ & -$6 $ & $ \tfrac{155}{8} $ & $ \tfrac{35}{8} $ & $ 
-\tfrac{37}{4} $ & -$\tfrac{37}{4} $ & -$\tfrac{45}{8} $ & $ 2 $ & $ \overline{\phi }_7, \overline{\phi }_8 $ \\ $
 1 $ & $ 1 $ & $ 1 $ & $ 1 $ & $ 1 $ & $ 0 $ & -$\tfrac{1}{2} $ & -$\tfrac{1}{2} $ & $ 1 $ & $ 0 $ & $ 0 $ & $ 0 $ & $ 1 $ & $ 1 $ & $ 0 $ & $ 2 $ & -$\tfrac{1}{4} $ & -$13 $ & -$\tfrac{7}{2} $ & $ \tfrac{27}{4} $ & $ 3 $ & $ 3 $ & $ 
-\tfrac{15}{2} $ & $ 3 $ & $ \overline{n}_1 $ \\ $
 1 $ & $ 1 $ & $ 1 $ & $ 1 $ & $ 1 $ & $ 0 $ & -$\tfrac{1}{2} $ & -$\tfrac{1}{2} $ & $ 1 $ & $ 0 $ & $ 0 $ & $ 0 $ & $ 1 $ & $ 1 
$ & $ 0 $ & $ 0 $ & -$\tfrac{1}{4} $ & $ 15 $ & -$\tfrac{1}{2} $ & $ \tfrac{51}{4} $ & -$3 $ & -$3 $ & $ 
\tfrac{15}{2} $ & $ 3 $ & $ \overline{n}_2 $ \\ $
 1 $ & $ 1 $ & $ 1 $ & $ 1 $ & $ 1 $ & $ 0 $ & -$\tfrac{1}{2} $ & -$\tfrac{1}{2} $ & $ 1 $ & $ 0 $ & $ 0 $ & $ 1 $ & $ 0 $ & $ 1 
$ & $ 0 $ & $ 2 $ & $ \tfrac{1}{4} $ & -$\tfrac{41}{2} $ & -$\tfrac{13}{4} $ & $ \tfrac{29}{4} $ & $ -\tfrac{29}{4} $ & -$\tfrac{29}{4} $ & $ \tfrac{15}{4} $ & $ 3 $ & $ \overline{n}_3 $ \\ $
 1 $ & $ 1 $ & $ 1 $ & $ 1 $ & $ 1 $ & $ 0 $ & -$\tfrac{1}{2} $ & -$\tfrac{1}{2} $ & $ 1 $ & $ 0 $ & $ 0 $ & $ 1 $ & $ 1 $ & $ 0 
$ & $ 0 $ & $ 0 $ & -$\tfrac{1}{4} $ & $ \tfrac{45}{2} $ & -$\tfrac{3}{4} $ & -$\tfrac{61}{4} $ & $ 
-\tfrac{17}{4} $ & -$\tfrac{17}{4} $ & -$\tfrac{15}{4} $ & $ 3 $ & $ \overline{n}_4 $ \\ $
 1 $ & $ 1 $ & $ 1 $ & $ 1 $ & $ 1 $ & -$\tfrac{1}{2} $ & $ 0 $ & -$\tfrac{1}{2} $ & $ 0 $ & $ 1 $ & $ n_1 $ & $ 0 $ & $ 0 $ & $ 1 
$ & $ 0 $ & $ \tfrac{5}{2} $ & -$\tfrac{1}{8} $ & $ \tfrac{35}{2} $ & $ \tfrac{15}{8} $ & $ 
-\tfrac{25}{8} $ & $ 8 $ & $ 8 $ & -$\tfrac{45}{8} $ & $ 3 $ & $ \overline{n}_5, \overline{n}_6 $ \\ $
 1 $ & $ 1 $ & $ 1 $ & $ 1 $ & $ 1 $ & -$\tfrac{1}{2} $ & -$\tfrac{1}{2} $ & $ 0 $ & $ 1 $ & $ 1 $ & $ n_1 $ & $ 1 $ & $ 0 $ & $ 1 $ & $ 0 $ & $ \tfrac{3}{2} $ & $ \tfrac{1}{4} $ & -$21 $ & $ \tfrac{25}{2} $ & -$\tfrac{5}{2} $ & $ 
-\tfrac{29}{4} $ & -$\tfrac{29}{4} $ & $ \tfrac{15}{4} $ & $ 3 $ & $ \overline{n}_7, \overline{n}_8$
\end{longtable}

\end{landscape}

\end{appendix}

\clearpage
\thispagestyle{empty}

\addcontentsline{toc}{chapter}{Bibliography}
\bibliographystyle{TitleAndArxiv}
\bibliography{dissertation}

\end{document}